\documentclass[review]{elsarticle}

\usepackage{lineno,hyperref}
\modulolinenumbers[5]

\journal{CAGD - New Progress in Geometry for Applications}

\usepackage{color,soul}
\usepackage{amsmath}
\usepackage{amssymb}
\usepackage{mathtools}
\usepackage{algorithmic}
\usepackage{algorithm}
\usepackage[marginclue,footnote,silent,draft]{fixme}
\usepackage{t1enc,dfadobe}
\usepackage{cite}

\usepackage{url}       
\usepackage{amsmath}
\usepackage{amssymb}
\usepackage[skins,breakable]{tcolorbox}
\newtcolorbox{mybox}[1][]{enhanced jigsaw,breakable,pad at break=1mm,
  oversize,left=8mm,interior hidden,colframe=black,nobeforeafter=,#1}

\usepackage[marginclue,footnote,silent,draft]{fixme}

\begin{document}
\begin{frontmatter}

\title{Laplacian Spectral Basis Functions}
\author[G. Patan\`e]{G. Patan\`e\\ CNR-IMATI, Italy}
\cortext[]{G. Patan\'e, Consiglio Nazionale delle Ricerche, Istituto di Matematica Applicata e Tecnologie Informatiche, Via De Marini 6, 16149 Genova, Italy}\ead{patane@ge.imati.cnr.it}

\begin{abstract}
Representing a signal as a linear combination of a set of basis functions is central in a wide range of applications, such as approximation, de-noising, compression, shape correspondence and comparison. In this context, our paper addresses the main aspects of signal approximation, such as the definition, computation, and comparison of basis functions on arbitrary 3D shapes. Focusing on the class of basis functions induced by the Laplace-Beltrami operator and its spectrum, we introduce the diffusion and Laplacian spectral basis functions, which are then compared with the harmonic and Laplacian eigenfunctions. As main properties of these basis functions, which are commonly used for numerical geometry processing and shape analysis, we discuss the partition of the unity and non-negativity; the intrinsic definition and invariance with respect to shape transformations (e.g., translation, rotation, uniform scaling); the locality, smoothness, and orthogonality; the numerical stability with respect to the domain discretisation; the computational cost and storage overhead. Finally, we consider geometric metrics, such as the area, conformal, and kernel-based norms, for the comparison and characterisation of the main properties of the Laplacian basis functions. 
%
\end{abstract}  
\end{frontmatter}
%

%

%
%
\section{Introduction\label{sec:INTRODUCTION}}
In Computer Graphics, scalar functions are ubiquitous to represent the values of a physical phenomenon (e.g., the heat that diffuses from one or more source points), a shape descriptor (e.g., curvature, spherical harmonics), a distance (e.g., geodesic, bi-harmonic, diffusion distance) from a set of source points, or the pixels of a surface texture. 

In all these cases, representing the input function in terms of a basis allows us to address a large number of applications, such as \emph{multi-resolution representations} by selecting a set of multi-scale basis functions (e.g., Laplacian eigenfunctions, diffusion basis functions), \emph{sparse representations} by choosing a low number of basis functions in order to achieve a target approximation accuracy, or \emph{compression} by quantising the representation coefficients. We can also address \emph{deformation} by modifying the coefficients that express the geometry of the input surface in terms of geometry-driven or shape-intrinsic basis functions, \emph{smoothing} by neglecting the coefficients associated with large Laplacian frequencies, and the definition of \emph{Laplacian spectral kernels and distances} as a filtered combination of the Laplacian spectral eigenfunctions.

\begin{figure}
\centering
\includegraphics[height=210pt]{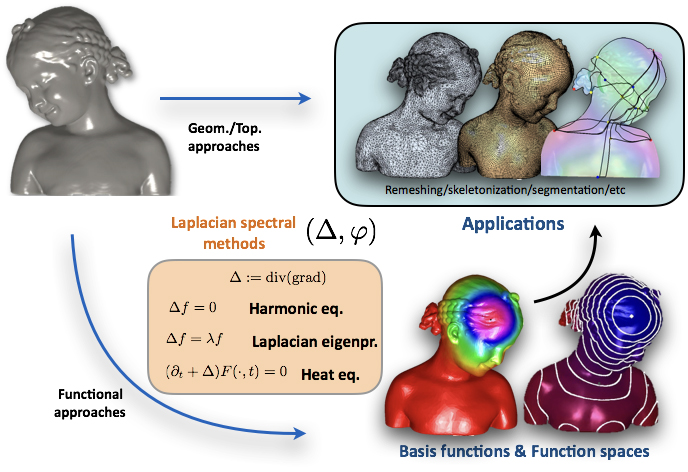}
\caption{Overview of the proposed approach for the definition of Laplacian spectral basis functions based on the solution to the harmonic equation, the Laplacian eigenproblem, and the diffusion equation, and on the filtering of the Laplacian spectrum.\label{fig:SPECTRAL-BASIS-OVERVIEW}}
\end{figure}

In this context, our paper focuses on the main aspects of signal approximation on arbitrary 3D shapes, such as the definition, computation, and comparison of basis functions for applications in numerical geometry processing and shape analysis (Fig.~\ref{fig:SPECTRAL-BASIS-OVERVIEW}). To define the functional space associated with an input 3D shape, we select a \emph{set of its basis functions} and then represent any signal as a linear combination of these functions. In this work, we focus on the class of bases induced by the Laplace-Beltrami operator and its spectrum; i.e., the \emph{harmonic basis functions} (Sect.~\ref{sec:HARMONIC-MAPS}), as solution to the Laplace equation; the \emph{Laplacian eigenfunctions} (Sect.~\ref{sec:LAPLACIAN-EIGENMAPS}) associated with the spectrum of the Laplace-Beltrami operator; and the \emph{Laplacian spectral basis functions} (Sect.~\ref{sec:LAPL-SPECT-KER-DIST}), which are defined by filtering the Laplacian spectrum and include the \emph{diffusion basis functions}.
%

The diffusion and Laplacian spectral functions are novel classes of functions, which will be defined by further developing our recent results on the Laplacian spectral distances~\citep{PATANE2016,PATANE-STAR2016,PATANE2014-PRL}. Then, these functions will be compared with the harmonic functions and the Laplacian eigenfunctions. As a main contribution with respect to the previous work, the diffusion and Laplacian spectral basis functions are intrinsic to the input shape and local; i.e., they have a compact support that can be tuned easily by selecting the diffusion scale and the decay of the filter function to zero, respectively.

Defining basis functions with a compact support is particularly important to localise their behaviour around feature points and to reduce their storage overhead in the discrete setting. In fact, given a discrete domain~$\mathcal{M}$ with~$n$ points, the space of functions on~$\mathcal{M}$ has dimension~$n$ and its basis functions are encoded as a \mbox{$n\times n$} matrix. This matrix can be stored only if we work with compactly-supported or analytically defined (e.g., radial basis or polynomial) functions, or if we select a subset of~$k$ basis functions, with~$k$ much smaller than~$n$.

For the \emph{comparison of functions} on the same surface (Sect.~\ref{sec:SAME-SHAPE-COMPARISON}), we review different metrics (e.g., area, conformal, kernel-based metrics), which measure differential properties of the input scalar function and geometric properties of the underlying domain. As main \emph{properties}, we discuss the partition of the unity and non-negativity; the intrinsic definition and invariance with respect to shape transformations (e.g., translation, rotation, uniform scaling); the locality, smoothness, and orthogonality; the numerical stability with respect to the domain discretisation; the computational cost and storage overhead. 

Our experiments (Sect.~\ref{sec:GENERAL-EXAMPLES-DISCUSSION},~\ref{sec:CONCLUSION}) show that the diffusion and the Laplacian spectral functions provide a valid alternative to the harmonic functions and Laplacian eigenfunctions. In fact, the diffusion basis functions can be centred at any point of the input domain, have a compact support, and encode local/global shape details according to the values of the temporal parameter or to the decay of the selected filter to zero. Similar properties also apply to the Laplacian spectral basis functions induced by a filter with a strong (e.g., exponential) decay to zero.
%
%
%
%
\section{Laplace-Beltrami operator\label{sec:LAPLACE-BELTRAMI}}
Let~$\mathcal{N}$ be a smooth surface, possibly with boundary, equipped with a Riemannian metrics and let us consider the \emph{inner product} \mbox{$\langle f,g\rangle_{2}:=\int_{\mathcal{N}}f(\mathbf{p})g(\mathbf{p})\textrm{d}\mathbf{p}$} defined on the space \mbox{$\mathcal{L}^{2}(\mathcal{N})$} of square integrable functions on~$\mathcal{N}$ and the corresponding norm \mbox{$\|\cdot\|_{2}$}. The \emph{Laplace-Beltrami operator} \mbox{$\Delta:=-\textrm{div}(\textrm{grad})$} is \emph{self-adjoint} \mbox{$\langle\Delta f,g\rangle_{2}=\langle f,\Delta g\rangle_{2}$}, and \emph{positive semi-definite}; i.e., \mbox{$\langle\Delta f,f\rangle_{2}\geq 0$}, \mbox{$\forall f,g$},~\citep{ROSENBERG1997}. Finally, the value \mbox{$\Delta f(\mathbf{p})$} does not depend on \mbox{$f(\mathbf{q})$}, for any couple of distinct points~$\mathbf{p}$,~$\mathbf{q}$ (\emph{locality}). 

According to~\citep{PATANE-STAR2016}, we represent the Laplace-Beltrami operator on surface and volume meshes in a unified way as~$\mbox{$\tilde{\mathbf{L}}:=\mathbf{B}^{-1}\mathbf{L}$}$, where the \emph{mass matrix}~$\mathbf{B}$ is sparse, symmetric, positive definite, and the \emph{stiffness matrix}~$\mathbf{L}$ is sparse, symmetric, positive semi-definite, and \mbox{$\mathbf{L}\mathbf{1}=\mathbf{0}$}. Analogously to the continuous case, the Laplacian matrix satisfies the following properties:
\begin{itemize}
\item$\mathbf{B}$-\emph{adjointness}:~$\tilde{\mathbf{L}}$ is adjoint with respect to the~$\mathbf{B}$-inner product \mbox{$\langle\mathbf{f},\mathbf{g}\rangle_{\mathbf{B}}:=\mathbf{f}^{\top}\mathbf{B}\mathbf{g}$}; i.e., \mbox{$\langle\tilde{\mathbf{L}}\mathbf{f},\mathbf{g}\rangle_{\mathbf{B}}=\langle\mathbf{f},\tilde{\mathbf{L}}\mathbf{g}\rangle_{\mathbf{B}}
=\mathbf{f}^{\top}\mathbf{L}\mathbf{g}$}. If \mbox{$\mathbf{B}:=\mathbf{I}$}, then this property reduces to the symmetry of~$\mathbf{L}$;
\item \emph{positive semi-definiteness}: \mbox{$\langle\tilde{\mathbf{L}}\mathbf{f},\mathbf{f}\rangle_{\mathbf{B}}=\mathbf{f}^{\top}\mathbf{L}\mathbf{f}\geq 0$}. In particular, the Laplacian eigenvalues are positive, with one \emph{null eigenvalue} associated with a constant eigenvector;
\item \emph{locality}: assuming that~$\mathbf{B}$ is diagonal, the value \mbox{$(\tilde{\mathbf{L}}\mathbf{f})_{i}$} depends only on the~$f$-values at~$\mathbf{p}_{i}$ and its~$1$-star neighbourhood.
\end{itemize}
On \emph{triangle meshes}, for the stiffness matrix we can consider the linear FEM Laplacian weights~\citep{reuter:cad06,VALLET2008} that reduce to the \emph{Voronoi-cotangent weights}~\citep{DESBRUN1999}, by lumping the FEM mass matrix, and extend the cotangent weights~\citep{PINKALL1993}. The \emph{mean-value weights}~\citep{FLOATER2003} have been derived from the mean-value theorem for harmonic functions and are always positive. In~\citep{CHUANG2009}, the weak formulation of the Laplacian eigenproblem is achieved by selecting a set of volumetric test functions, which are defined as \mbox{$k\times k\times k$} B-splines (e.g., \mbox{$k:=4$}) and restricted to the input shape. For the \emph{anisotropic Laplacian}~\citep{ANDREUX2014}, the entries of~$\mathbf{L}$ are a variant of the cotangent weights (i.e., with respect to different angles) and the entries of the diagonal mass matrix are the areas of the Voronoi regions. 

On \emph{polygonal meshes}, the Laplacian discretisation in~\citep{ALEXA2011,HERHOLZ2015} generalises the Laplacian matrix with cotangent weights to surface meshes with non-planar, non-convex faces. An approximation of the Laplace-Beltrami operator with point-wise convergence has been proposed in~\citep{BELKIN2008-MESH}. 

In~\citep{BELKIN2003,BELKIN2006,BELKIN2008,BELKIN2009}, the Laplace-Beltrami operator on a \emph{point set} has been defined as the Gram matrix associated with the exponential kernel. Starting from this approach, a new discretisation~\citep{LIU2012} has been achieved through a finer approximation of the local geometry of the surface at each point through its Voronoi cell. The discretisation of the Laplace-Beltrami operator on volumes represented as a tetrahedral mesh has been addressed in~\citep{ALLIEZ2005,LIAO2009,TONG2003}.

Finally, in the paper examples we apply the linear FEM weights and the level-sets of a given function are associated with iso-values uniformly sampled in its range.

\section{Harmonic basis\label{sec:HARMONIC-MAPS}}
Given a set \mbox{$\mathcal{S}:=\{\mathbf{p}_{i}\}_{i\in\mathcal{I}}$} of seed points (Sect.~\ref{sec:SEED-SELECTION}), we compute the \emph{harmonic basis functions} \mbox{$\mathcal{B}:=\{\psi_{i}\}_{i\in\mathcal{I}}$}, as solution to the harmonic equation \mbox{$\Delta\psi_{i}=0$}, \mbox{$\psi_{i}(\mathbf{p}_{j})=\delta_{ij}$}, \mbox{$i,j\in\mathcal{I}$}. Each harmonic function, and in particular any harmonic basis function,
\begin{itemize}
\item minimizes the \emph{Dirichlet energy} \mbox{$\mathcal{E}(u):=\int_{\mathcal{N}}\|\nabla u(\mathbf{p})\|_{2}^{2}\textrm{d}\mathbf{p}$};
\item satisfies the \emph{locality property}; i.e., if~$\mathbf{p}$ and~$\mathbf{q}$ are two distinct points, then \mbox{$\Delta u(\mathbf{p})$} is not affected by the value of~$u$ at~$\mathbf{q}$;
\item verifies the \emph{mean-value theorem} 
\begin{equation*}
u(\mathbf{p})=(2\pi R)^{-1}\int_{\Gamma}u(s)\textrm{d}s=(\pi R^{2})^{-1}\int_{\mathcal{B}}u(\mathbf{q})\textrm{d}\mathbf{q},
\end{equation*}
where \mbox{$\mathcal{B}\subseteq\mathcal{N}$} is a disc of center~$\mathbf{p}$, radius~$R$, and boundary~$\Gamma$ (\emph{mean-value theorem});
\item satisfies the \emph{strong maximum principle}~\citep{ROSENBERG1997}. More precisely, if~$\Omega$ is a connected open set, \mbox{$u\in\mathcal{C}^{2}(\Omega)$},~$u$ is harmonic and attains either a global minimum or maximum in \mbox{$\Omega$}, then~$u$ is constant. 
%
%
\end{itemize}
The maximum principle states that the values of a harmonic function in a compact domain are bounded by its maximum and minimum values on the boundary. In particular, the values of the harmonic basis function~$\psi_{i}$ belong to the interval \mbox{$[0,1]$} and its maximum is attained at the selected seed point~$\mathbf{p}_{i}$.
\begin{figure}[t]
\centering
\begin{tabular}{ccc}
\includegraphics[height=150pt]{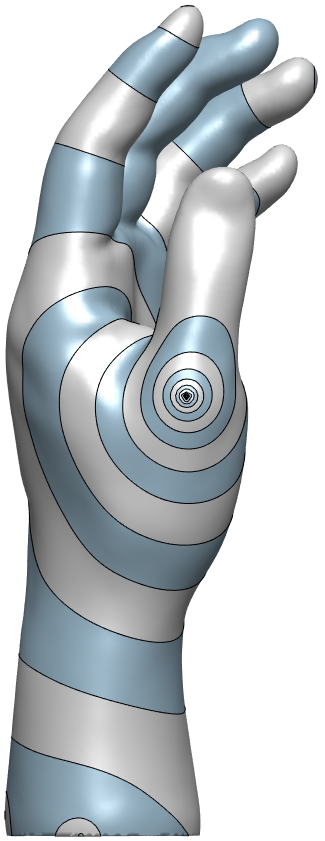}
&\includegraphics[height=150pt]{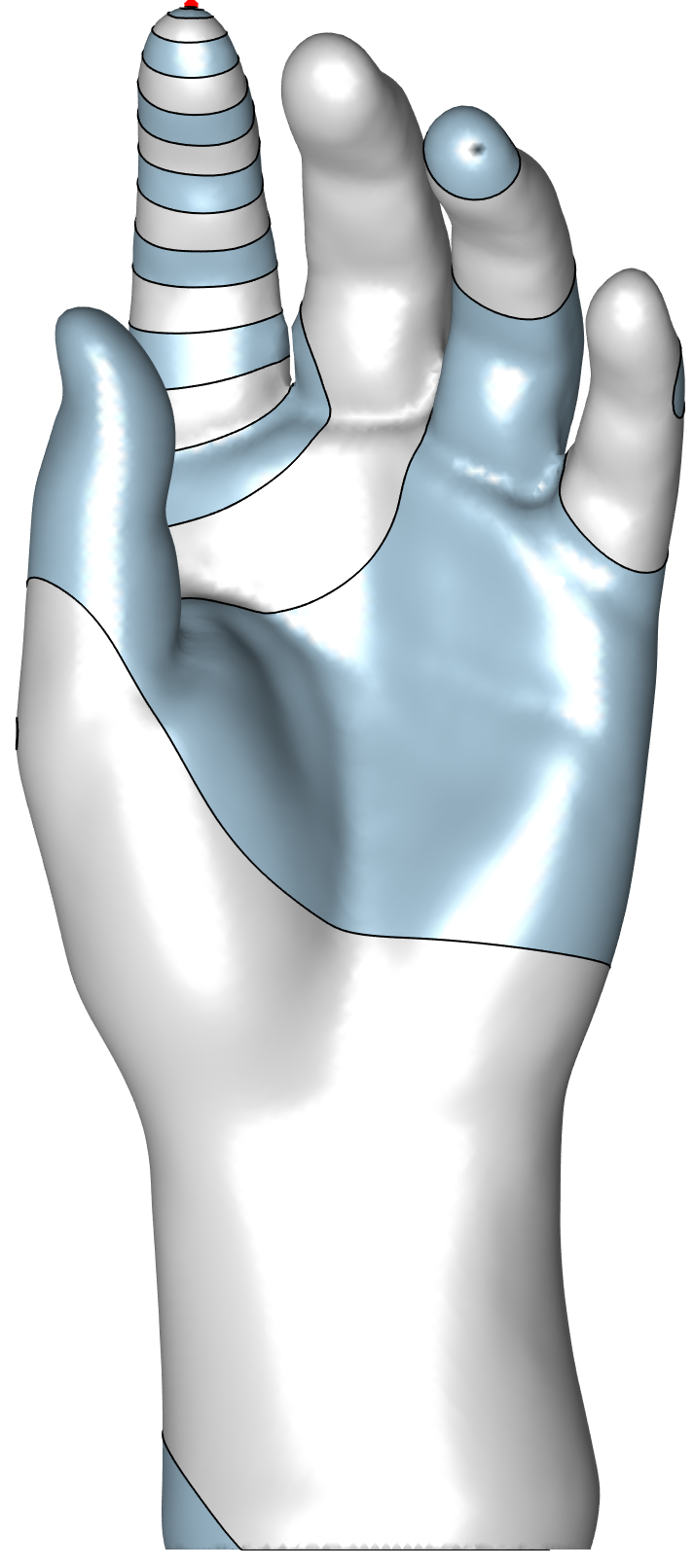}
&\includegraphics[height=150pt]{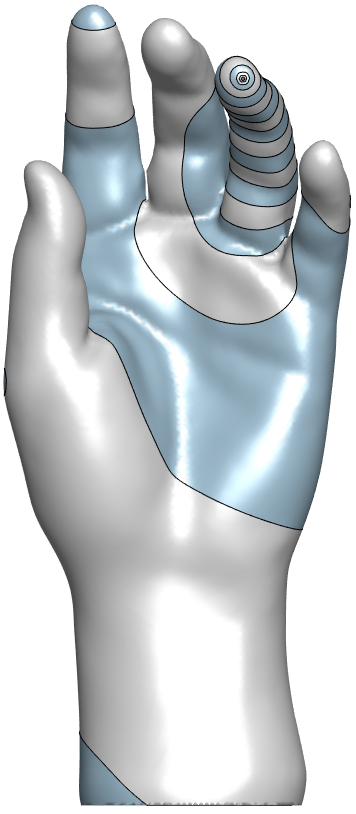}\\
\end{tabular}
\caption{Distribution and shape of the level-sets of harmonic basis functions, centred at seed points.\label{fig:3D-DIFFUSIVE-COMPARISON}}
\end{figure}

The discrete harmonic basis associated with a seed point~$\mathbf{p}_{i}$ is the solution to the linear system \mbox{$\mathbf{L}^{\star}\mathbf{g}=\mathbf{e}_{i}$}, where~$\mathbf{L}^{\star}$ is achieved by replacing the~$i$-th row of the stiffness matrix with the vector~$\mathbf{e}_{i}$, where \mbox{$\mathbf{e}_{i}^{(j)}:=\delta_{ij}$} (Fig.~\ref{fig:3D-DIFFUSIVE-COMPARISON}). Harmonic functions are efficiently computed in \mbox{$\mathcal{O}(n)$} time with iterative solvers of sparse linear systems; their computation is stable for the mean-value weights while negative Voronoi cotangent weights might induce local undulations in the resulting harmonic function. Finally, the (globally-supported) \emph{geometry-aware basis}~\citep{SORKINE2005} has been defined by minimising the Dirichlet energy, with interpolating and least-squares constraints.

\paragraph*{Pre-conditioners and fast solvers of elliptic problems}
In the context of the iso-geometric analysis of elliptic PDEs, previous work has proposed (i) \emph{additive multilevel preconditioners}~\citep{BUFFA2013}, whose spectral conditioning number is independent of the grid spacing, and (ii) \emph{preconditioning methods} based on the solution of a Sylvester-like equation~\citep{SANGALLI2016}, which are robust with respect to the mesh size and the spline degree. In the context of Computer Graphics, specialised pre-conditioners for the Laplacian matrix~\citep{KRISHNAN2013} can be applied to further attenuate numerical instabilities.

As solvers, we can select (i) \emph{robust and efficient multi-grid methods} for single-patch iso-geometric discretisations of elliptic problems~\citep{HOFREITHER2017}, based on tensor product B-splines of maximum smoothness, and (ii) \emph{multi-grid methods}~\citep{DONATELLI2017} for the Galerkin iso-geometric analysis based on  B-splines and with optimal convergence rate (i.e., bounded independently of the matrix size). As alternatives, we mention the \emph{fast solvers for large linear systems} arising from the Galerkin approximation based on B-splines~\citep{DONATELLI2015}, which are linear with respect to the matrix size and have a convergence speed that is independent of the matrix size, the spline degree, and the dimensionality of the input problem. 

For the analysis of the main spectral properties (e.g., non-singularity, conditioning, spectral distribution of the eigenvalue) of the stiffness matrix related to the iso-geometric analysis of elliptic problems, we refer the reader to~\citep{GARONI2014}.

\section{Laplacian and Hamiltonian eigenbasis\label{sec:LAPLACIAN-EIGENMAPS}}
Recalling that the Laplace-Beltrami operator is self-adjoint and positive semi-definite (Sect.~\ref{sec:LAPLACE-BELTRAMI}), it has an \emph{orthonormal eigensystem} \mbox{$\mathcal{B}:=\{(\lambda_{n},\phi_{n})\}_{n=0}^{+\infty}$}, \mbox{$\Delta\phi_{n}=\lambda_{n}\phi_{n}$}, with \mbox{$0=\lambda_{0}\leq\lambda_{n}\leq\lambda_{n+1}$}, in \mbox{$\mathcal{L}^{2}(\mathcal{N})$} (Sect.~\ref{sec:LAPLACE-BELTRAMI}). Any function~$f$ in \mbox{$\mathcal{L}^{2}(\mathcal{N})$} satisfies the following relations 
\begin{equation*}
\left\{
\begin{array}{ll}
f=\sum_{n=0}^{+\infty}\alpha_{n}\phi_{n},
&\alpha_{n}:=\langle f,\phi_{n}\rangle_{2};\\
\|f\|_{2}^{2}=\sum_{n=0}^{+\infty}\alpha_{n}^{2},
&\|\nabla f\|_{2}^{2}=\sum_{n=0}^{+\infty}\alpha_{n}^{2}\lambda_{n}.
\end{array}
\right.
\end{equation*}
The Laplacian eigenfunctions are intrinsic to the input shape and those ones related to smaller eigenvalues correspond to smooth and slowly-varying functions. Increasing the eigenvalues, the corresponding eigenfunctions generally show rapid oscillations. If two shapes are isometric, then they have the same Laplacian spectrum (\emph{iso-spectral property}); however, the viceversa does not hold~\citep{GORDON2002,ZENG2012} and we cannot recover the metric of a given surface. 
\begin{figure}[t]
\centering
\begin{tabular}{ccc}
$\phi_{1}$\includegraphics[height=150pt]{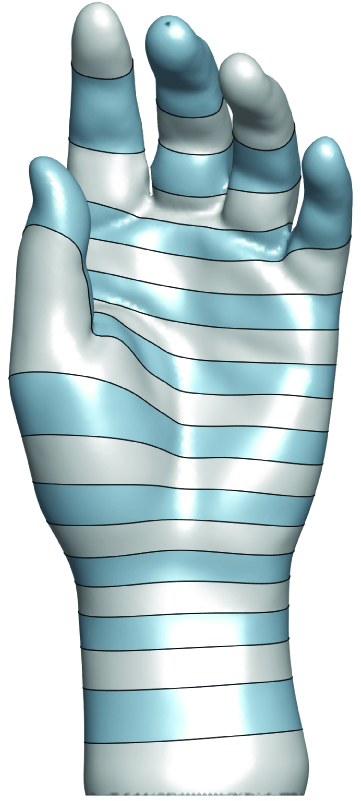}
&$\phi_{2}$\includegraphics[height=150pt]{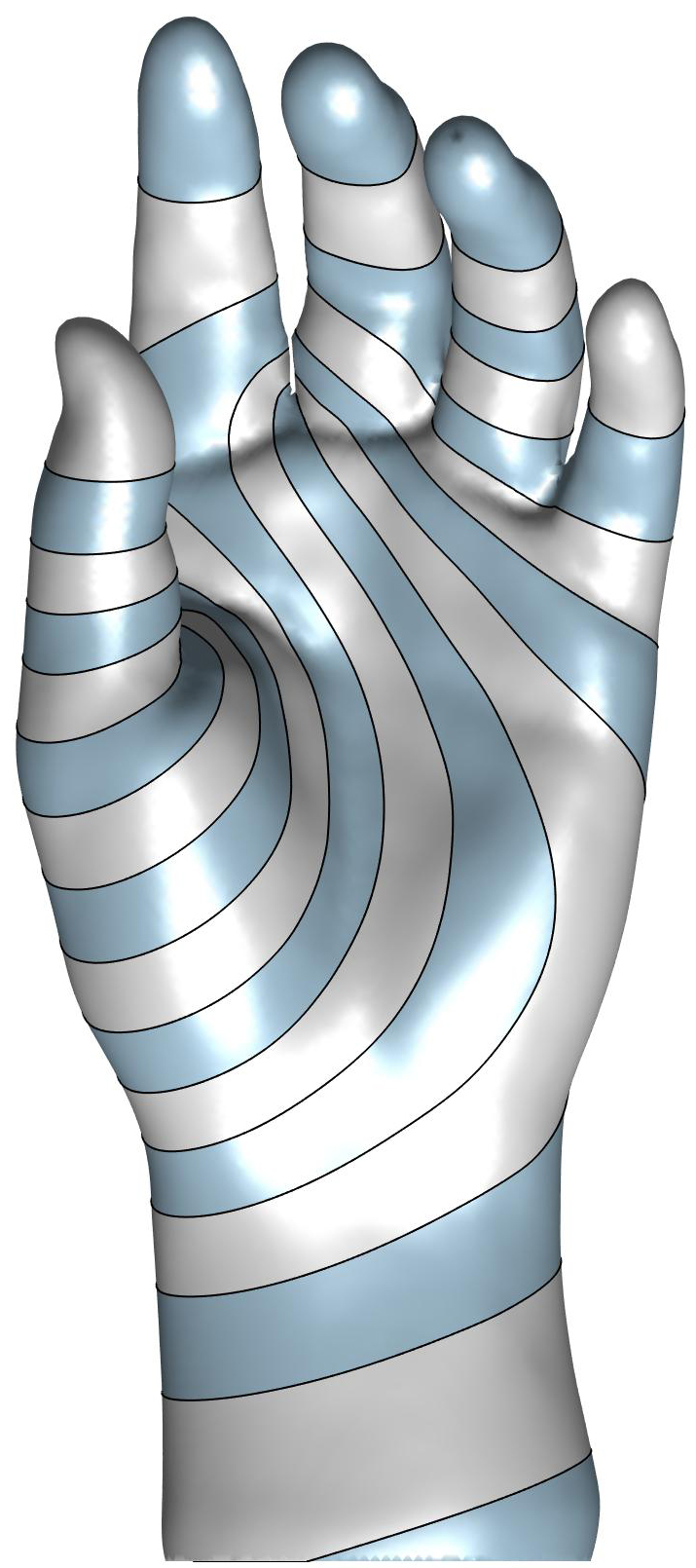}
&$\phi_{3}$\includegraphics[height=150pt]{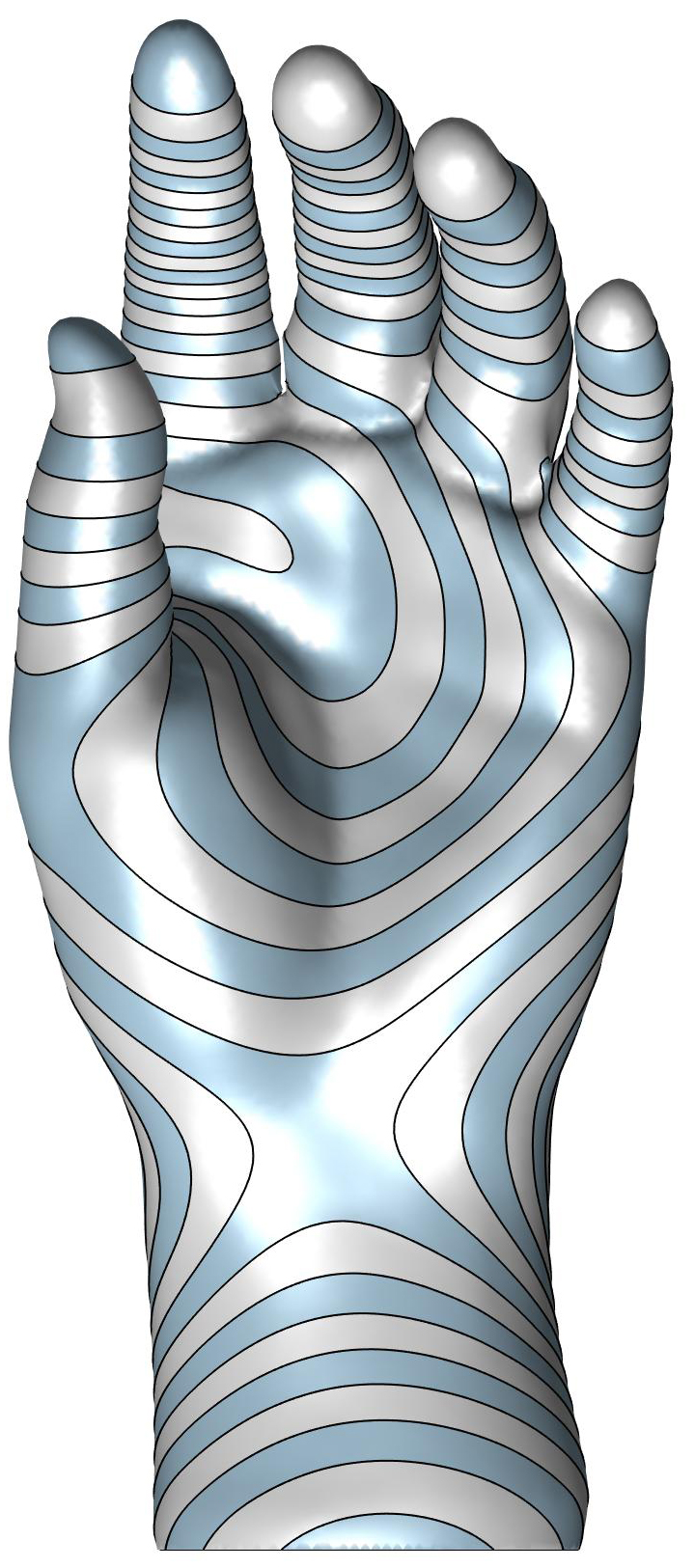}
\end{tabular}
\caption{Level-sets of three Laplacian eigenfunctions.\label{fig:LAPLACIANS}}
\end{figure}
\paragraph*{Hamiltonian basis and eigenbasis}
A \emph{Hamiltonian operator} \mbox{$\mathcal{H}:\mathcal{L}^{2}(\mathbb{R})\rightarrow\mathbb{R}$} is defined as \mbox{$\mathcal{H}f=-\Delta f+\mu V f$}, where \mbox{$V:\mathcal{N}\rightarrow\mathbb{R}$} is a potential function and~$\mu$ is a real parameter that controls the influence of the potential on the Hamiltonian operator and defines the trade-off between the local and global support of the corresponding eigenfunctions. Analogously to the harmonic basis and Laplacian eigenbasis, we can define the \emph{Hamiltonian basis}, as the solution to the problem \mbox{$\mathcal{H}\psi_{i}=0$}, \mbox{$\psi_{i}(\mathbf{p}_{j})=\delta_{ij}$}, and the \emph{Hamiltonian eigenbasis} \mbox{$\mathcal{H}\psi_{n}=\sigma_{n}\psi_{n}$}, \mbox{$n\in\mathbb{N}$}.
\begin{figure*}[t]
\centering
(a)\includegraphics[height=105pt]{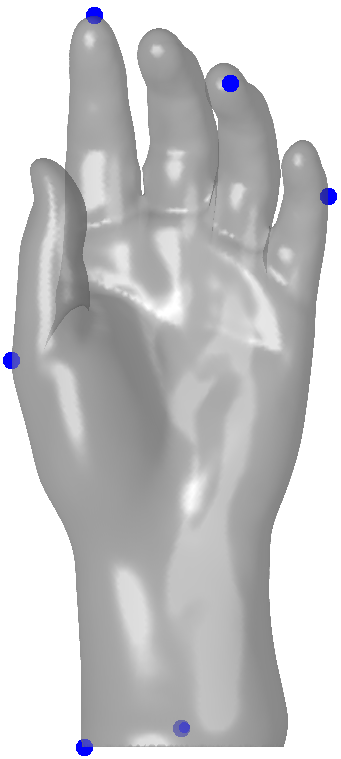}
\begin{tabular}{ccccccc}
(b)\includegraphics[height=105pt]{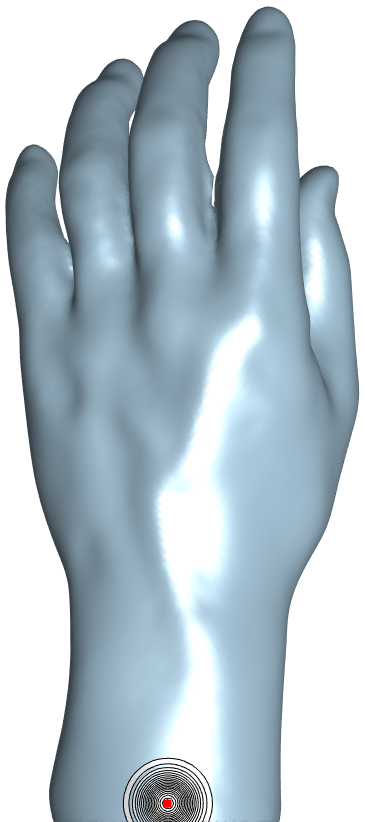}
&\includegraphics[height=105pt]{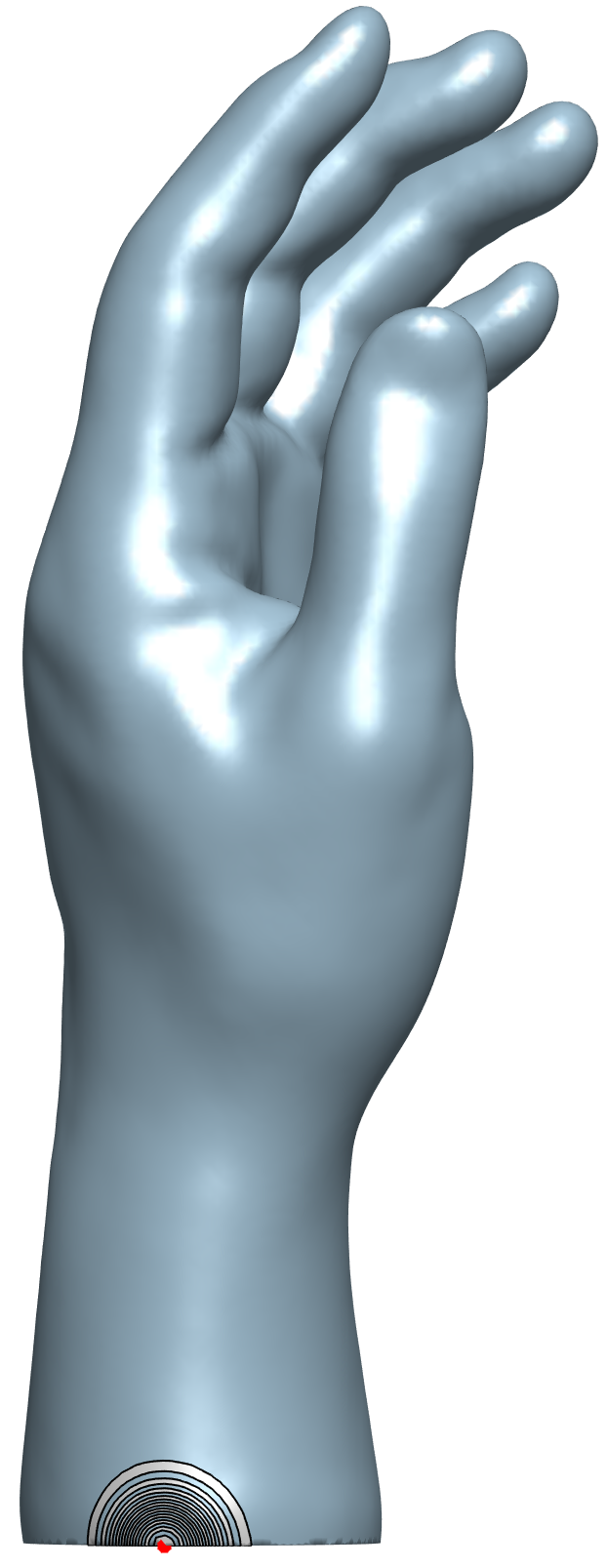}
&\includegraphics[height=105pt]{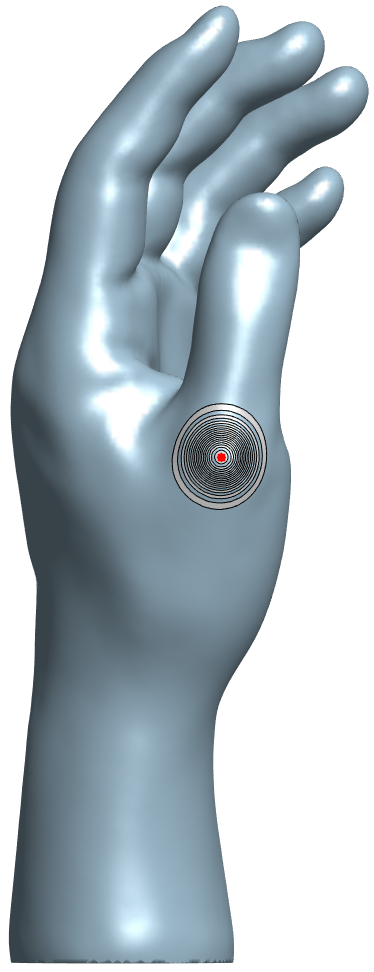}
&\includegraphics[height=105pt]{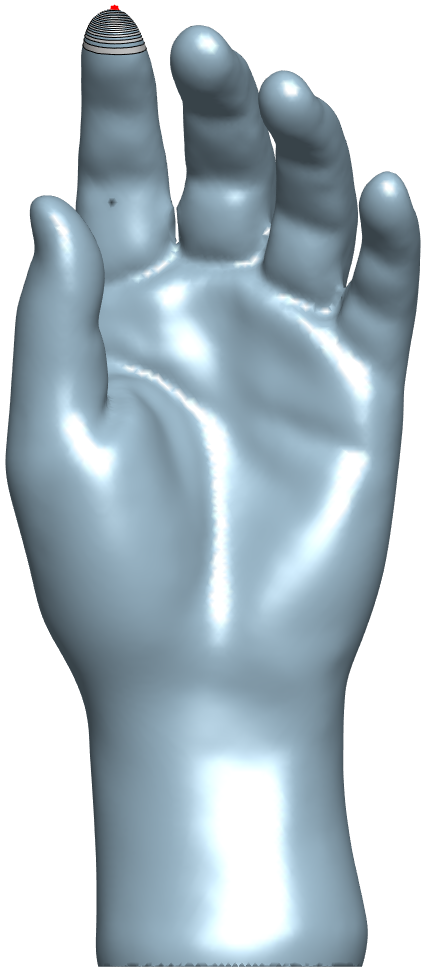}
&\includegraphics[height=105pt]{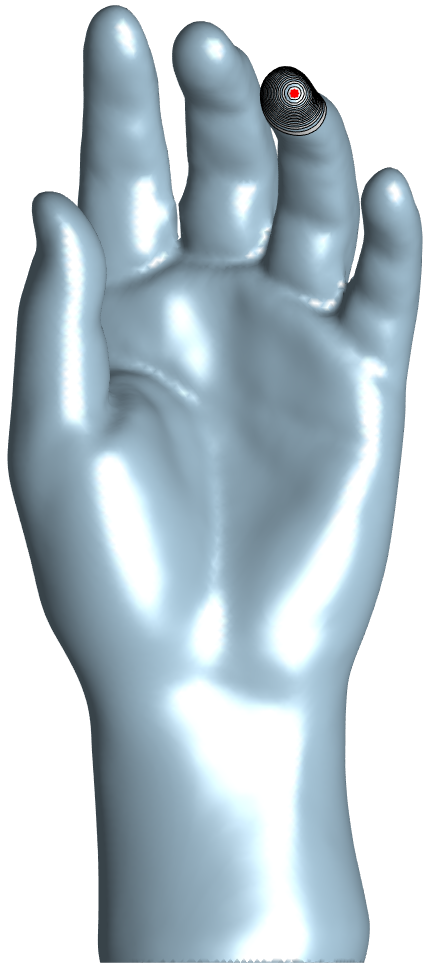}
&\includegraphics[height=105pt]{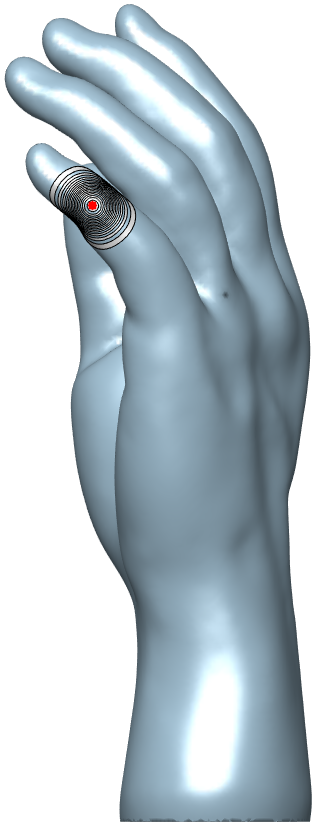}\\
\hline
(c)\includegraphics[height=105pt]{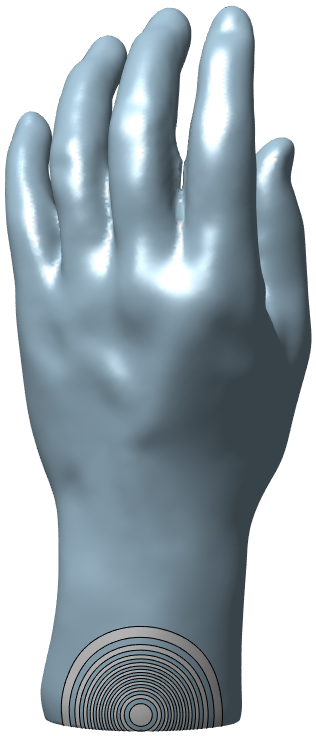}
&\includegraphics[height=105pt]{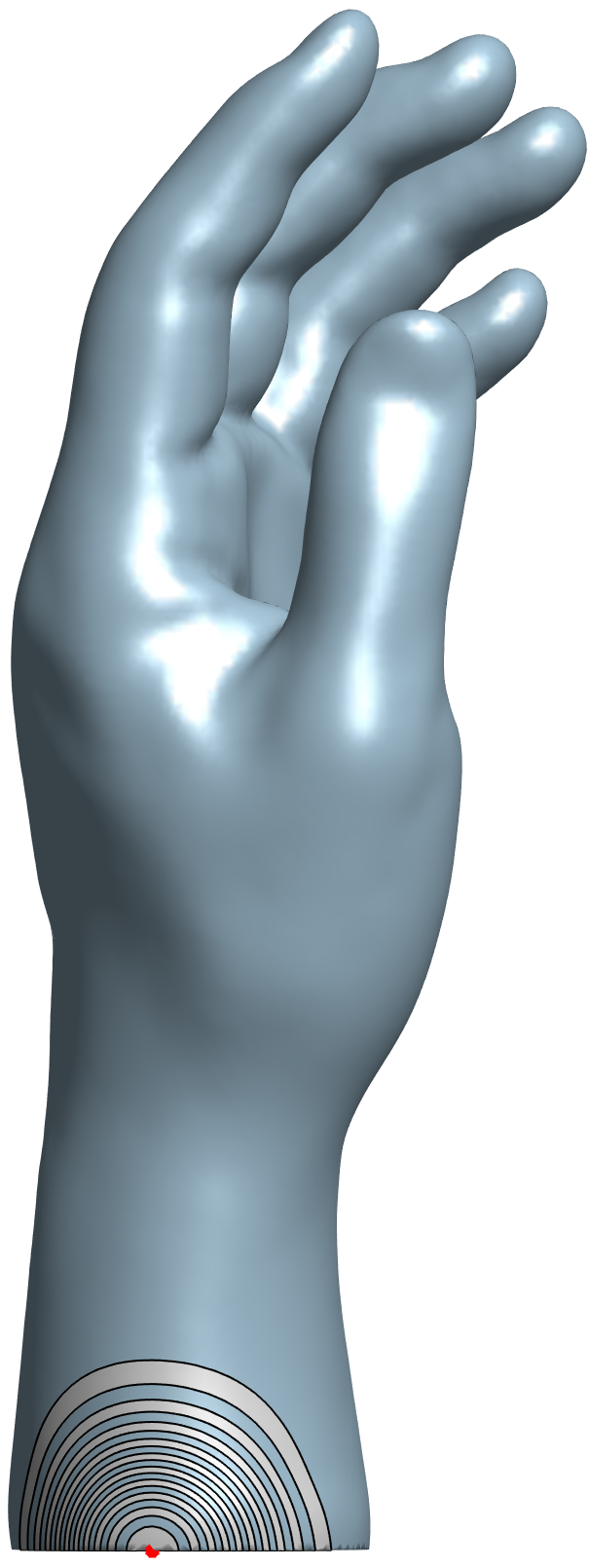}
&\includegraphics[height=105pt]{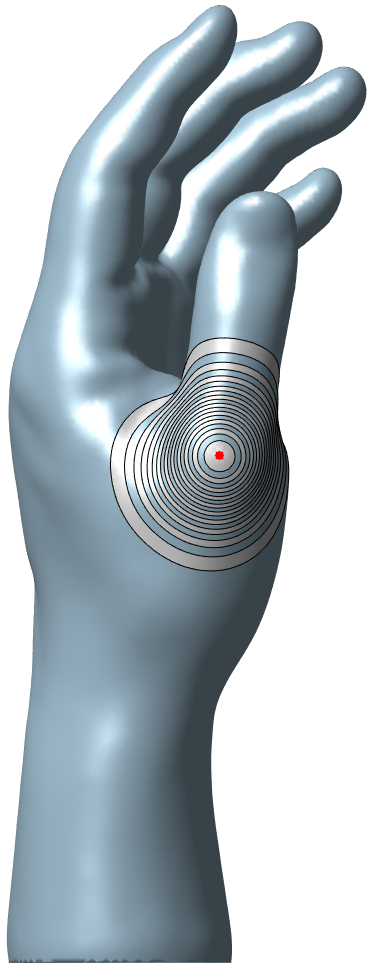}
&\includegraphics[height=105pt]{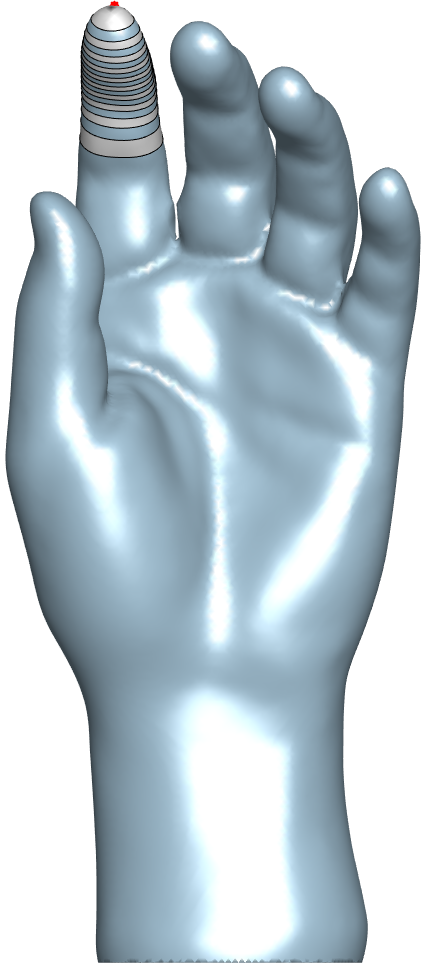}
&\includegraphics[height=105pt]{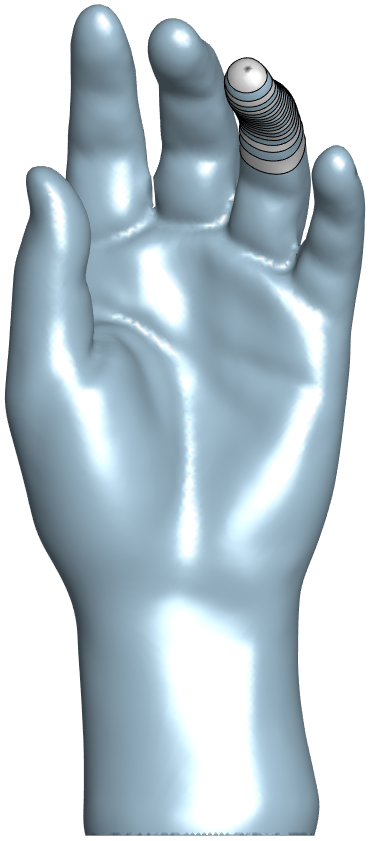}
&\includegraphics[height=105pt]{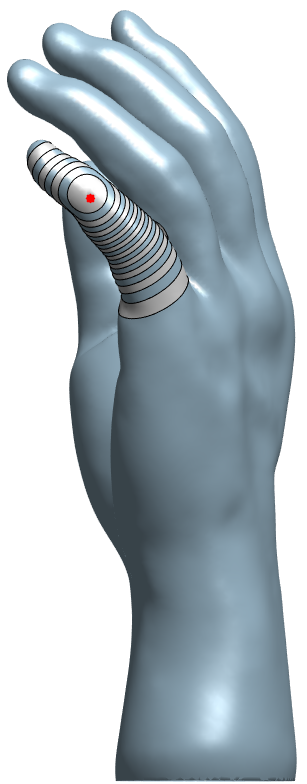}\\
\end{tabular}
\caption{(b,c) Level-sets of the diffusion basis functions at different scales and centered at different seed points in (a). The example confirms the locality, smoothness, and shape-awareness of the diffusion basis functions, computed with the spectrum-free method (Sect.~\ref{sec:SPEC-DIST-COMP}).\label{fig:HEAT-DIFFUSION}}
\end{figure*}
\paragraph*{Optimality of the Laplacian eigenbasis}
According to~\citep{AFLALO2015}, the Laplacian eigenfunctions represent an optimal basis for the representation of signals with bounded gradient magnitude. In fact,
\begin{itemize}
\item the spectral decomposition \mbox{$f=\sum_{i=0}^{n}\langle f,\phi_{i}\rangle_{2}\phi_{i}$} is optimal in approximating functions with~$\mathcal{L}^{2}$ bounded gradient magnitude; i.e., the residual error \mbox{$r_{n}:=f-f_{n}$} is bounded as
\begin{equation}\label{eq:LS-SPECTRAL-ERROR}
\|r_{n}\|_{2}^{2}
\leq\frac{\|\nabla f\|_{2}^{2}}{\lambda_{n+1}};
\end{equation}
\item the spectral decomposition is optimal in approximating functions with respect to the error estimates in (\ref{eq:LS-SPECTRAL-ERROR}). In fact, for any \mbox{$0\leq\alpha<1$} there is no integer~$n$ and no sequence \mbox{$(\psi_{i})_{i=0}^{n}$} of linearly independent functions in~$\mathcal{L}^{2}$ such that
\begin{equation*}\label{eq:LAPL-OPT-BOUND}
\left\|f-\sum_{i=0}^{n}\langle f,\psi_{i}\rangle_{2}\psi_{i}\right\|_{2}
\leq\alpha\frac{\|\nabla f\|_{2}}{\lambda_{n+1}},
\qquad\forall f.
\end{equation*}
%
\end{itemize}
\paragraph*{Discrete Laplacian eigenbasis}
The \emph{generalised Laplacian eigensystem} \mbox{$(\lambda_{i},\mathbf{x}_{i})_{i=1}^{n}$}, with \mbox{$0=\lambda_{1}<\lambda_{2}\leq\cdots\leq\lambda_{n}$}, satisfies the identity \mbox{$\mathbf{L}\mathbf{x}_{i}=\lambda_{i}\mathbf{B}\mathbf{x}_{i}$} and the eigenvectors are orthonormal with respect to the~$\mathbf{B}$-inner product; i.e., \mbox{$\langle\mathbf{x}_{i},\mathbf{x}_{j}\rangle_{\mathbf{B}}=\mathbf{x}_{i}^{\top}\mathbf{B}\mathbf{x}_{j}=\delta_{ij}$}. In particular, \mbox{$\tilde{\mathbf{L}}=\mathbf{X}\Gamma\mathbf{X}^{\top}\mathbf{B}$} is the \emph{spectral decomposition} of the Laplacian matrix, where~$\mathbf{X}$ is the eigenvectors' matrix and~$\Gamma$ is the diagonal matrix of the eigenvalues. Fig.~\ref{fig:LAPLACIANS} shows the level-sets of three eigenfunctions. In~\citep{BAREKAT2017,NEUMANN2014}, Laplacian eigenfunctions with a compact support (\emph{compressed manifold modes}) have been defined by solving an orthogonality-constrained optimisation problem with a~$\mathcal{L}^{1}$ penalty term. In~\citep{HUSKA2018}, the compact support of the Laplacian eigenfunctions is enforced with a~$\mathcal{L}^{p}$, \mbox{$0<p<1$}, penalisation term; then, these eigenfunctions are computed by combining splitting strategies with an ADMM-based (\emph{Alternating Direction Method of Multipliers}) iterative algorithm. Finally, the optimality of the Laplacian eigenfunctions for spectral geometry processing and for signal approximation has been addressed in~\citep{BENCHEN2005} and in~\citep{AFLALO2016-BB,AFLALO2015}, respectively.
\begin{figure}[t]
\centering
\begin{tabular}{ccc}
\includegraphics[width=110pt]{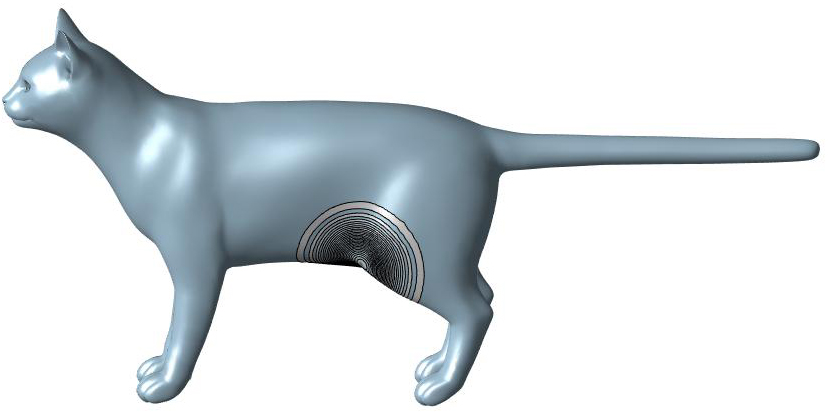}
&\includegraphics[width=110pt]{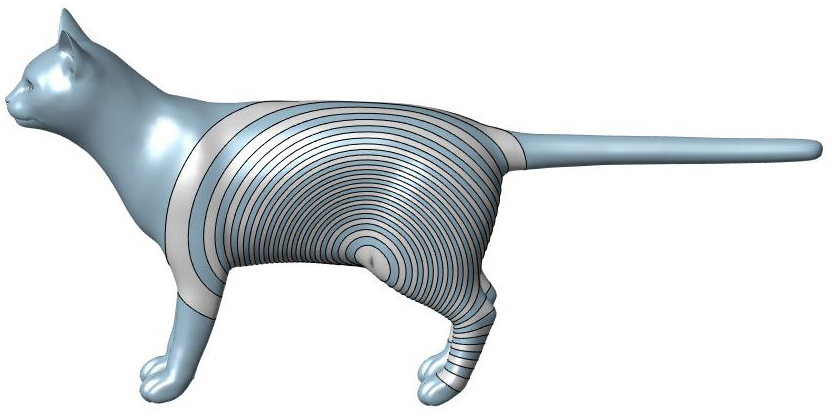}
&\includegraphics[width=110pt]{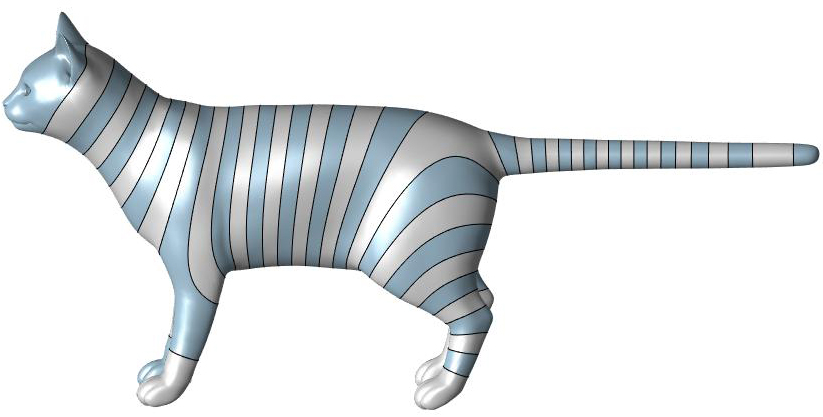}\\
\includegraphics[width=70pt]{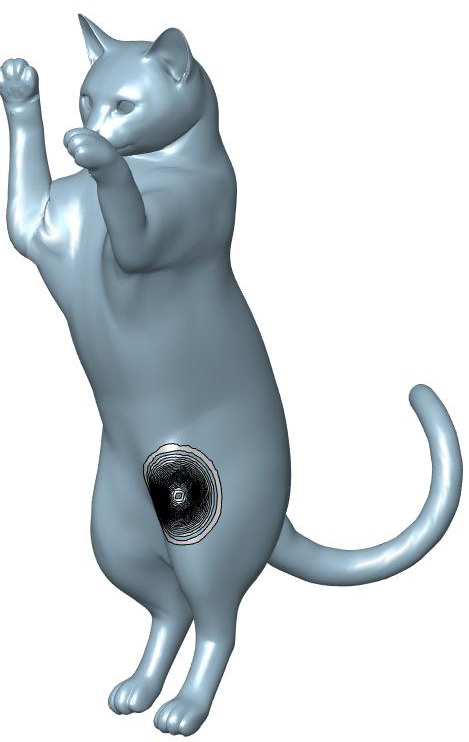}
&\includegraphics[width=70pt]{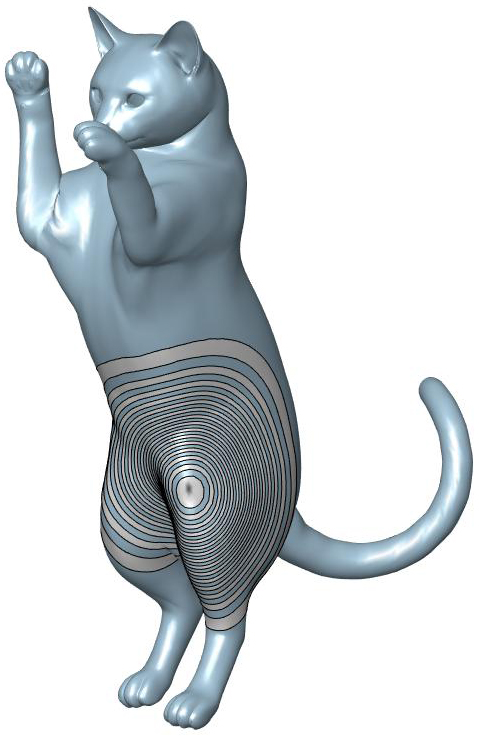}
&\includegraphics[width=70pt]{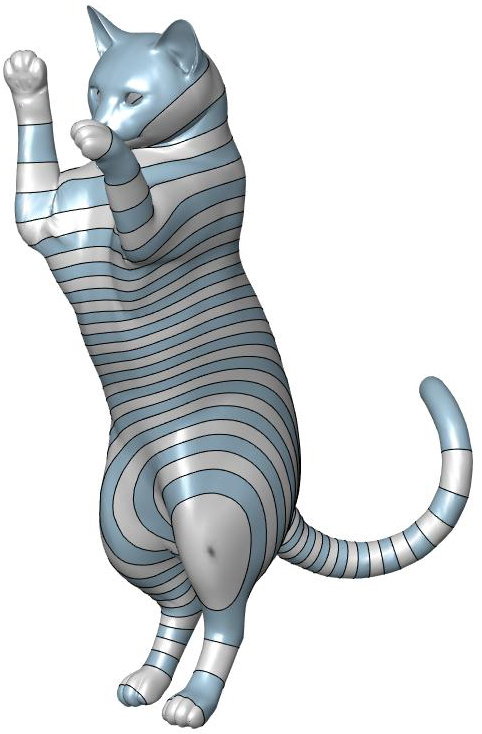}\\
$t=0.001$ &$t=0.01$ &$t=0.1$
\end{tabular}
\caption{Level-sets of diffusion basis functions at three scales and centred at the same seed point.\label{fig:CAT-DIFFUSION}}
\end{figure}
\paragraph*{Computation of the Laplacian eigenbasis}
For the computation of the Laplacian eigenfunctions, numerical methods generally exploit the sparsity of the Laplacian matrix and reduce the high-dimensional eigenproblem to one of lower dimension, by applying a coarsening step. Main examples include the algebraic multi-grid method~\citep{FALGOUT2006}, Arnoldi iterations~\citep{LEHOUCQ1996,SORENSEN1992}, and the Nystrom method~\citep{FOWLKES2004}. Even though the eigenvalues and eigenvectors are computed in super-linear time~\citep{VALLET2008}, this computational cost and the required \mbox{$\mathcal{O}(n^{2})$} storage are expensive for densely sampled domains. Indeed, modifications of the Laplacian eigenproblem are applied to evaluate specific sub-parts of the Laplacian spectrum (e.g., shift, power, and inverse methods). Finally, the Laplacian eigenvectors are computed only for a small set of eigenvalues and do not provide a flexible alignment of the function behaviour to specific shape features.

\paragraph*{Stability of the Laplacian eigenbasis}
Theoretical results on the sensitivity of the Laplacian spectrum against geometry changes, irregular sampling density and connectivity have been presented in~\citep{HILDEBRANDT2006,XU2004}. Here, we briefly recall that the instability of the computation of the Laplacian eigenbasis is generally due to repeated or close eigenvalues, with respect to the numerical accuracy of the solver of the eigen-equation~\citep{PATANE-STAR2016}. While repeated eigenvalues are quite rare and typically associated with symmetric shapes, numerically close or switched eigenvalues can be present in the spectrum, in spite of the regularity of the input discrete surface.

\section{Laplacian spectral basis\label{sec:LAPL-SPECT-KER-DIST}}
We address the definition (Sect.~\ref{sec:SPECTRAL-BASIS}) and computation (Sect.~\ref{sec:SPEC-DIST-COMP}) of the Laplacian spectral basis by filtering the Laplacian spectrum. Then, we study the relations between the Laplacian spectral basis and the Green kernel (Sect.~\ref{sec:GREEN-KERNEL}).
\begin{figure}[t]
\centering
\begin{tabular}{ccc||c}
\hline
\multicolumn{3}{c||}{\emph{Truncated  spectral approximation}}
&\multicolumn{1}{c}{\emph{P.C. approximation}}\\
\hline
(a)~$k=100$		&(b)~$k=1000$		&(c)~$k=2K$		&(d)~$\epsilon_{\infty}=10^{-6}$\\
\hline\hline
$s=10^{-1}$\includegraphics[height=85pt]{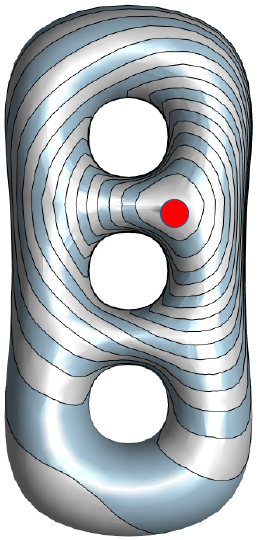}
&\includegraphics[height=85pt]{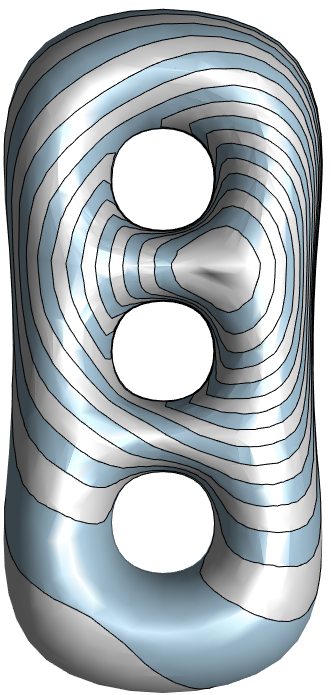}
&\includegraphics[height=85pt]{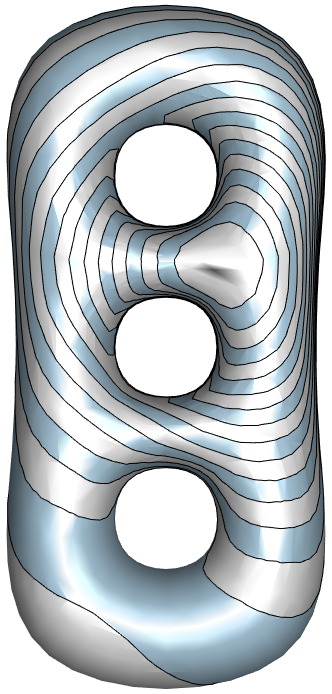}
&\includegraphics[height=85pt]{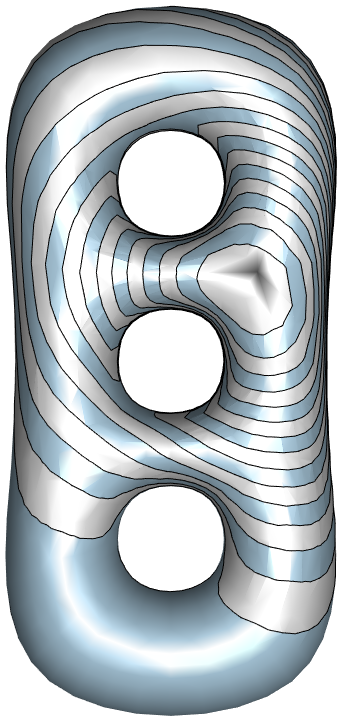}\\
\hline
$s=10^{-2}$\includegraphics[height=85pt]{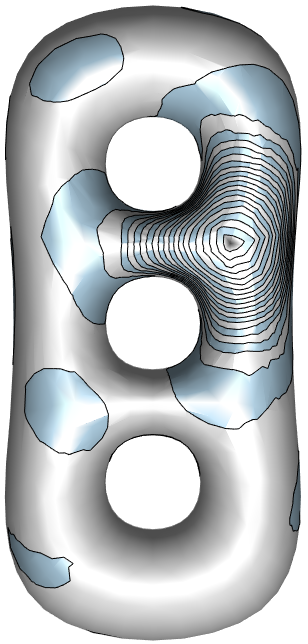}
&\includegraphics[height=85pt]{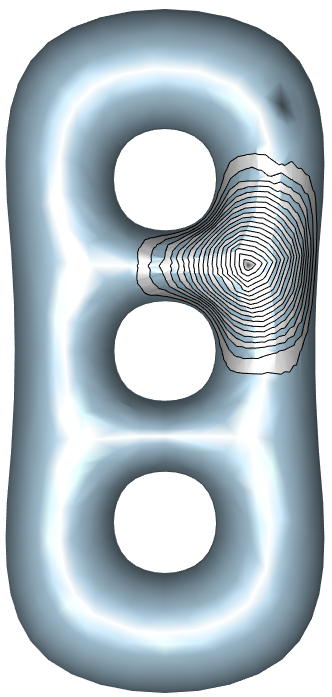}
&\includegraphics[height=85pt]{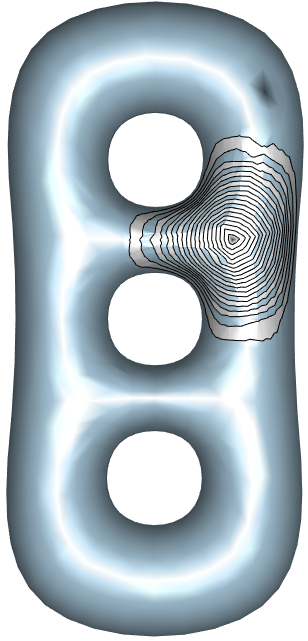}
&\includegraphics[height=85pt]{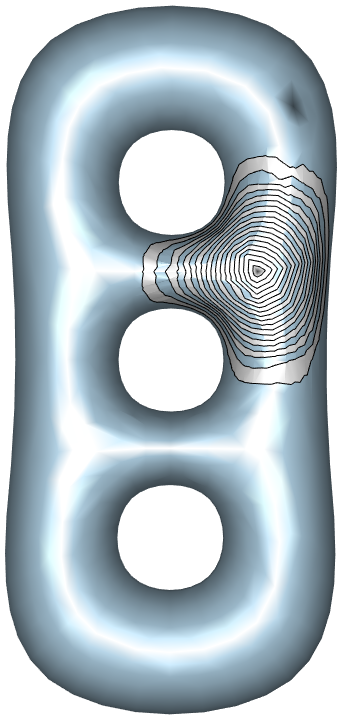}\\
\hline
$s=10^{-3}$\includegraphics[height=85pt]{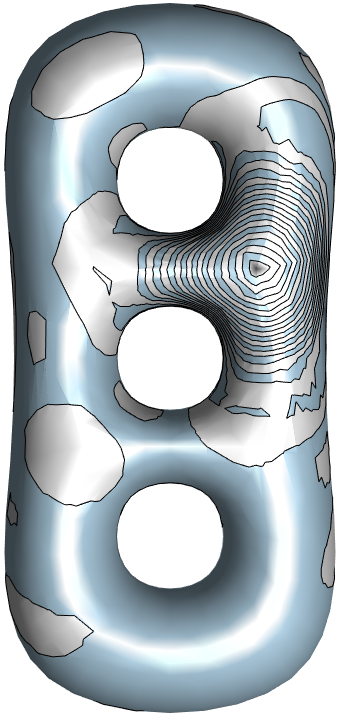}
&\includegraphics[height=85pt]{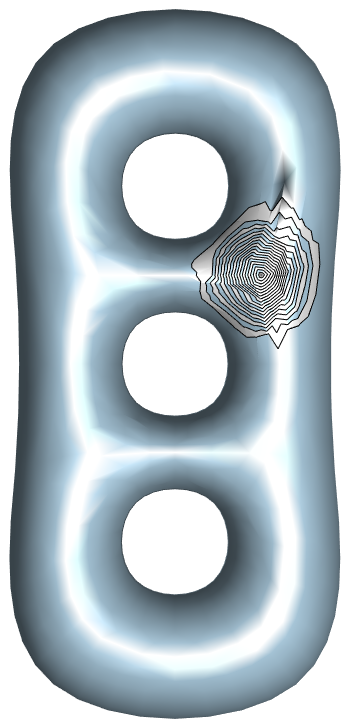}
&\includegraphics[height=85pt]{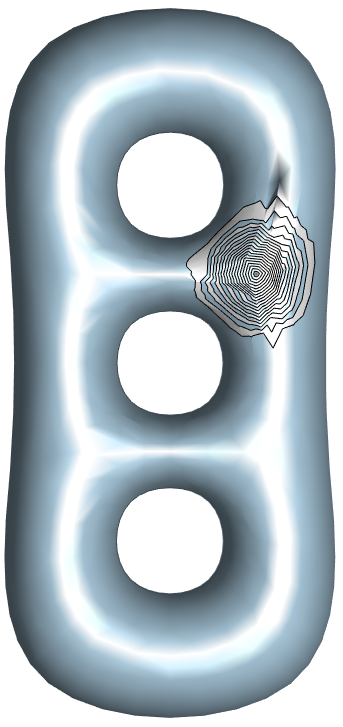}
&\includegraphics[height=85pt]{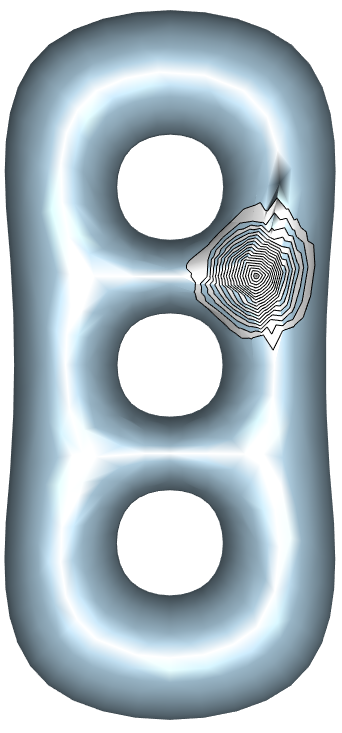}
\end{tabular}
\caption{Smoothness and locality of the diffusion basis functions at a seed point (red dot) and at different scales, which have been computed with (a-c) the truncated spectral approximation (i.e.,~$k$ Laplacian eigenpairs, linear FEM Laplacian weights) and (d) the Pad\`e-Chebyshev approximation. Since the input shape has~$2K$ vertices, the spectral approximation (c) provides the ground-truth.\label{fig:DIFFUSIVE-COMPUTATION}}
\end{figure}
\subsection{Spectral basis\label{sec:SPECTRAL-BASIS}}
Given a positive \emph{filter} \mbox{$\varphi:\mathbb{R}^{+}\rightarrow\mathbb{R}^{+}$}, let us consider the power series \mbox{$\varphi(s)=\sum_{n=0}^{+\infty}\alpha_{n}s^{n}$}. Noting that \mbox{$\Delta^{i}f=\sum_{n=0}^{+\infty}\lambda_{n}^{i}\langle f,\phi_{n}\rangle_{2}\phi_{n}$} and under a few hypotheses on the decay of the filter to zero~\citep{PATANE2016}, we define the \emph{spectral operator} as
\begin{equation}\label{eq:FUNCT-OPER}
\Phi(f)
=\sum_{n=0}^{+\infty}\alpha_{n}\Delta^{n}f
=\sum_{n=0}^{+\infty}\varphi(\lambda_{n})\langle f,\phi_{n}\rangle_{2}\phi_{n},
\end{equation}
which is linear, continuous, and \mbox{$\Phi(f)=\langle K_{\varphi},f\rangle_{2}$}, where
\begin{equation}\label{eq:SPECTRAL-KERNEL}
K_{\varphi}(\mathbf{p},\mathbf{q})=\sum_{n=0}^{+\infty}\varphi(\lambda_{n})\phi_{n}(\mathbf{p})\phi_{n}(\mathbf{q}),
\end{equation} 
is the \emph{spectral kernel}. Then, the \emph{spectral basis function} centred at a point~$\mathbf{p}$ is defined as the action of the spectral operator on~$\delta_{\mathbf{p}}$ or equivalently as the spectral kernel centred at~$\mathbf{p}$; in fact, \mbox{$\Phi(\delta_{\mathbf{p}})=K_{\varphi}(\mathbf{p},\cdot)$}.

The filters \mbox{$\varphi_{t}(s):=\exp{(-ist)}$}, \mbox{$\varphi(s):=s^{-k/2}$},  \mbox{$\varphi(s):=s^{-1/2}$}  induce the \emph{wave}~\citep{BRONSTEIN-PAMI2011,AUBRY2011}, \emph{poly-harmonic}, and \emph{commute-time basis functions}, respectively. Mexican hat wavelets~\citep{HOU2012} are generated by the filter \mbox{$\varphi(s):=s^{1/2}\exp(-s^{2})$} and the filter function \mbox{$\varphi(s):=\exp(is)$}, \mbox{$s\in [0,2\pi]$}, defines the wave kernel. In \mbox{$\varphi(s):=t^{k}s^{\alpha}\exp(-ts^{\alpha})$}, the parameter~$k$ scales the rate of diffusion and~$\alpha$ controls the decay of the Laplacian eigenvalues to zero, similarly to random walks~\citep{SINHA2013}.
\begin{figure}[t]
\centering
\begin{tabular}{cccc}
\includegraphics[height=70pt]{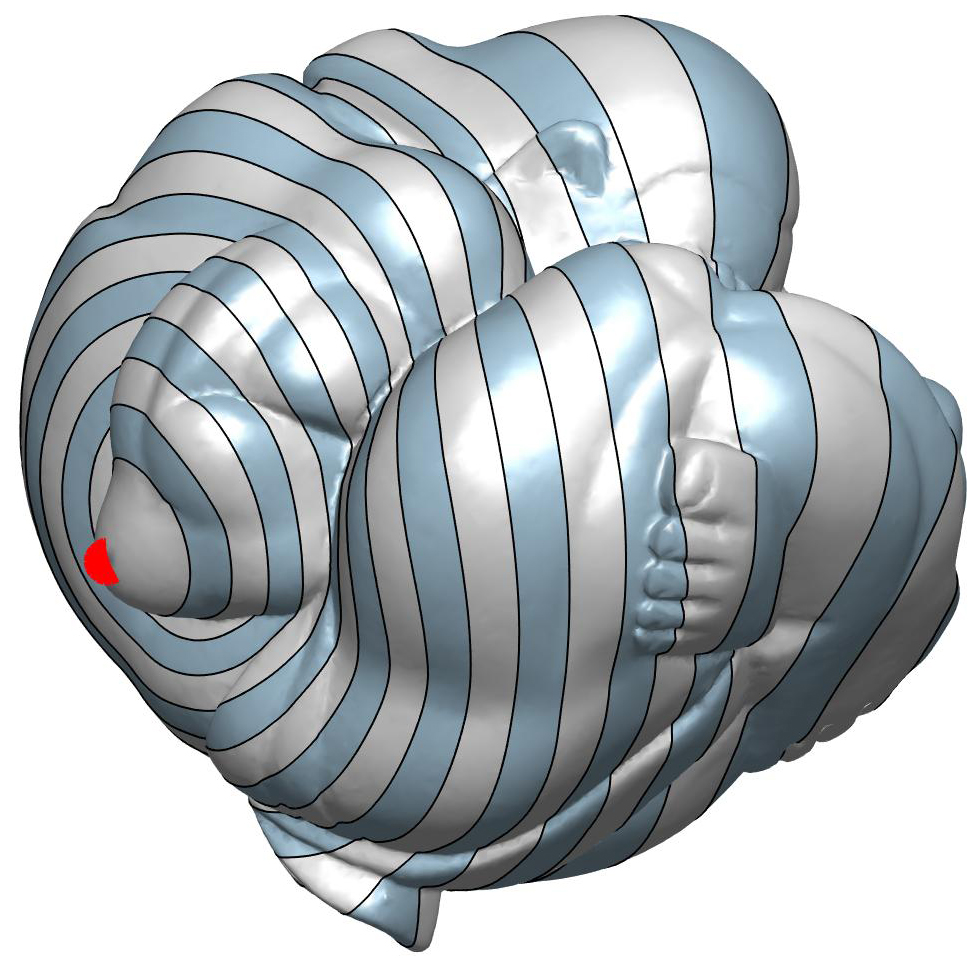}
&\includegraphics[height=70pt]{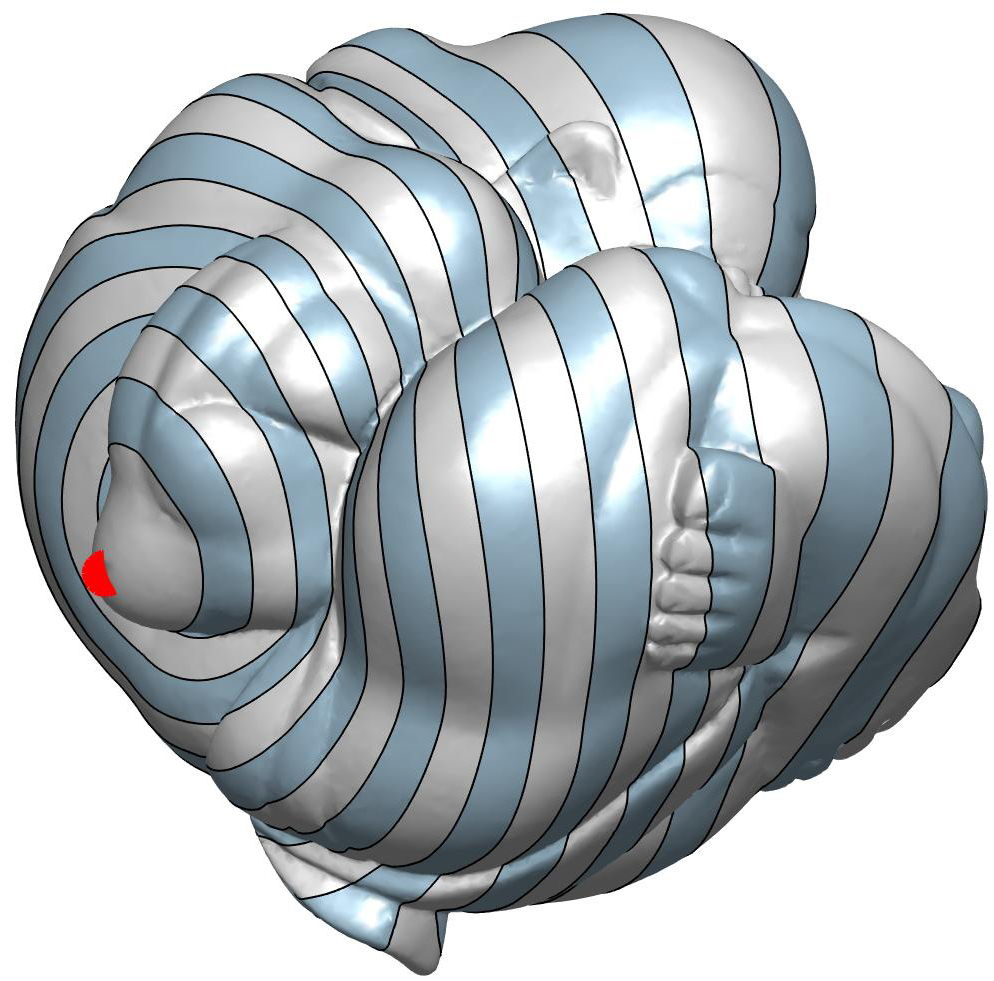}
&\includegraphics[height=70pt]{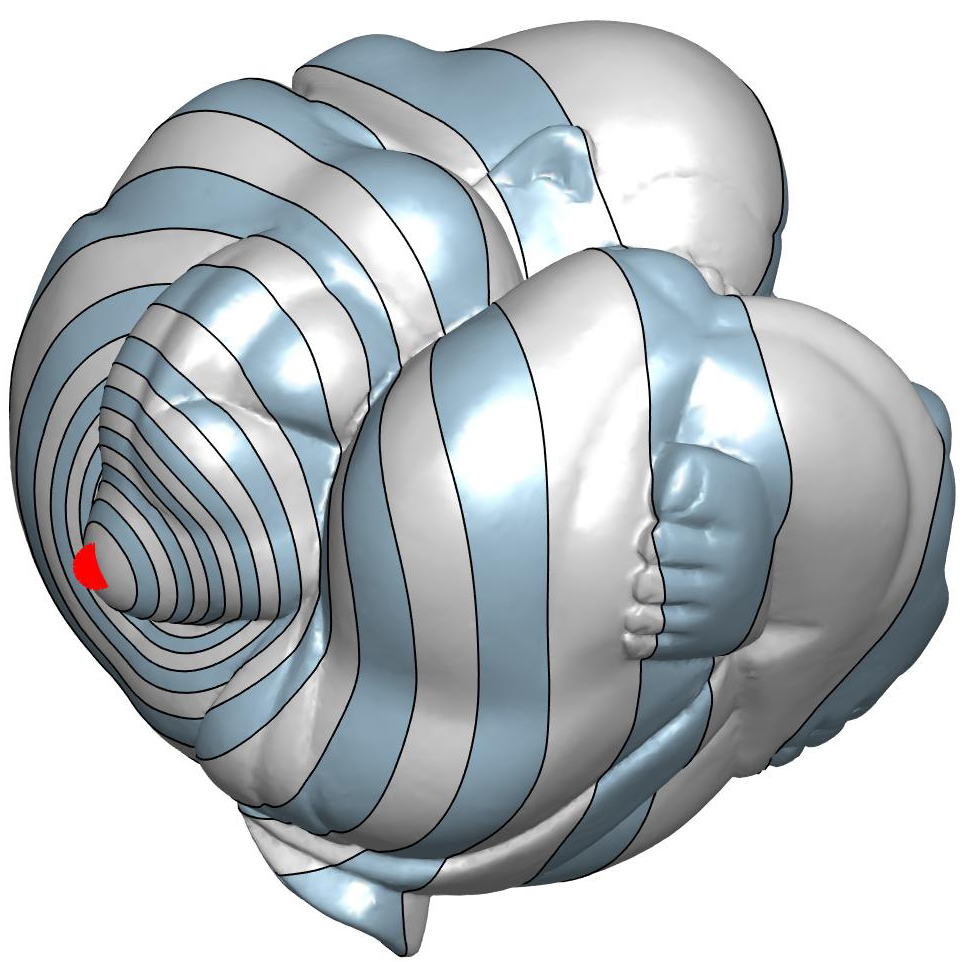}
&\includegraphics[height=70pt]{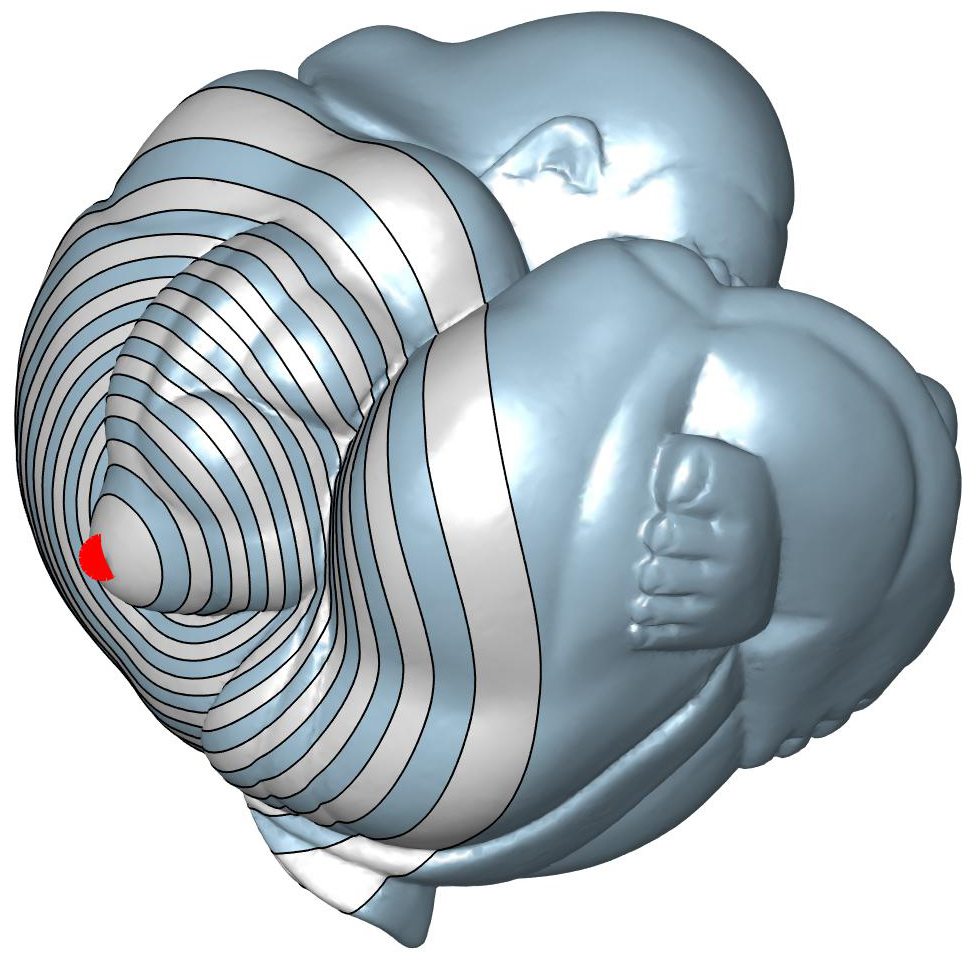}\\
~$\varphi(s)=(1+s)^{-2}$ &$\varphi(s)=(1+s^{2})^{-1}$
&$\varphi(s)=\frac{1+s}{(1+s^{2})}$ &$\varphi_{t}(s)=\exp(-ts)$\\
& & &\mbox{$t=0.1$}
\end{tabular}
\caption{Level-sets of Laplacian spectral basis functions induced by different filters.\label{fig:OMOTONDO-RATIONAL}}
\end{figure}
\paragraph*{Diffusion basis}
Selecting the filter \mbox{$\varphi_{t}(s):=\exp(-ts)$}, the spectral operator reduces to the diffusion operator \mbox{$\Phi_{t}:=\exp(-t\Delta)$} and the \emph{diffusion basis function} at~$\mathbf{p}$ (Fig.~\ref{fig:HEAT-DIFFUSION}, Fig.~\ref{fig:CAT-DIFFUSION}) is the heat kernel \mbox{$K_{t}(\cdot,\mathbf{p})$} centred at~$\mathbf{p}$, which solves the \emph{heat diffusion equation} \mbox{$(\partial_{t}+\Delta) H(\cdot,t)=0$}, \mbox{$H(\cdot,0)=\delta_{\mathbf{p}}$}; i.e., 
\begin{equation}\label{eq:HK-SPECTRAL-REPR}
K_{t}(\mathbf{p},\cdot)
=\sum_{n=0}^{+\infty}\exp(-\lambda_{n}t)\phi_{n}(\mathbf{p})\phi_{n}.
\end{equation}
The spectral representation (\ref{eq:HK-SPECTRAL-REPR}) shows the smoothing effect on the initial condition; as the scale increases, the component of~$\delta_{\mathbf{p}}$ along the eigenfunctions associated with the larger Laplacian eigenvalue becomes null.

Assuming that~$\Omega$ is an open and bounded set and that the solution \mbox{$F(\cdot,t)$} to the heat equation is sufficiently smooth, we define the parabolic cylinder \mbox{$\Omega_{T}:=\Omega\times (0,T]$} and the parabolic boundary \mbox{$\Gamma_{T}:=\overline{\Omega}_{T}\backslash\Omega_{T}$}. According to the \emph{strong maximum principle} for the heat diffusion equation~\citep{LAWRENCE1997} (Ch.~$2$), if~$\Omega$ is connected  and there exists a point \mbox{$(\mathbf{p}_{0},t_{0})$} in~$\Omega_{T}$ such that \mbox{$H(\mathbf{p}_{0},t_{0})=\max_{\overline{\Omega_{T}}}H(\mathbf{p},t)$}, then \mbox{$H(\cdot,t)$} is constant in \mbox{$\overline{\Omega_{T}}$}. In particular, the diffusion basis functions are positive.
\begin{figure*}[t]
\centering
\begin{tabular}{|c|c|c|c|}
\hline
$t=10^{-3}$ &$t=10^{-2}$ &$t=10^{-1}$ &$t=1$\\
\hline\hline
\multicolumn{4}{|c|}{\textbf{Area metrics}}\\
\hline
\includegraphics[height=60pt]{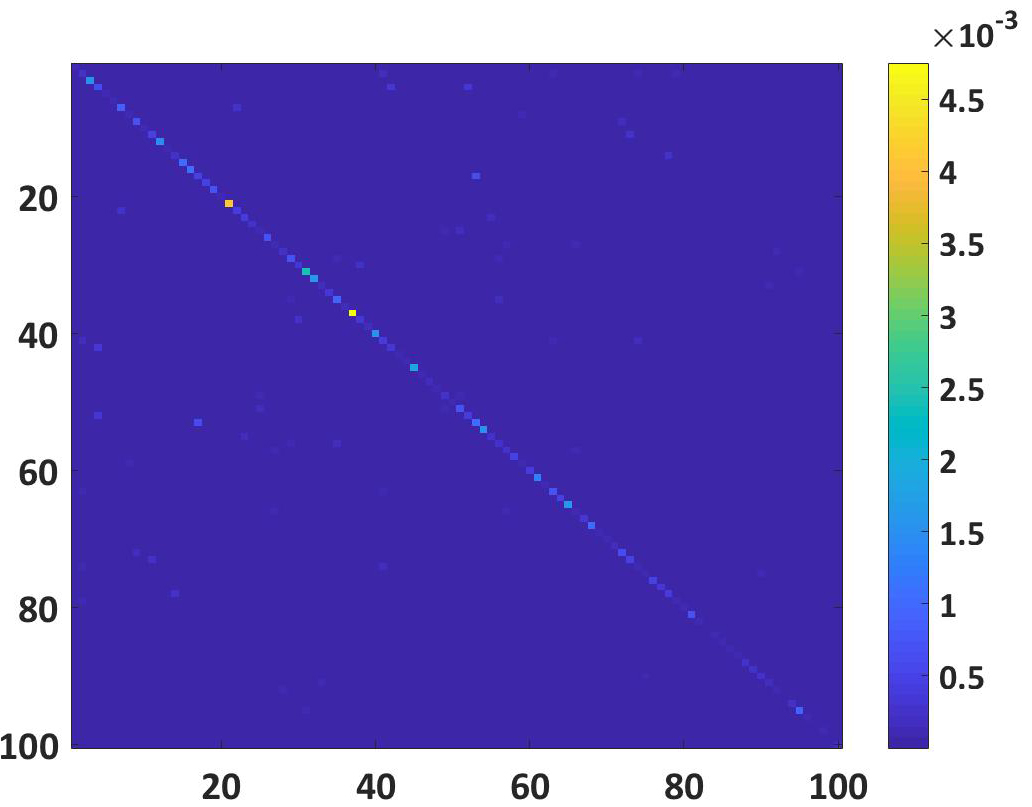}
&\includegraphics[height=60pt]{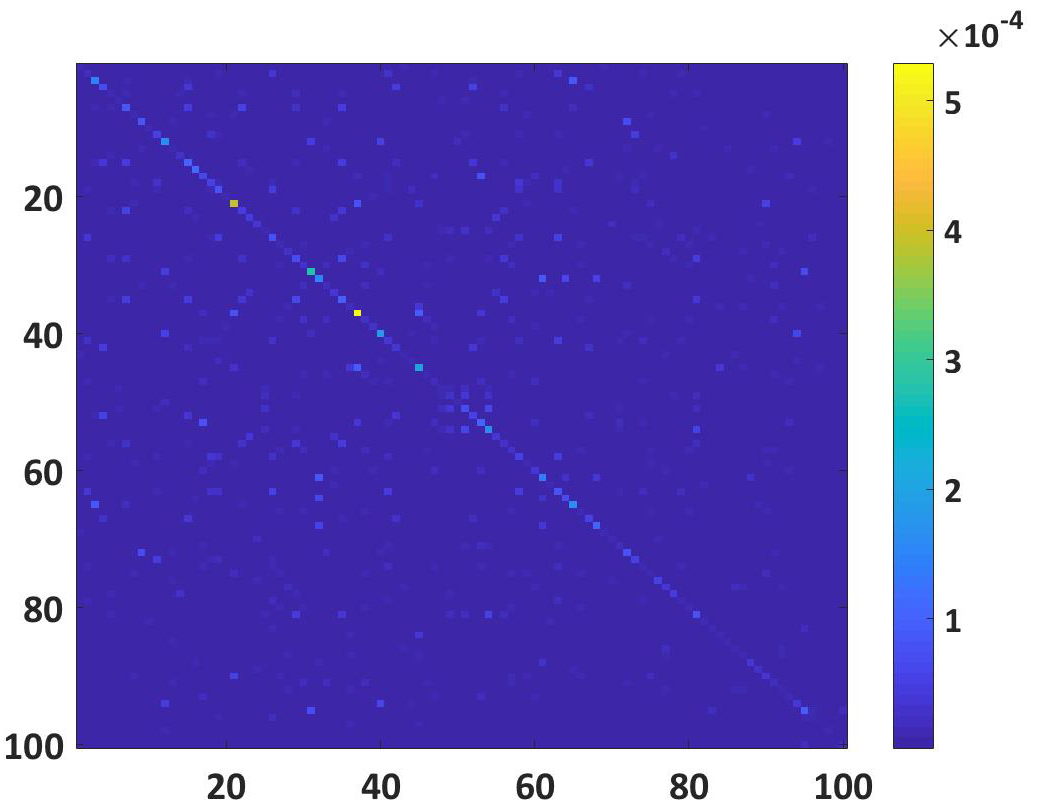}
&\includegraphics[height=60pt]{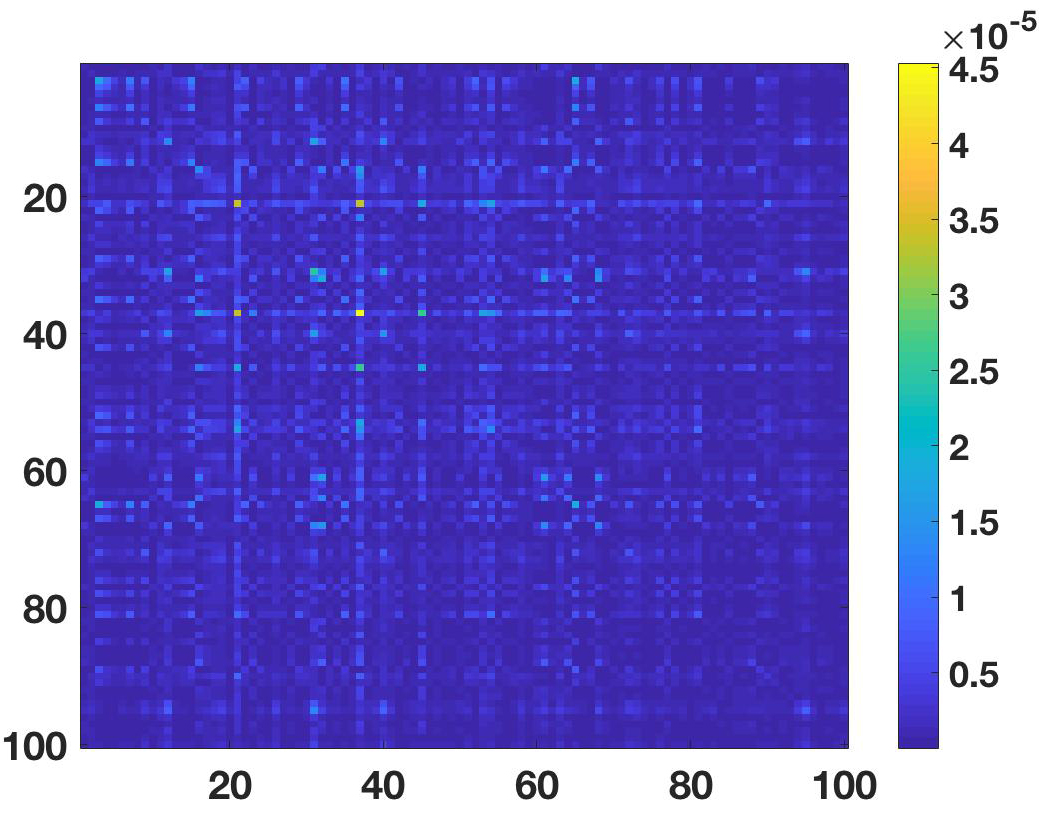}
&\includegraphics[height=60pt]{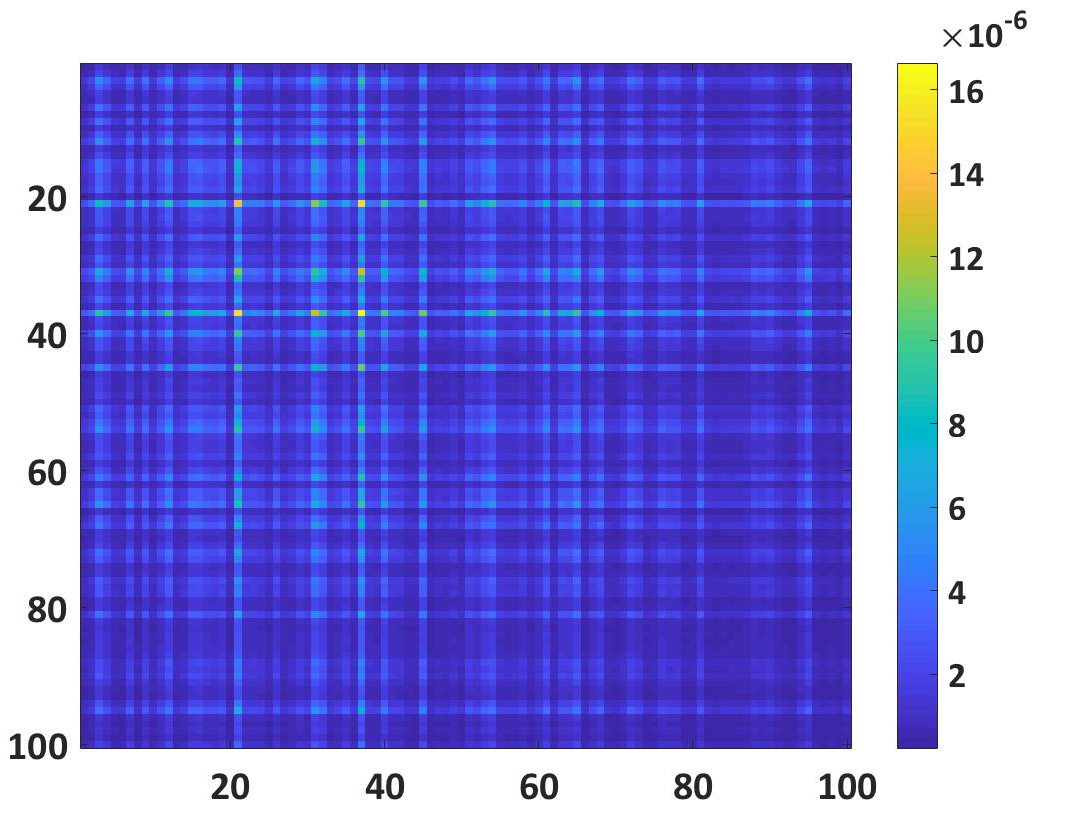}\\
\hline\hline
\multicolumn{4}{|c|}{\textbf{Conformal metrics}}\\
\hline
\includegraphics[height=60pt]{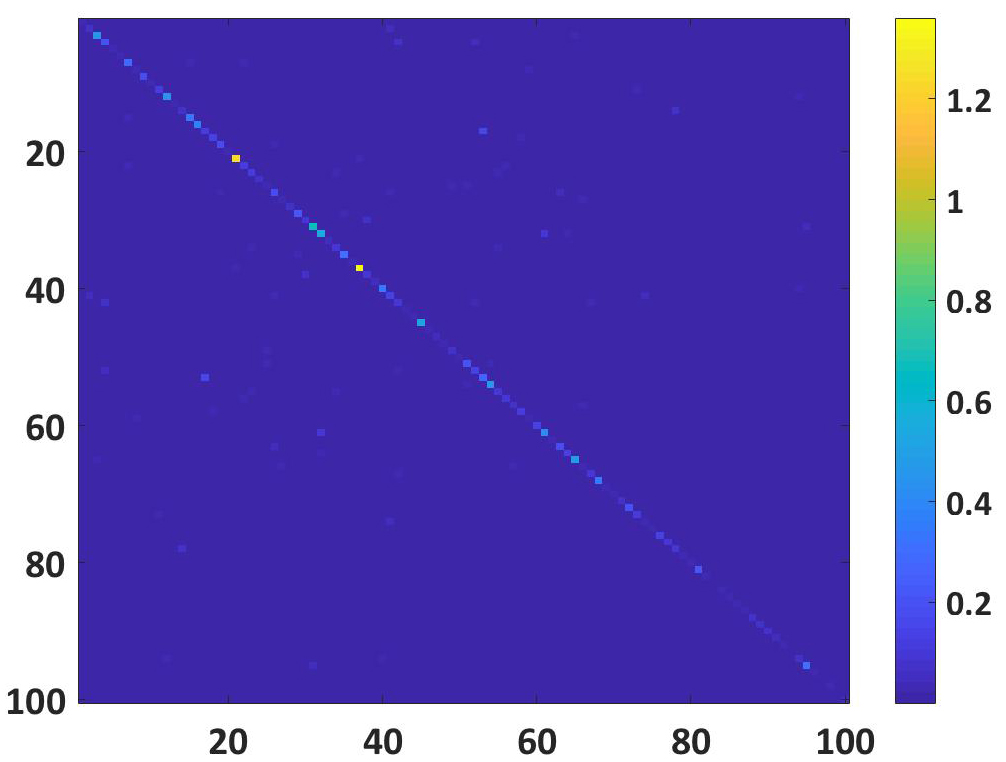}
&\includegraphics[height=60pt]{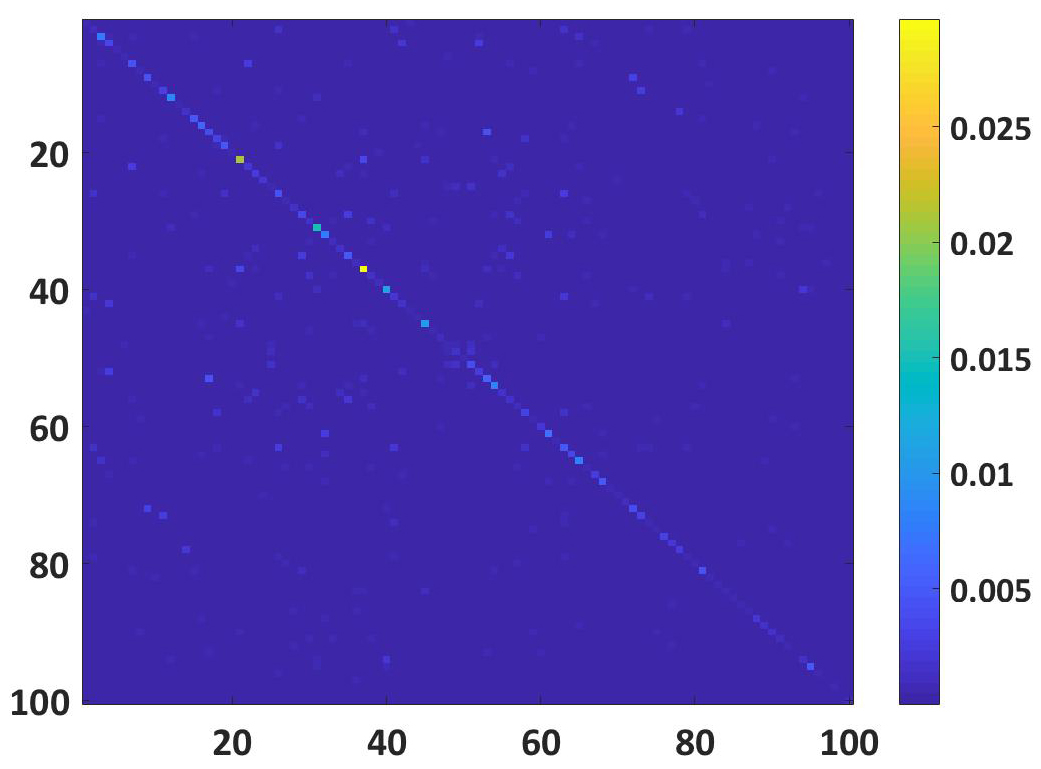}
&\includegraphics[height=60pt]{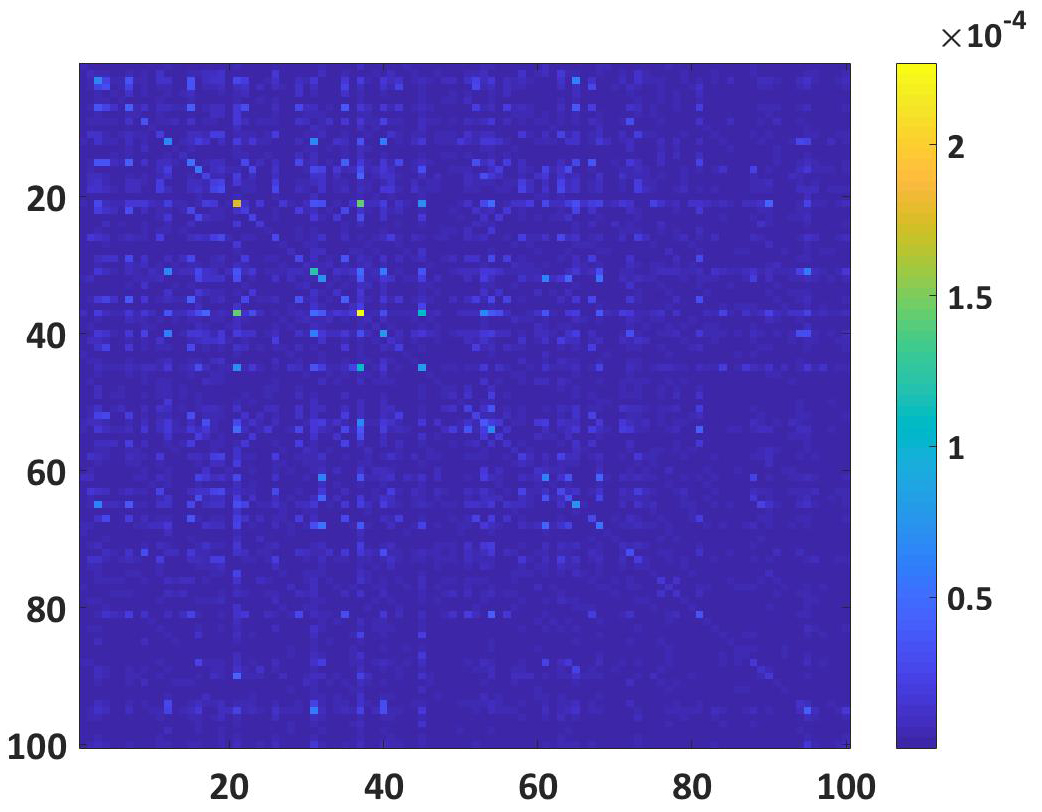}
&\includegraphics[height=60pt]{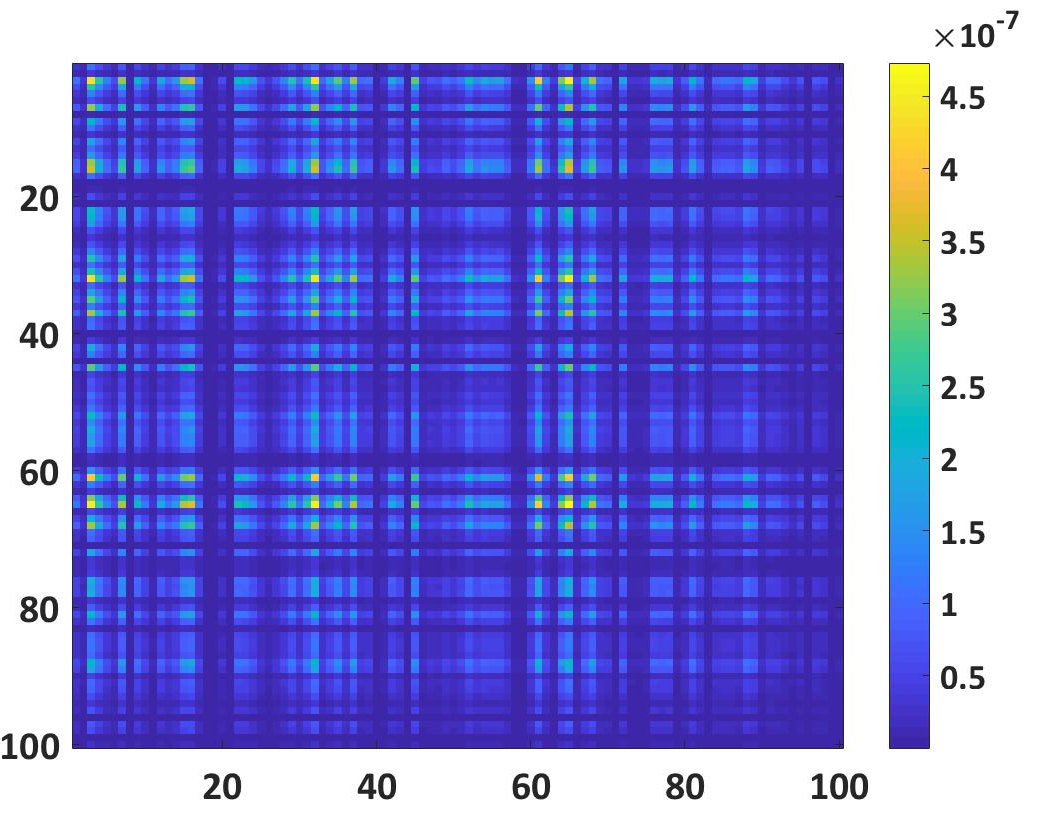}\\
\hline
\includegraphics[height=60pt]{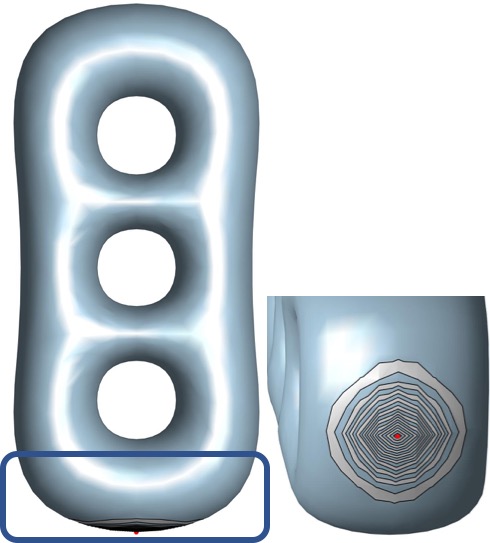}
&\includegraphics[height=60pt]{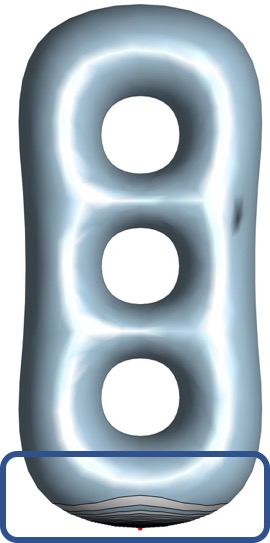}
&\includegraphics[height=60pt]{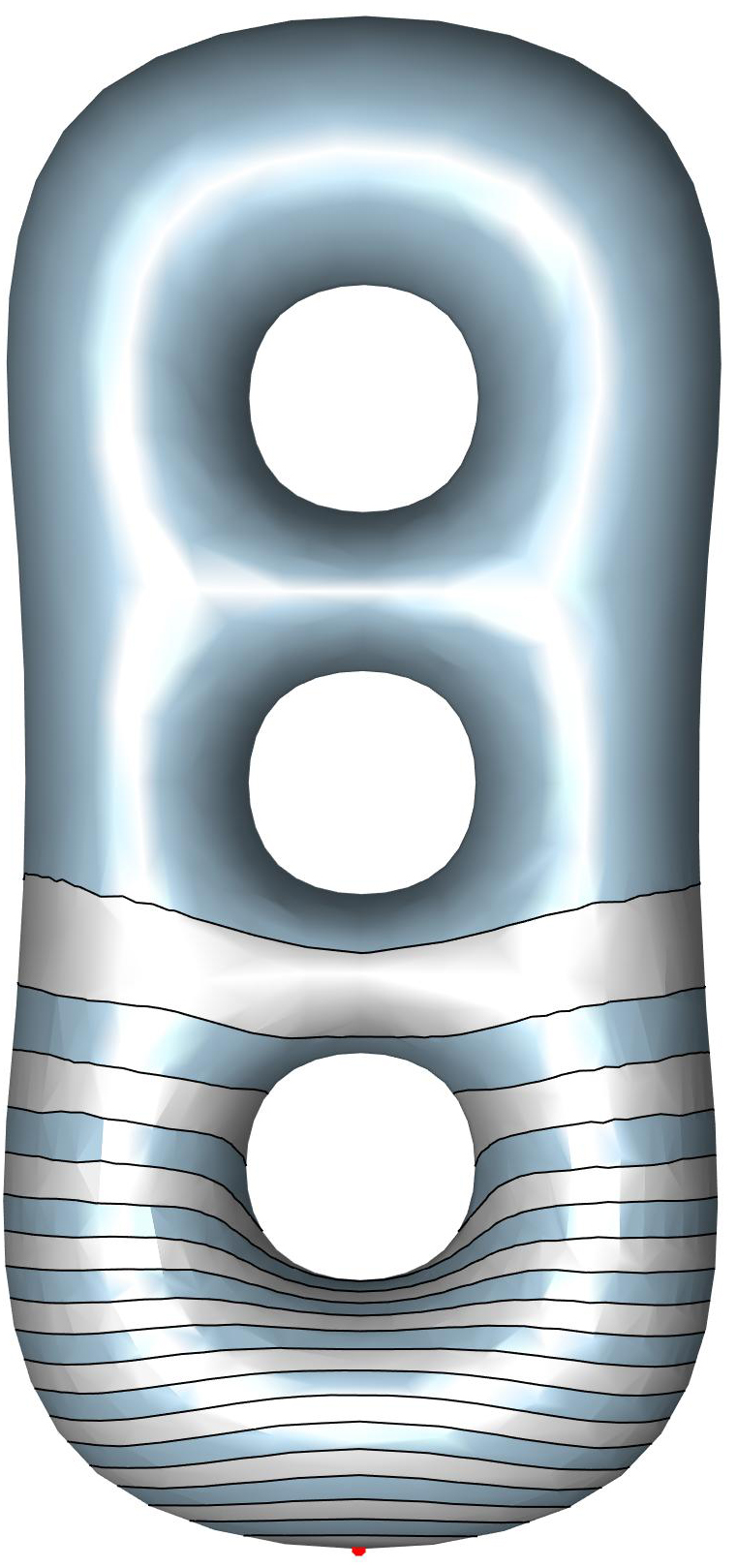}
&\includegraphics[height=60pt]{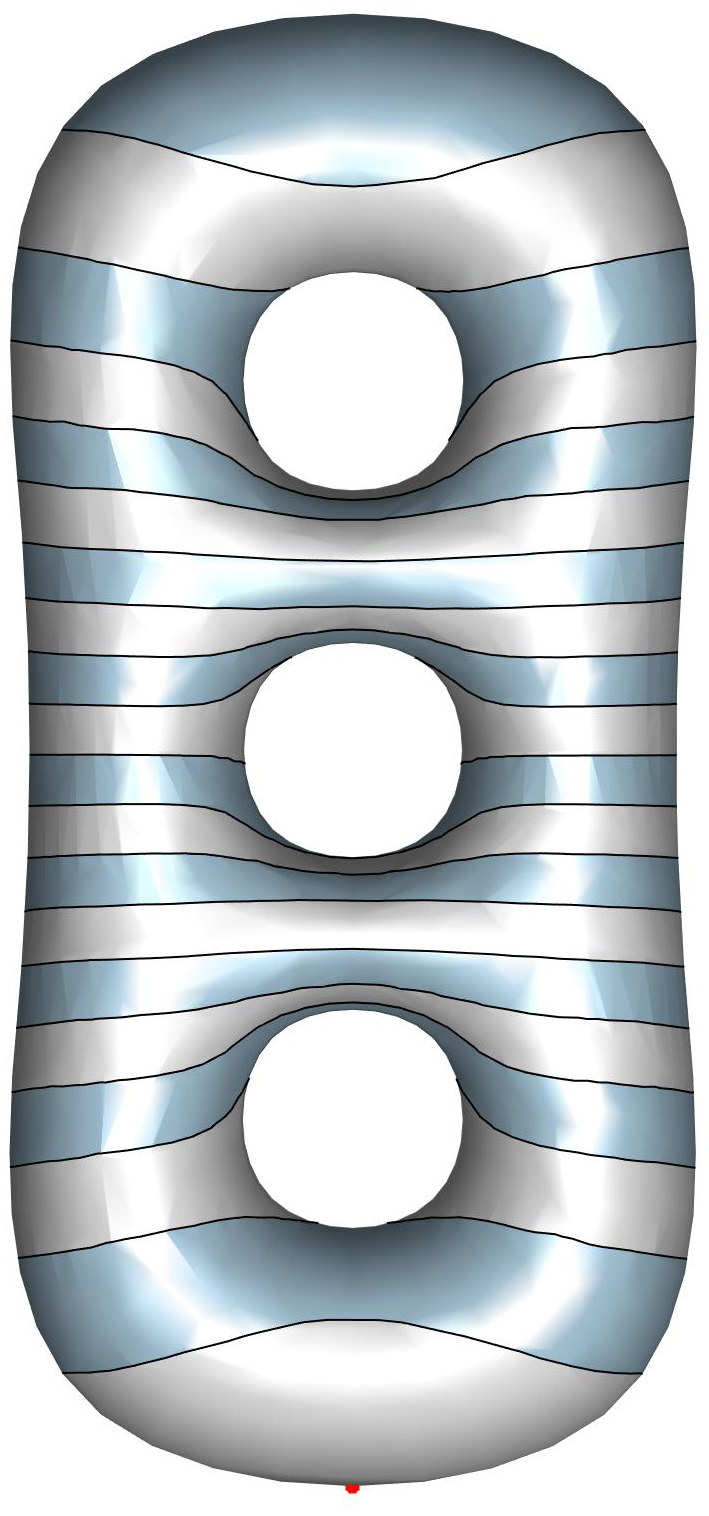}\\
\hline
\end{tabular}
\caption{The pixel \mbox{$(i,j)$} of each \mbox{$100\times 100$} image indicates the area and conformal metrics of the couples \mbox{$(\mathbf{K}_{t}\mathbf{e}_{i},\mathbf{K}_{t}\mathbf{e}_{j})$}, \mbox{$i,j=1,\ldots,100$}, of diffusion basis functions centred at 100 randomly-sampled seed points and at 4 different scales (one for each column). At small scales (first, second columns), the corresponding diffusion functions have small, non-overlapping supports and are almost orthogonal (i.e., null non-diagonal entries of the comparison matrices) with respect to the inner product induced by the mass and stiffness matrices. At larger scales, the comparison matrices are no more diagonal, as a matter of a larger support of the diffusion basis functions. Level-sets (third row) of the diffusion basis functions at different scales and centred at the same seed point (red dot at the bottom).\label{fig:3TORUS-METIRCS}}
\end{figure*}
\subsection{Computation of the spectral basis\label{sec:SPEC-DIST-COMP}}
Considering the generalised eigensystem \mbox{$\mathbf{L}\mathbf{X}=\mathbf{B}\mathbf{X}\Gamma$}, with orthonormal eigenvectors \mbox{$\mathbf{X}^{\top}\mathbf{B}\mathbf{X}=\mathbf{I}$}, the discretisation of the spectral operator (\ref{eq:FUNCT-OPER}) is \mbox{$\mathbf{K}_{\varphi}=\varphi(\tilde{\mathbf{L}})=\mathbf{X}\varphi(\Gamma)\mathbf{X}^{\top}\mathbf{B}$}, \mbox{$\varphi(\Gamma):=\textrm{diag}(\varphi(\lambda_{i}))_{i=1}^{n}$}; i.e.,
\begin{equation}\label{eq:EIGS-SPECTRAL-OP}
\mathbf{K}_{\varphi}\mathbf{f}
=\sum_{i=1}^{n}\varphi(\lambda_{i})\langle\mathbf{f},\mathbf{x}_{i}\rangle_{\mathbf{B}}\mathbf{x}_{i}.
\end{equation}
In particular,~$\tilde{\mathbf{L}}$ and~$\mathbf{K}_{\varphi}$ have the same eigenvectors and \mbox{$(\varphi(\lambda_{i}))_{i=1}^{n}$} are the (filtered) Laplacian eigenvalues of~$\mathbf{K}_{\varphi}$. Assuming that \mbox{$\mathbf{K}_{\varphi}:=\varphi(\tilde{\mathbf{L}})$} is invertible (i.e., the filter does not vanish at the Laplacian eigenvalues), the vectors in \mbox{$\mathcal{B}:=\{\mathbf{K}_{\varphi}\mathbf{e}_{i}\}_{i=1}^{n}$} (i.e., \mbox{$\mathbf{f}:=\mathbf{e}_{i}$} in Eq. (\ref{eq:EIGS-SPECTRAL-OP})) are linearly independent; in fact, \mbox{$\sum_{i=1}^{n}\alpha_{i}\mathbf{K}_{\varphi}\mathbf{e}_{i}=\mathbf{0}$} is verified if and only if \mbox{$\mathbf{K}_{\varphi}\alpha=\mathbf{0}$}; i.e., \mbox{$\alpha=\mathbf{0}$}. The coefficients \mbox{$\alpha:=(\alpha_{i})_{i=1}^{n}$}, which express \mbox{$\mathbf{f}:=\sum_{i=1}^{n}\alpha_{i}\mathbf{K}_{\varphi}\mathbf{e}_{i}$} as a linear combination of the basis~$\mathcal{B}$, are computed by solving the linear system \mbox{$\mathbf{K}_{\varphi}\alpha=\mathbf{f}$}. We notice that \mbox{$\alpha=\mathbf{K}_{\varphi}^{\dag}\mathbf{f}=\mathbf{X}\varphi^{\dag}(\Gamma)\mathbf{X}^{\top}\mathbf{B}\mathbf{f}$}, where \mbox{$\varphi^{\dag}(\Gamma)$} is the diagonal matrix whose non-null values are \mbox{$(1/\varphi(\lambda_{i}))_{i=1}^{n}$}.
\begin{figure}[t]
\centering
\includegraphics[width=300pt]{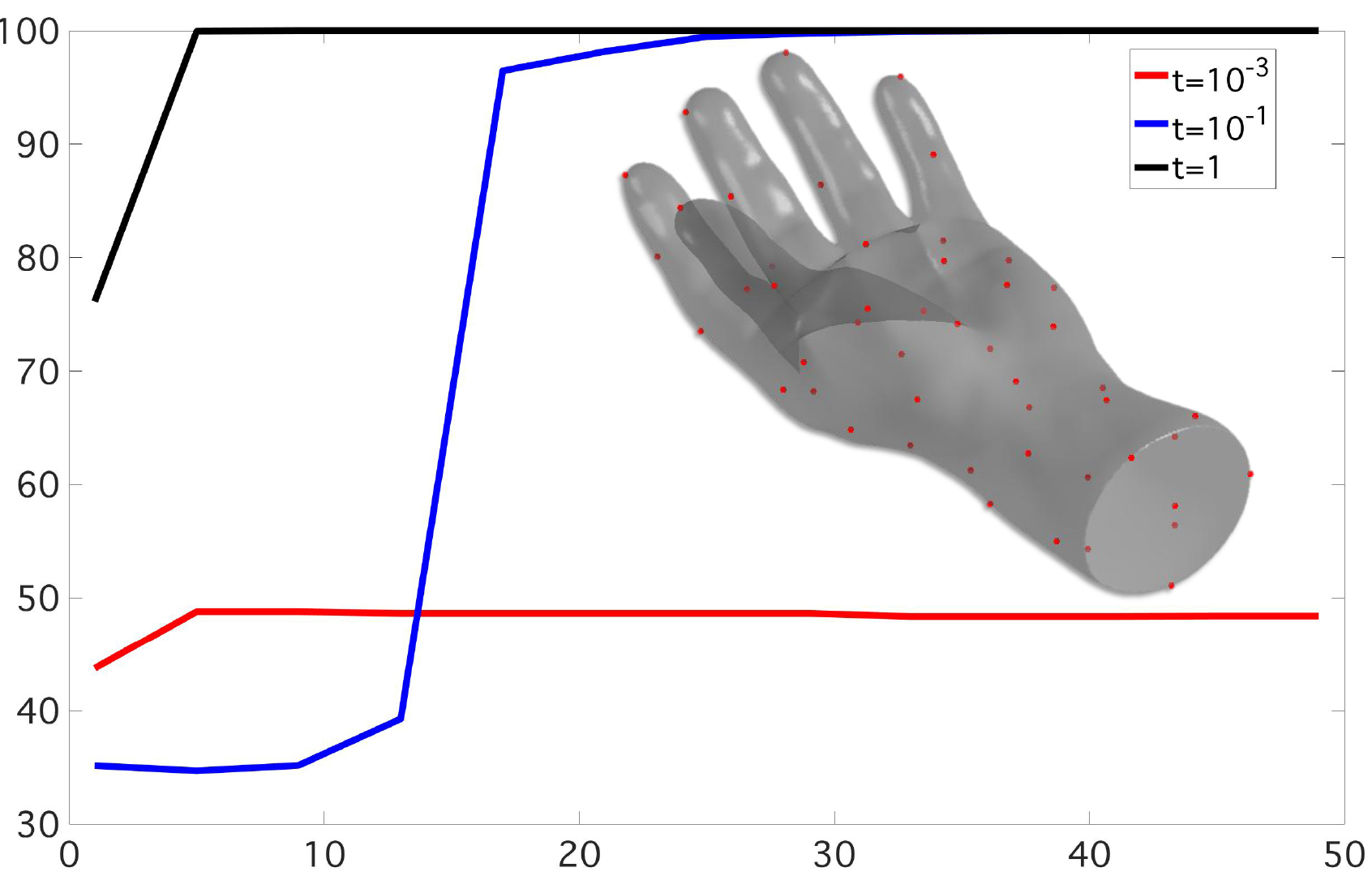}
\caption{Percentage ($y$-axis) of vertices belonging to the support of~$k$ ($x$-axis) diffusion basis functions at three different scales \mbox{$t=0.001, 0.1, 1$} and centred at a set of seed points (red dots) computed by the Euclidean farthest point sampling. At large/medium scales (\mbox{$t=1$}, \mbox{$t=0.1$}), the supports of a few (\mbox{$k=7$}, \mbox{$k=26$}, respectively) diffusion basis functions cover the whole surface. At small scales (\mbox{$t=0.001$}), the diffusion basis functions centred at 50 seed points have no overlapping supports, as they cover around the \mbox{$50\%$} of the whole surface.\label{fig:SEEDS}}
\end{figure}

Recalling that the computation of the Laplacian eigenpairs is numerically unstable in case of repeated eigenvalues~\citep{PATANE-STAR2016}, the filter function should be chosen in such a way that the filtered Laplacian matrix does not have additional (if any) repeated eigenvalues. This condition is generally satisfied by choosing an injective filter. The selection of periodic filters, the expensive cost of the computation of the Laplacian spectrum, and the sensitiveness of multiple Laplacian eigenvalues to surface discretisation motivate the definition of alternative approaches for the evaluation of the spectral basis functions. Among them, we discuss the truncated and spectrum-free approximations. 
\begin{figure*}[t]
\centering
\begin{tabular}{cc}
\includegraphics[height=100pt]{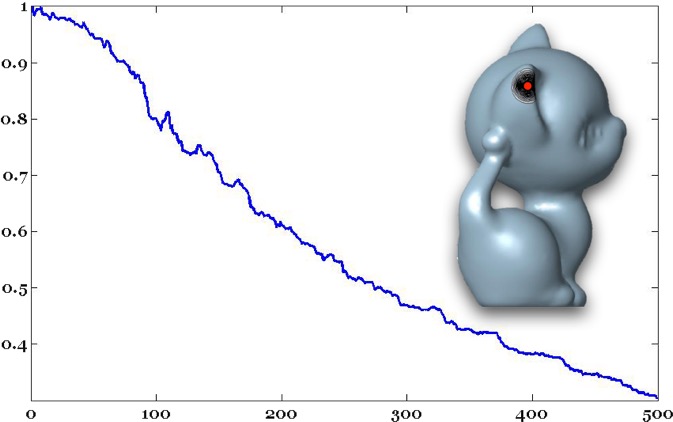}
&\includegraphics[height=100pt]{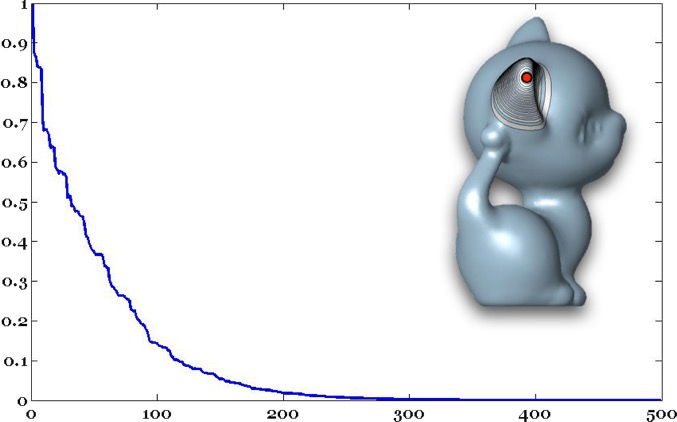}\\
$t=10^{-2}$ &$t=10^{-1}$\\
\end{tabular}
\includegraphics[height=100pt]{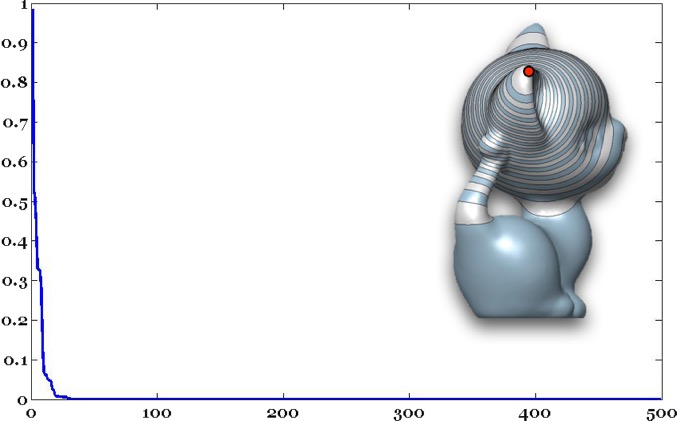}\quad~$t=1$
\caption{$\ell_{\infty}$ error ($y$-axis) between the Pad\'e-Chebyshev approximation of the diffusion basis functions at different scales and the truncated spectral approximation with a different number ($x$-axis) of Laplacian eigenpairs. \label{fig:KITTEN-CHEB-VS-SPECTRAL}}
\end{figure*}
\paragraph*{Truncated approximation}
The computational limits for the evaluation of the whole Laplacian spectrum and the decay of the filtered coefficients are the main reasons behind the approximation of a spectral basis function as a truncated sum; i.e., applying the relation \mbox{$\Phi_{k}\mathbf{e}_{i}
=\sum_{j=1}^{k}\varphi(\lambda_{j})\langle\mathbf{e}_{i},\mathbf{x}_{j}\rangle_{\mathbf{B}}\mathbf{x}_{j}$}, where~$k$ is the selected number of eigenpairs. Even though the first~$k$ Laplacian eigenpairs are computed in super-linear time~\citep{VALLET2008}, the evaluation of the whole Laplacian spectrum is unfeasible for storage and computational cost, which are quadratic in the number of surface samples. The selection of filters that are periodic or do not decrease to zero motivates the need of defining a spectrum-free computation of the corresponding kernels, which cannot be accurately approximated with the contribution of only a subpart of the Laplacian spectrum. Furthermore, the number of selected eigenpairs is heuristically adapted to the decay of the filter function and the approximation accuracy cannot be estimated without computing the whole spectrum.

\paragraph*{Pad\'e-Chebyshev (spectrum-free) approximation}
Truncated spectral representations apply a low pass filter and the resulting reconstructed signal generally preserves only its global features. Local features, which are associated with a high frequency, are generally missing in the approximated signals and can be recovered by considering only a high number of Laplacian eigenpairs. However, the evaluation of a large part of the Laplacian spectrum is computationally unfeasible, requires a large memory overhead, and is numerically unstable in case of duplicated or numerically close eigenvalues~\citep{PATANE-STAR2016}. 

To avoid these drawbacks, we introduce the spectrum-free evaluation of the basis functions, which is based on a rational approximation of the filter.
\begin{figure*}[t]
\centering
\begin{tabular}{ccc}
\includegraphics[height=80pt]{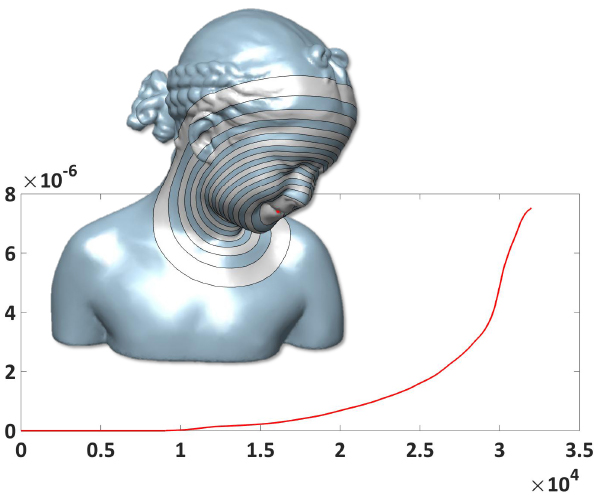}
&\includegraphics[height=80pt]{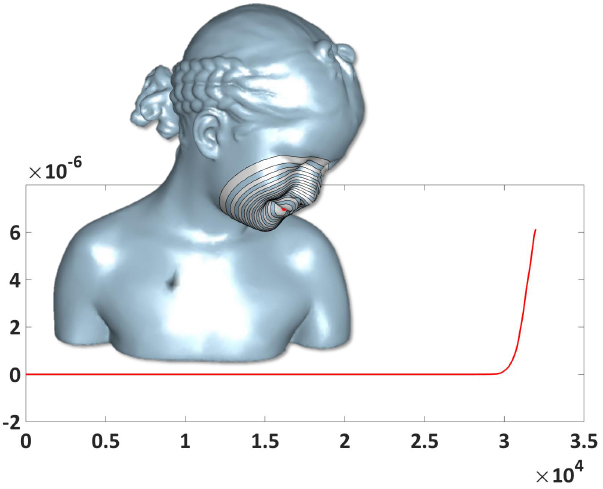}
&\includegraphics[height=80pt]{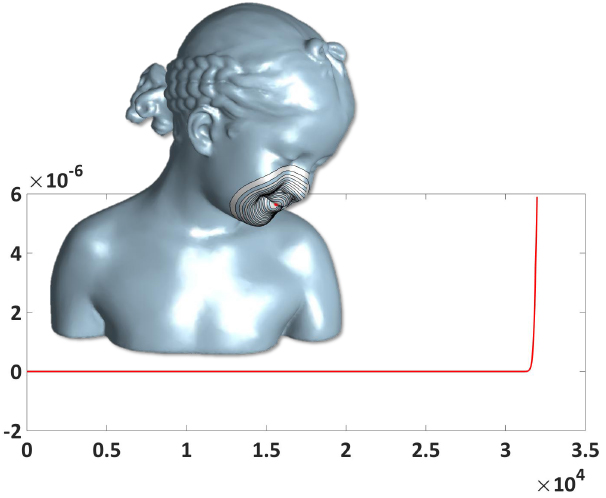}\\
(a)~$t=10^{-1}$ &(b)~$t=10^{-2}$ &(c)~$t=10^{-3}$\\
\includegraphics[height=80pt]{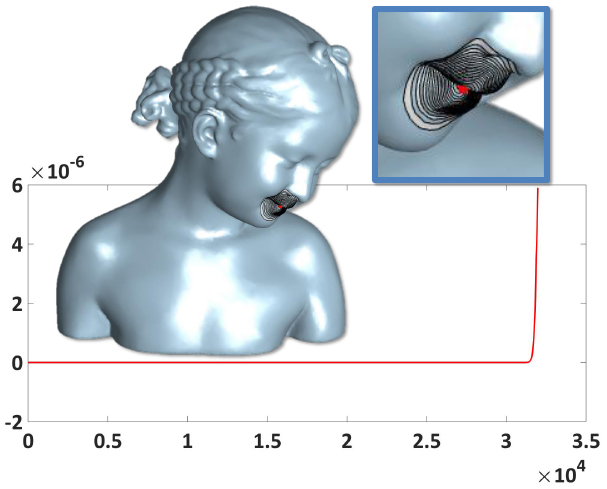}
&\includegraphics[height=80pt]{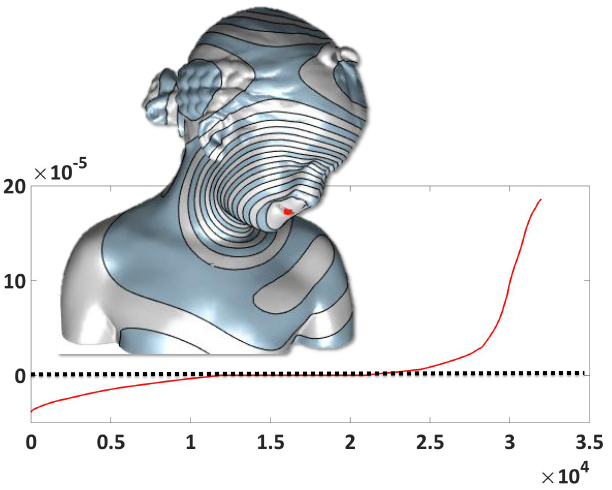}
&\includegraphics[height=80pt]{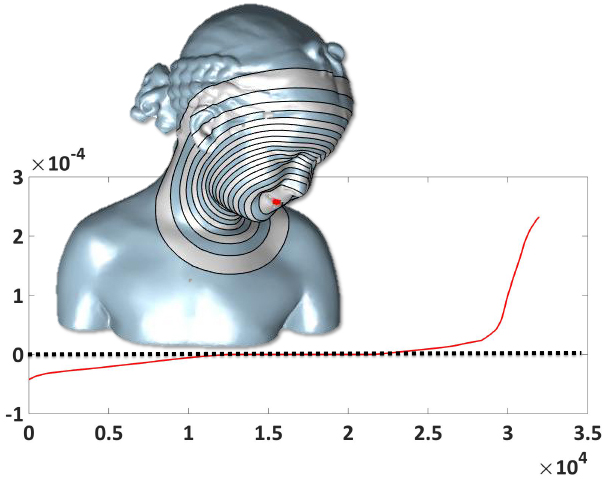}\\
(d)~$t=10^{-4}$ &(e)~$t=10^{-1}$ &(f)~$t=10^{-2}$
\end{tabular}
\caption{(a-d) Robustness of the Pad\'e-Chebyshev approximation of the diffusion basis functions and (e,f) sensitiveness of truncated spectral approximation to the Gibbs phenomenon. At all scales (a-d), the function values (red curve) computed with the Pad\'e-Chebyshev approximation are positive; at large scales (e,f), the truncated spectral approximation is affected by the Gibbs phenomenon, as represented by the part of the plot below the zero line (black curve).\label{fig:GIBBS-PHENOMENON}}
\end{figure*}
For an arbitrary filter, we consider the rational Pad\'e-Chebyshev approximation \mbox{$p_{r}(s)=\frac{a_{r}(s)}{b_{r}(s)}$} of~$\varphi$~\citep{GOLUB1989} (Ch.~$11$) with respect to the~$\mathcal{L}^{\infty}$ norm. Here,~\mbox{$a_{r}(\cdot)$} and \mbox{$b_{r}(\cdot)$} are polynomials of degree equal to or lower than~$r$. Let \mbox{$p_{r}(s)=\sum_{i=1}^{r}\alpha_{i}(1+\beta_{i}s)^{-1}$} be the partial form of the Pad\'e-Chebyshev approximation, where \mbox{$(\alpha_{i})_{i=1}^{r}$} are the weights and \mbox{$(\beta_{i})_{i=1}^{r}$} are the nodes of the~$r$-point Gauss-Legendre quadrature rule~\citep{GOLUB1989} (Ch.~$11$). The weights and nodes are precomputed for any degree of the rational polynomial~\citep{CARPENTER1984}. Applying this approximation to the spectral kernel, we get that
\begin{equation*}
\mathbf{K}_{\varphi}\mathbf{e}_{i}\\
\approx p_{r}(\tilde{\mathbf{L}})\mathbf{e}_{i}\\
=\sum_{j=1}^{r}\alpha_{j}\left(\mathbf{I}+\beta_{j}\tilde{\mathbf{L}}\right)^{-1}\mathbf{e}_{i}\\
=\sum_{j=1}^{r}\alpha_{j}\mathbf{g}_{j},
\end{equation*}
where~$\mathbf{g}_{j}$ solves the symmetric and sparse linear system
\begin{equation}\label{eq:GEN-CHEB-APPROX}
\left(\mathbf{B}+\beta_{j}\mathbf{L}\right)\mathbf{g}_{j}=\mathbf{B}\mathbf{e}_{i}, \quad j=1,\ldots,r.
\end{equation}
The Pad\'e-Chebyshev approximation generally provides an accuracy higher than the polynomial approximation, as a matter of its uniform convergence to the filter. In the paper examples, we have selected \mbox{$r:=5$} as degree of the rational polynomial approximation of the filter function.

We now derive the spectral representation of the Pad\`e-Chebyshev approximation of the function \mbox{$\mathbf{K}_{\varphi}\mathbf{f}$}. Re-writing the solution to \mbox{$(\mathbf{B}+\beta_{j}\mathbf{L})\mathbf{g}=\mathbf{B}\mathbf{f}$} (i.e., Eq. (\ref{eq:GEN-CHEB-APPROX}) with \mbox{$\mathbf{f}=\mathbf{e}_{i}$}) as \mbox{$\mathbf{g}=\mathbf{X}\alpha$}, we get that \mbox{$\alpha=(\mathbf{I}+\beta_{j}\Gamma)^{\dag}\mathbf{X}^{\top}\mathbf{B}\mathbf{f}$} and the Pad\`e-Chebyshev approximation is
\begin{equation*}
\mathbf{K}_{\varphi}\mathbf{f}
\approx\alpha_{0}\mathbf{f}+\sum_{i=1,\ldots,n}^{j=1,\ldots,r}\alpha_{j}\frac{1}{1+\beta_{j}\lambda_{i}}\langle\mathbf{f},\mathbf{x}_{i}\rangle_{\mathbf{B}}\mathbf{x}_{i}.
\end{equation*}
Setting \mbox{$\mathbf{f}:=\mathbf{e}_{i}$}, we obtain that the spectral representation of the Pad\`e-Chebyshev approximation of the spectral basis function \mbox{$\mathbf{K}_{\varphi}\mathbf{e}_{i}$} centred at~$\mathbf{p}_{i}$.

According to~\citep{MOLER2003}, the approximation of the matrix \mbox{$\varphi(\tilde{\mathbf{L}})$} might be numerically unstable if \mbox{$\|\tilde{\mathbf{L}}\|_{2}$} is large. From the bound \mbox{$\|\mathbf{B}^{-1}\mathbf{L}\|_{2}\leq {\lambda_{\min}^{-1}(\mathbf{B})}\lambda_{\max}(\mathbf{L})$}, a well-conditioned mass matrix~$\mathbf{B}$ guarantees that \mbox{$\|\mathbf{B}^{-1}\mathbf{L}\|_{2}$} is bounded. Noting that \mbox{$(1+\beta_{i}\lambda_{j},\mathbf{x}_{j})_{j=1}^{n}$} are the eigenpairs of \mbox{$\mathbf{B}+\beta_{i}\mathbf{L}$}, the corresponding conditioning number is bounded as
\begin{equation*}\label{eq:GENERAL-COND-NUM}
\kappa_{2}(\mathbf{B}+\beta_{i}\mathbf{L})
=\frac{\max_{j}\{1+\beta_{i}\lambda_{j}\}}{\min_{j}\{1+\beta_{i}\lambda_{j}\}}
=\left\{
\begin{array}{ll}
1+\beta_{i}\lambda_{n}			&\beta_{i}>0,\\
(1+\beta_{i}\lambda_{n})^{-1}	&\beta_{i}<0,
\end{array}
\right.
\end{equation*}
which is always greater than 1. The numerical solution to the linear system in Eq. (\ref{eq:GEN-CHEB-APPROX}) is generally stable; as reviewed in Sect.~\ref{sec:HARMONIC-MAPS}, pre-conditioners can be applied to further attenuate numerical instabilities.

The truncated spectral approximation (Fig.~\ref{fig:DIFFUSIVE-COMPUTATION}) is affected by small geometric undulations (especially at small scales), the use of heuristics for the selection of the number of Laplacian eigenpairs with respect to the target approximation accuracy, and the scale of features of the input shape. The spectrum-free computation generally provides better results in terms of smoothness, regularity, and accuracy of the computed spectral basis. Changing the filter function and its decay to zero allows us to define different basis functions with a different behaviour and support (Fig.~\ref{fig:OMOTONDO-RATIONAL}), thus showing the flexibility and simplicity of our approach. For further comparison examples, we refer the reader to Sect.~\ref{sec:GENERAL-EXAMPLES-DISCUSSION}.

\subsection{Green kernel and basis functions\label{sec:GREEN-KERNEL}}
To study the relation between the Green kernel and the basis previously defined, let us consider a linear differential operator~$\mathcal{L}$ and the corresponding \emph{Green kernel} \mbox{$G:\mathcal{N}\times\mathcal{N}\rightarrow\mathbb{R}$} such that \mbox{$\mathcal{L}G(\mathbf{p},\mathbf{q})=\delta(\mathbf{p}-\mathbf{q})$}, where \mbox{$\delta(\cdot)$} is the delta function. Then, the solution to the differential equation \mbox{$\mathcal{L}u=f$} is expressed in terms of the Green kernel as \mbox{$u(\mathbf{p})=\langle G(\mathbf{p},\cdot),f\rangle_{2}$}. Noting that the eigensystem \mbox{$(\lambda_{n},\phi_{n})_{n=0}^{+\infty}$} of the integral operator \mbox{$\mathcal{K}_{G}f:=\langle G(\mathbf{p},\cdot),f\rangle_{2}$} induced by the Green kernel satisfies the relations \mbox{$\mathcal{L}\phi_{n}=\lambda_{n}^{-1}\phi_{n}$}, \mbox{$\lambda_{n}\neq 0$}, we have that the eigensystem of~$\mathcal{L}$ is \mbox{$(\lambda_{n}^{-1},\phi_{n})_{n=0}^{+\infty}$}. Indeed,~$\mathcal{L}$,~$\mathcal{K}_{G}$ have the same eigenfunctions and reciprocal eigenvalues. In particular, the spectral representation of the Green kernel \mbox{$G(\mathbf{p},\mathbf{q})=\sum_{n=0}^{+\infty}\lambda_{n}\phi_{n}(\mathbf{p})\phi_{n}(\mathbf{q})$} is uniquely defined by the spectrum of~$\mathcal{L}$. 

Combining the previous result with the fact that the spectral basis functions have been defined as solution to different differential equations that involve the Laplace-Beltrami operator, we get the following relations
\begin{equation*}
\left\{
\begin{array}{ll}
G_{\Delta}(\mathbf{p},\mathbf{q})=\sum_{n=0}^{+\infty}\frac{1}{\lambda_{n}}\phi_{n}(\mathbf{p})\phi_{n}(\mathbf{q}) 				&\textrm{harmonic oper. } (\mathcal{L}=\Delta);\\
K_{t}(\mathbf{p},\mathbf{q})=\sum_{n=0}^{+\infty}\exp(-t\lambda_{n})\phi_{n}(\mathbf{p})\phi_{n}(\mathbf{q})					&\textrm{diffusion oper. } (\mathcal{L}=\exp(t\Delta));\\
G_{\varphi}(\mathbf{p},\mathbf{q})=\sum_{n=0}^{+\infty}\varphi(\lambda_{n})\phi_{n}(\mathbf{p})\phi_{n}(\mathbf{q})					&\textrm{spectral oper. } (\mathcal{L}=\varphi^{\dag}(\Delta)).			
\end{array}
\right.
\end{equation*}
\section{Comparison metrics for scalar functions\label{sec:SAME-SHAPE-COMPARISON}}
Since the level-sets and critical points (i.e., maxima, minima, saddles) of a scalar function \mbox{$f:\mathcal{N}\rightarrow\mathbb{R}$} are independent of positive re-scalings of~$f$, we can assume that the function values have been normalised in such a way that \mbox{$\textrm{Image}(f)=[0,1]$} and compare two functions by evaluating the corresponding~$\mathcal{L}_{\infty}$ or~$\mathcal{L}_{2}$ norm. However, these norms do not measure the differential properties of the underlying functions and are better suited to evaluate the approximation accuracy. Indeed, in the following we disucss the area, conformal, and kernel-based metrics.
\begin{figure}[t]
\centering
\begin{tabular}{ccc}
\includegraphics[height=90pt]{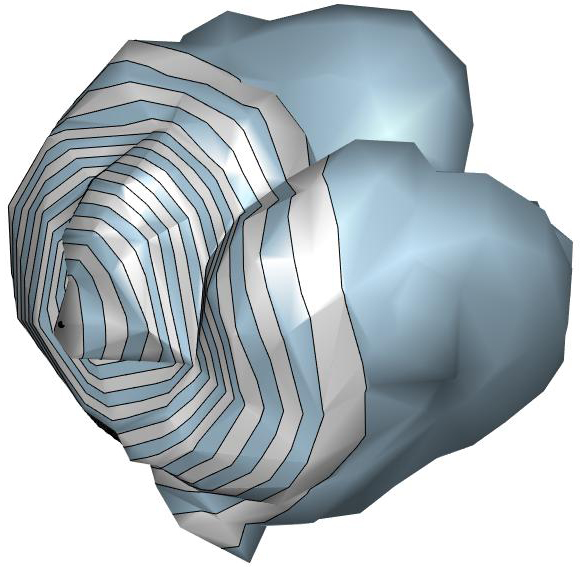}
&\includegraphics[height=90pt]{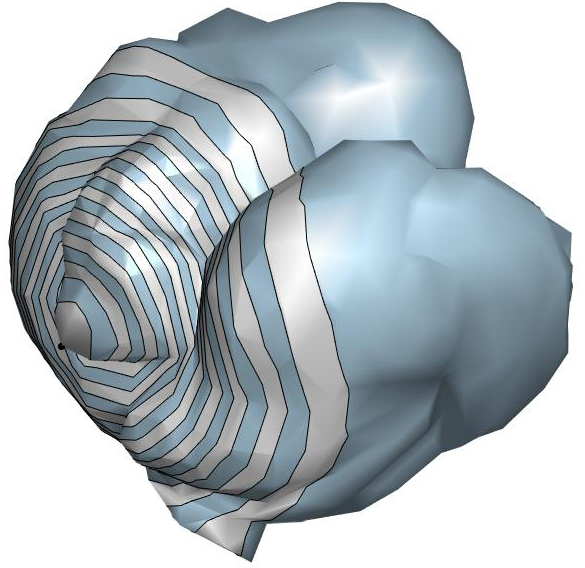}
&\includegraphics[height=90pt]{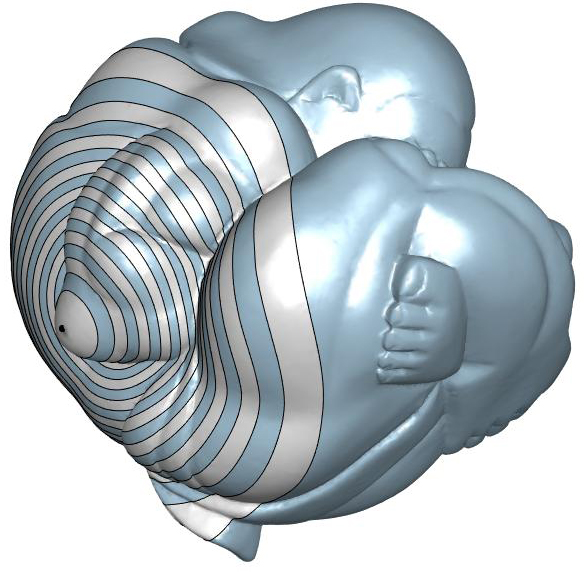}\\
$n=200$ &$n=500$ &$n=60K$
\end{tabular}
\caption{Robustness of the Pad\'e-Chebyshev approximation of the diffusion basis functions with respect to different resolutions ($n$ vertices).\label{fig_OMOTONDO-SIMPLIFIED}}
\end{figure}
\paragraph*{Area and conformal metric}
To compare two functions \mbox{$f,g:\mathcal{N}\rightarrow\mathbb{R}$} on the same surface~$\mathcal{N}$, we consider the
\begin{itemize}
\item \emph{area distortion}: \mbox{$h_{a}(f,g):=\int_{\mathcal{N}}f(\mathbf{p})g(\mathbf{p})\textrm{d}\mathbf{p}$};
\item \emph{conformal distortion}: \mbox{$h_{c}(f,g):=\int_{\mathcal{N}}\langle \nabla f(\mathbf{p}),\nabla g(\mathbf{p})\rangle_{2}\textrm{d}\mathbf{p}$}. 
\end{itemize}
The area and conformal distortions are invariant with respect to area-preserving and conformal transformations, respectively. In fact, let~$\mathcal{N}$,~$\mathcal{Q}$ be 2 surfaces and \mbox{$\mathcal{F}(\mathcal{Q})$} the space of scalar functions defined on~$\mathcal{Q}$. Given a map \mbox{$T:\mathcal{N}\rightarrow\mathcal{Q}$}, we have that~\citep{RUSTAMOV2011} for all \mbox{$f,g\in\mathcal{F}(\mathcal{Q})$}, \mbox{$h_{a}^{\mathcal{Q}}(f,g)=h_{a}^{\mathcal{N}}(f\circ T,g\circ T)$} if and only if~$T$ is locally area-preserving and \mbox{$h_{c}^{\mathcal{Q}}(f,g)=h_{c}^{\mathcal{N}}(f\circ T,g\circ T)$} if and only if~$T$ is conformal. For instance, for the Laplacian eigenbasis we have that
\begin{equation*}\label{eq:METRIC-EIGS}
h_{a}(\phi_{i},\phi_{j})=\delta_{ij},\qquad
h_{c}(\phi_{i},\phi_{j})=\langle\phi_{i},\Delta\phi_{j}\rangle_{2}=\lambda_{j}\delta_{ij}.
\end{equation*}
The discrete area distortion is \mbox{$h_{a}(f,g)=\langle\mathbf{f},\mathbf{g}\rangle_{\mathbf{B}}=\mathbf{f}^{\top}\mathbf{B}\mathbf{g}$}, where~$\mathbf{B}$ is the area-driven mass matrix (Sect.~\ref{sec:LAPLACE-BELTRAMI}), and the discrete conformal distortion is
\begin{equation*}
h_{c}(f,g)
=\langle f,\Delta g\rangle_{2}
\approx\mathbf{f}^{\top}\mathbf{B}\tilde{\mathbf{L}}\mathbf{g}
=\langle\mathbf{f},	\mathbf{g}\rangle_{\mathbf{L}}.
\end{equation*}
Expressing the input functions as a linear combination of the Laplacian eigenbasis with coefficients \mbox{$\alpha=\mathbf{X}^{\top}\mathbf{B}\mathbf{f}$}, \mbox{$\beta=\mathbf{X}^{\top}\mathbf{B}\mathbf{g}$}, we have that \mbox{$h_{a}(f,g)=\alpha^{\top}\beta$} and \mbox{$h_{c}(f,g)=\alpha^{\top}\Gamma\beta$}. For example, the area and conformal distortion of the delta functions at~$\mathbf{p}_{i}$,~$\mathbf{p}_{j}$ reduce to the entry \mbox{$(i.j)$} of~$\mathbf{B}$ and~$\mathbf{L}$, respectively. For the Laplacian eigenvectors, we have that \mbox{$h_{a}(\mathbf{x}_{i},\mathbf{x}_{j})=\delta_{ij}$} and \mbox{$h_{c}(\mathbf{x}_{i},\mathbf{x}_{j})=\lambda_{j}\delta_{ij}$}. Fig.~\ref{fig:3TORUS-METIRCS} shows the area and conformal metrics of the diffusion basis functions at 100 randomly sampled seed points and at 4 scales.
\begin{figure}[t]
\centering
\begin{tabular}{cc|cc}
\includegraphics[height=75pt]{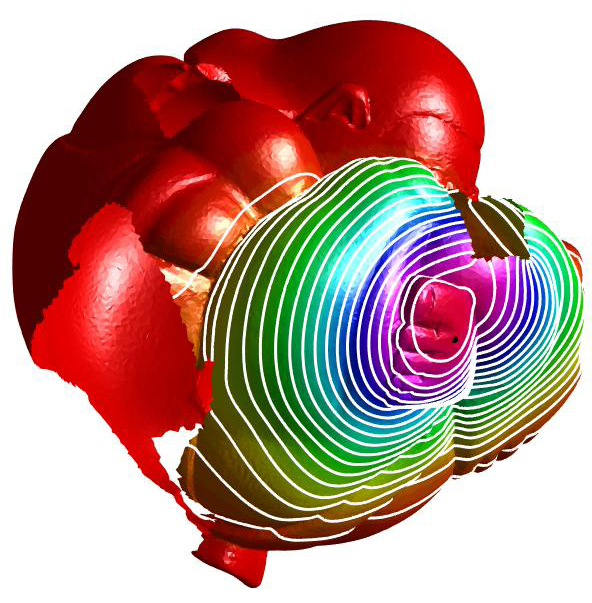}
&\includegraphics[height=75pt]{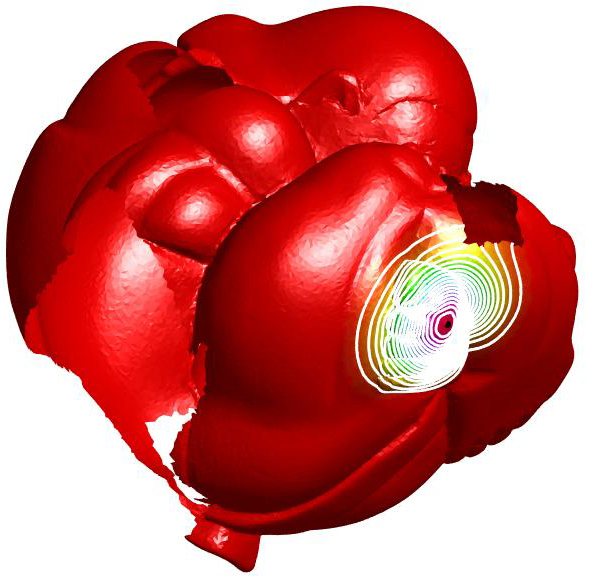}
&\includegraphics[height=75pt]{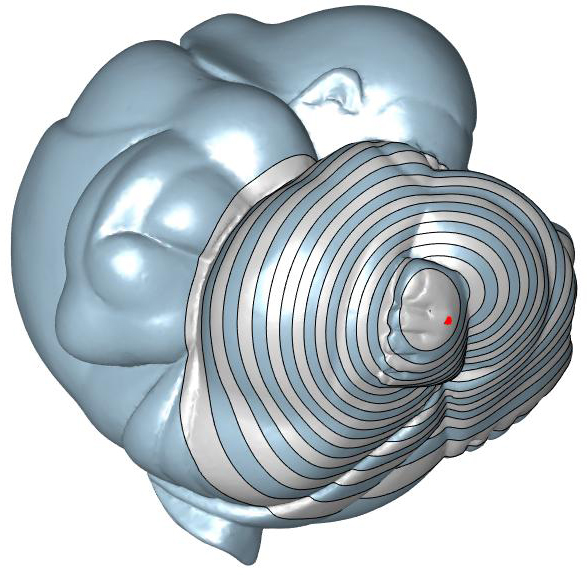}
&\includegraphics[height=75pt]{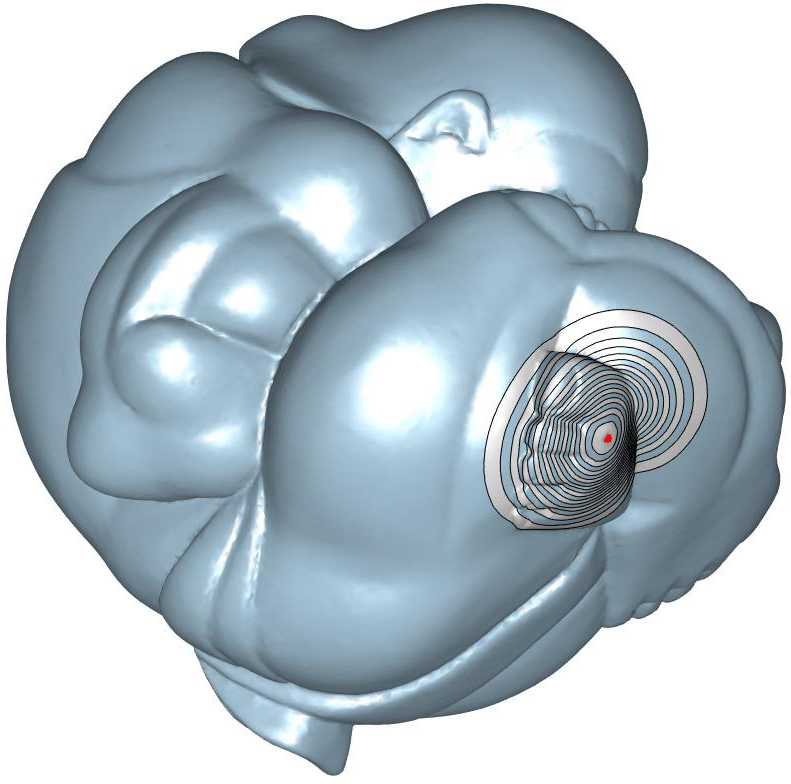}\\
$t=0.1$ &$t=0.01$ &$t=0.1$ &$t=0.01$
\end{tabular}
\caption{Stability of the Pad\'e-Chebyshev computation of the diffusion basis functions with respect to incomplete surfaces.\label{fig:HOLE-ROBUSTNESS}}
\end{figure}
\paragraph*{Kernel-based metric}
Any symmetric and positive kernel \mbox{$K:\mathcal{N}\times\mathcal{N}\rightarrow\mathbb{R}$} (e.g., the Gaussian kernel) induces the following \emph{kernel-based inner product}
\begin{equation}\label{eq:KERNEL-SP}
\langle f,g\rangle:=\int_{\mathcal{N}\times\mathcal{N}}K(\mathbf{p},\mathbf{q})f(\mathbf{p})g(\mathbf{q})\textrm{d}\mathbf{p}\textrm{d}\mathbf{q}.
\end{equation}
%
%
%
Combining the spectral representation (\ref{eq:SPECTRAL-KERNEL}) of~$K_{\varphi}$ with the definition of the kernel-based inner product (\ref{eq:KERNEL-SP}), we get that
\begin{equation*}
\langle f,g\rangle
=\sum_{n=0}^{+\infty}\varphi(\lambda_{n})\langle f,\phi_{n}\rangle_{2}\langle g,\phi_{n}\rangle_{2}.
\end{equation*}
%
%
The discretisation of (\ref{eq:KERNEL-SP}) is \mbox{$\langle\mathbf{f},\mathbf{g}\rangle=\mathbf{f}^{\top}\mathbf{B}\mathbf{K}\mathbf{g}\\
=\langle\mathbf{f},\mathbf{K}\mathbf{g}\rangle_{\mathbf{B}}$}, where \mbox{$\mathbf{K}:=(K(\mathbf{p}_{i},\mathbf{p}_{j}))_{i,j=1}^{k}$} is the Gram matrix associated with the kernel \mbox{$K(\cdot,\cdot)$}. To guarantee the symmetry of this inner product, it is enough to assume that~$\mathbf{K}$ is positive definite and ~$\mathbf{B}$-adjoint; i.e., \mbox{$\langle\mathbf{f},\mathbf{K}\mathbf{g}\rangle_{\mathbf{B}}=\langle\mathbf{K}\mathbf{f},\mathbf{g}\rangle_{\mathbf{B}}$}. As main examples of kernels, we mention the filtered spectral kernel (e.g., the diffusion kernel), which is~$\mathbf{B}$-adjoint, and the Laplacian matrix with cotangent weights (Sect.~\ref{sec:LAPLACE-BELTRAMI}), which is symmetric.
\begin{figure}[t]
\centering
\begin{tabular}{cc}
(a)\includegraphics[width=110pt]{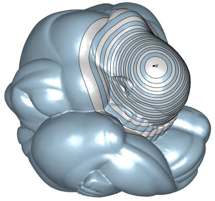}
&(b)\includegraphics[width=130pt]{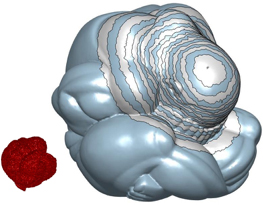}\\
(c)\includegraphics[width=130pt]{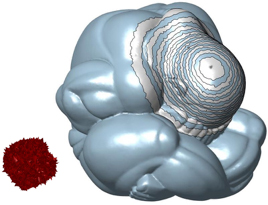}
&(d)\includegraphics[width=130pt]{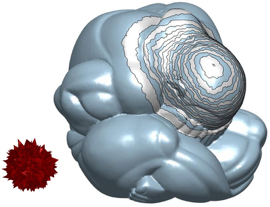}
\end{tabular}
\caption{Robustness of the computation of the diffusion basis functions centred at a seed point placed on the head. Level-sets of the diffusion basis functions on (b-d) the noisy shapes (bottom part, red surfaces) have been plotted on (a) the initial shape.\label{fig:SHAPE-NOISE}}
\end{figure}
\section{Discussion and examples\label{sec:GENERAL-EXAMPLES-DISCUSSION}}
We describe a simple criterion for the selection of the seed points of the spectral basis functions (Sect.~\ref{sec:SEED-SELECTION}); then, we discuss their main properties (Sect.~\ref{sec:PROPERTIES}) our experiments (Sect.~\ref{sec:EXAMPLES}) on 3D shapes.

\subsection{Selection of the seed points\label{sec:SEED-SELECTION}}
Since the diffusion basis functions at small scales and the spectral basis functions induced by filters with a strong decay to zero are compactly-supported, the coverage of the whole surface with their support depends on the selected seeds and on the support of each basis function. Increasing the number of seed points or the scale~$t$, or reducing the filter decay to zero, all the vertices of the input mesh will belong to the support of one or more basis functions (Fig.~\ref{fig:SEEDS}). The seed points can be selected as high-curvature points, or by uniformly sampling the input 3D shape, or by applying a sampling or clustering method (e.g., the farthest point sampling, the principal component analysis).

In our approach, at the first step we select a seed point (e.g., a point of maximum curvature) and define~$k$ (e.g., \mbox{$k:=10$}, in our experiments) seed points by applying the farthest point sampling method~\citep{ELDAR2007,MOENNING2003}. Then, we compute the corresponding set~$\mathcal{B}$ of basis functions and identify the points of the input surface that are not covered by the support of the basis functions in~$\mathcal{B}$. These points are then clustered and the representative points of these clusters provide the seeds for the new basis functions, which will be included in~$\mathcal{B}$. This process proceeds until all the vertices belong to the support of at least one basis function. At each iteration, the seed points are selected among the vertices of the input surface and are marked as visited.
\begin{table*}[t]
\caption{Overview of the main properties (Sect.~\ref{sec:INTRODUCTION}) of the basis functions discussed in this paper. The symbols~$\bullet$,~$\circ$,~$\otimes$ indicate that the corresponding property is satisfied, is not satisfied, or might be satisfied according to additional conditions, which are detailed in Sect.~\ref{sec:GENERAL-EXAMPLES-DISCUSSION}.\label{tab:OVERVIEW-BASIS-PROPERTIES}}
\centering
{\tiny{
\begin{tabular}{|l|l|l|l|l|l|l|l|}
\hline
\textbf{Property}									&\textbf{Harm.}		&\textbf{Hamil.} 	&\textbf{Lapl. }		&\textbf{Lapl. compr. }			&\textbf{Hamil.}		&\textbf{Spectral} 	&\textbf{Diff.}\\
													&\textbf{basis}		&\textbf{basis}		&\textbf{eigenf.}		&\textbf{modes}					&\textbf{eigenf.}		&\textbf{basis}		&\textbf{basis}\\
\hline
Partition of unity									&$\otimes$			&$\otimes$			&$\otimes$				&$\otimes$								&$\otimes$				&$\otimes$			&$\otimes$\\
\hline
Non-negativity										&$\bullet$			&$\circ$			&$\circ$				&$\circ$								&$\circ$				&$\circ$			&$\bullet$\\
\hline
Intrinsic def.										&$\bullet$			&$\bullet$			&$\bullet$				&$\bullet$								&$\bullet$				&$\bullet$			&$\bullet$\\
\hline
Locality											&$\circ$			&$\circ$			&$\circ$				&$\bullet$								&$\circ$				&$\otimes$			&$\otimes$\\
\hline
Orthogonality 										&$\otimes$			&$\otimes$			&$\bullet$				&$\bullet$								&$\bullet$				&$\otimes$			&$\otimes$\\
\hline
Isometry-inv.										&$\bullet$			&$\circ$			&$\bullet$				&$\bullet$								&$\circ$				&$\bullet$			&$\bullet$\\
\hline\
Numer. stability 									&$\bullet$			&$\bullet$			&$\otimes$				&$\otimes$								&$\bullet$				&$\bullet$			&$\bullet$\\
\hline\
Comput. cost~$\mathcal{O}(\cdot)$ 					&$n$				&$n$				&$kn\log n$				&$kn\log n$								&$kn\log n$				&$rn\log n$ 		&$rn\log n$\\
\hline		
Storage overhead~$\mathcal{O}(\cdot)$ 				&$n$				&$n$				&$kn^{2}$				&$kn$									&$kn^{2}$				&$n$				&$n$\\
\hline
\end{tabular}}}
\end{table*}
\subsection{Properties of the basis functions\label{sec:PROPERTIES}}
For each of type of basis, we discuss the following properties
\begin{itemize}
\item \emph{partition of the unity} \mbox{$\sum_{i}\psi_i(\mathbf{p})=1$} and \emph{non-negativity} \mbox{$\psi_i(\mathbf{p})\geq 0$};
\item \emph{orthogonality} with respect to an inner product on the space of scalar functions defined on the input shape;
\item \emph{locality}: the support  \mbox{$\textrm{supp}(\psi):=\overline{\{\mathbf{p}\in\mathcal{N}:\,\psi(\mathbf{p})\neq 0\}}$} of the basis function is compact;
\item \emph{smoothness}, \emph{intrinsic definition} with respect to the domain metric and/or encoding of local geometric properties (e.g., curvature, geodesic distance), and \emph{invariance} to isometric transformations (e.g., for shape analysis and comparison);
\item \emph{sparsity of the representation}: scalar functions are accurately approximated by a small number of basis functions;
\item \emph{functional stability of the basis} with respect to domain discretisation (e.g., connectivity, sampling, geometric/topological noise) and \emph{stability of the representation with respect to a basis} under small shape perturbations;
\item \emph{computational cost and storage overhead} feasible for applications in geometry processing and shape analysis.
\end{itemize}
Table~\ref{tab:OVERVIEW-BASIS-PROPERTIES} summarises the properties discussed in Sect.~\ref{sec:INTRODUCTION} for each set of basis functions \mbox{$(\psi_{i})_{i}$}, defined with one the methods previously described. For those functions that might satisfy a given property under some additional hypothesis, indicated with the symbol~$\otimes$, we notice that the \emph{partition of the unity} can be always induced by normalising the basis functions as \mbox{$\psi_{i}/\sum_{j}\psi_{j}$}. However, this normalisation generally alters other properties, such as the general Lagrange property. The \emph{locality} is satisfied by the diffusion basis function at small scales and by the spectral basis functions if the corresponding filter rapidly decays to zero. 

The \emph{non-negativity} is satisfied by the harmonic and diffusion bases, as a result of the maximum principle for the Laplace and heat diffusion equations. By definition, the compressed manifold modes are local and orthonormal; however, we cannot center these basis functions at a given seed point and easily control their support, as we can do with the diffusion basis functions. The \emph{orthogonality} can be induced by applying the Gram-Schmidt method, at the cost of altering other properties of the input functions. Two diffusion basis functions without overlapping supports are always orthogonal; the orthogonality of a set of diffusion basis functions with overlapping supports can be achieved with the Gram-Schmidt method at the cost of a (generally) larger support of the orthonormal diffusion basis functions, as a matter of the linear combination of functions with a different support size during the orthonormalisation iterations. The \emph{numerical stability} is generally satisfied by all those basis functions that are computed through the solution of a linear system (e.g., harmonic, Hamiltonian, spectral basis functions, etc.). Numerical instabilities are generally associated with the computation of the Laplacian and Hamiltonian eigenbasis associated with multiple or numerically close eigenvalues.
\begin{figure}[t]
\centering
\begin{tabular}{ccc}
\hline
$t=10^{-2}$ &$t=10^{-1}$ &$t=1$\\
\hline
\includegraphics[height=80pt]{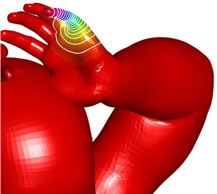}
&\includegraphics[height=80pt]{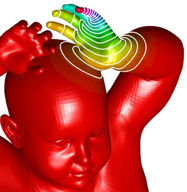}
&\includegraphics[height=80pt]{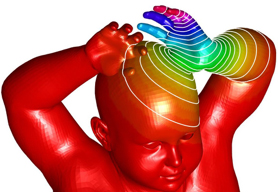}\\
\includegraphics[height=80pt]{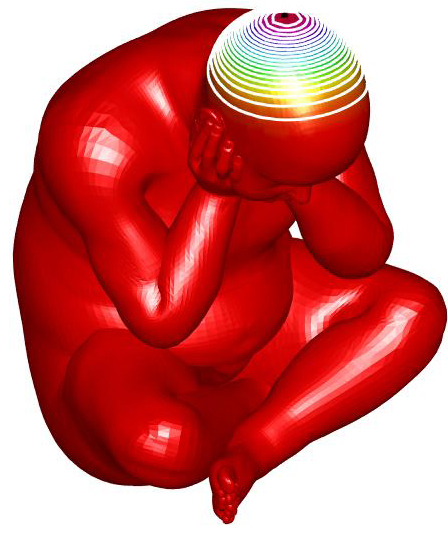}
&\includegraphics[height=80pt]{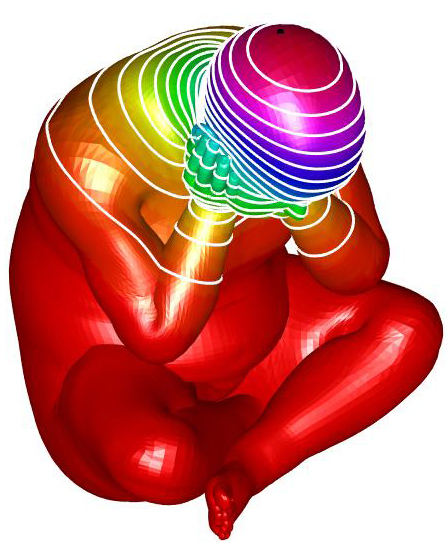}
&\includegraphics[height=80pt]{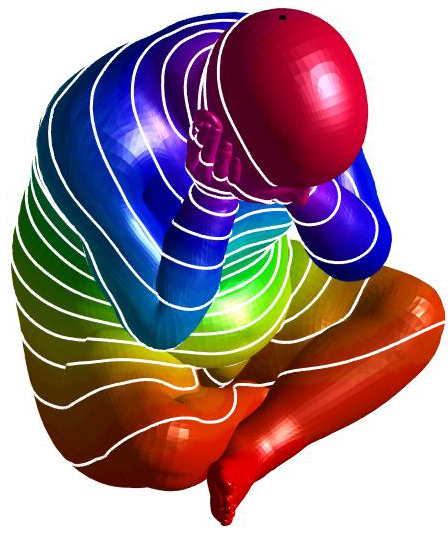}\\
\hline\hline
\end{tabular}
\begin{tabular}{cc}
\includegraphics[height=80pt]{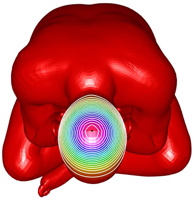}
&\includegraphics[height=80pt]{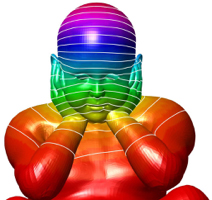}\\
$t=10^{-2}$ &$t=10^{-1}$\\
\end{tabular}
\caption{Robustness of the Pad\'e-Chebyshev computation of the diffusion basis functions at different scales with respect to (fist, second row) different poses. (third row) Zoom-in.\label{fig:TOPOLOGICAL-NOISE}}
\end{figure}
\subsection{Examples\label{sec:EXAMPLES}}
For an arbitrary 3D shape, we cannot compute the ground-truth diffusion basis functions at a given seed point through the analytic representation of its Laplacian eigenfunctions (e.g., as for the sphere and the cylinder). Recalling the upper bound to the accuracy of the Pad\'e-Chebyshev approximation of the diffusion basis with respect to the selected degree of the rational polynomial, we can analyse the discrepancy of this approximation with respect to the truncated spectral approximation. In Fig.~\ref{fig:KITTEN-CHEB-VS-SPECTRAL}, we report the~$\ell_{\infty}$ error ($y$-axis) between the Pad\'e-Chebyshev approximation of the diffusion basis at different seed points and scales, and the truncated spectral approximation with a different number ($x$-axis) of Laplacian eigenpairs. At large scales (e.g., \mbox{$t=10^{-1},1$}), the truncated spectral approximation with a low number of eigenpairs (e.g., \mbox{$k\approx 50$}) provides an approximation of the solution to the heat equation that is close to the one computed by the Pad\'e-Chebyshev approximation, with respect to the~$\ell_{\infty}$-norm (\mbox{$\epsilon_{\infty}=10^{-4}$}). 
\begin{figure*}[t]
\centering
\begin{tabular}{cccc}
\includegraphics[height=120pt]{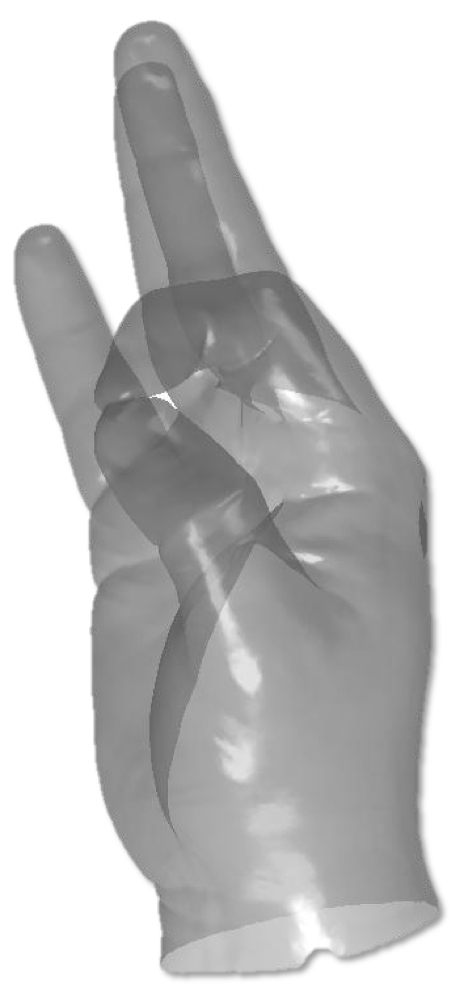}
&\includegraphics[height=120pt]{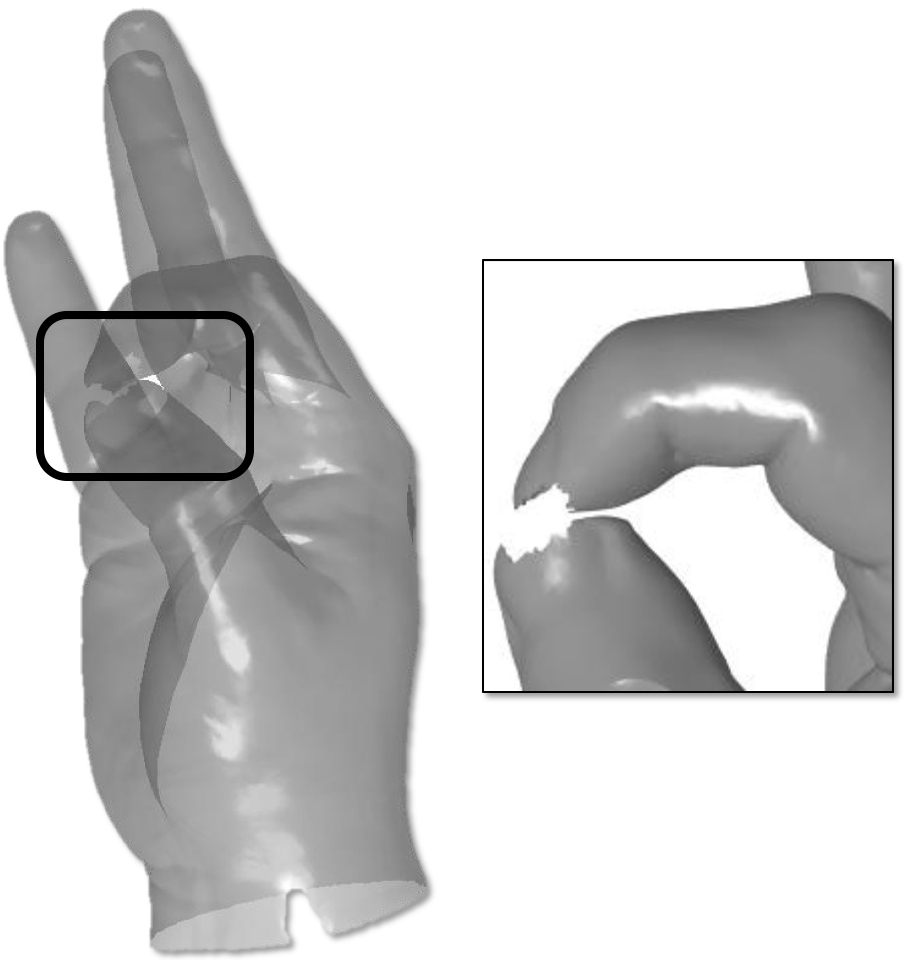}
&\includegraphics[height=120pt]{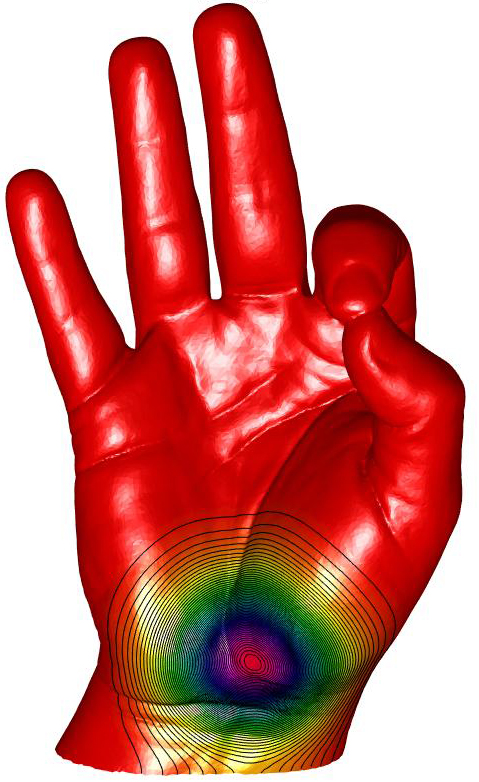}
&\includegraphics[height=120pt]{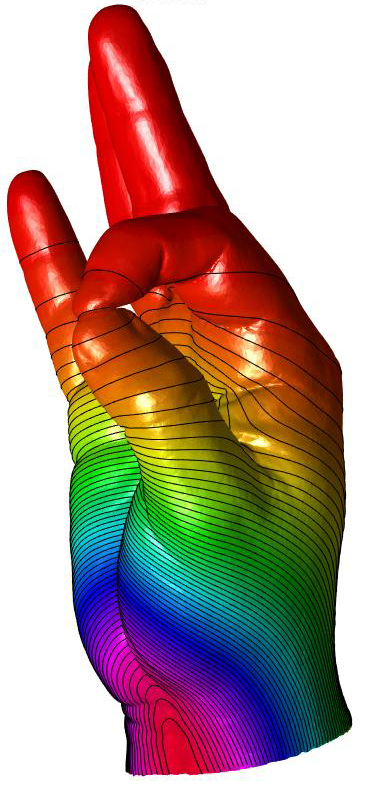}\\
(a) &(b) &(c)~$t=0.01$ &(d)~$t=0.1$\\
\includegraphics[height=120pt]{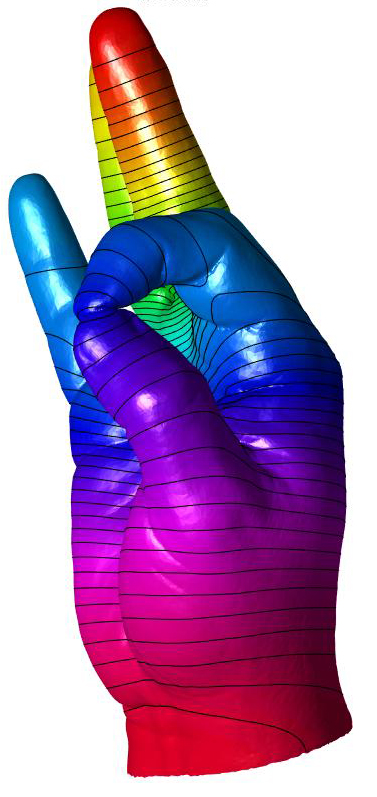}
&\includegraphics[height=120pt]{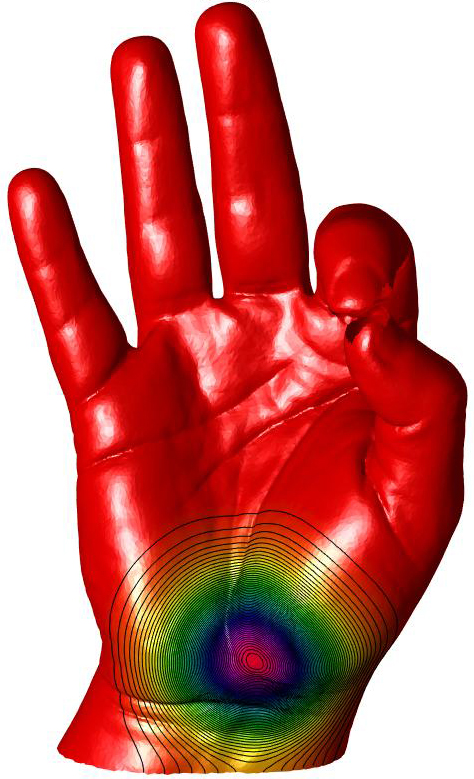}
&\includegraphics[height=120pt]{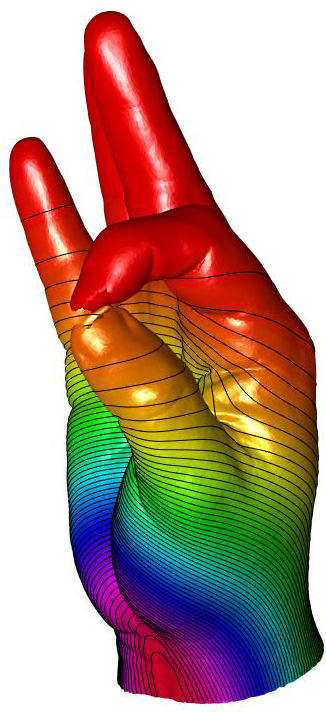}
&\includegraphics[height=120pt]{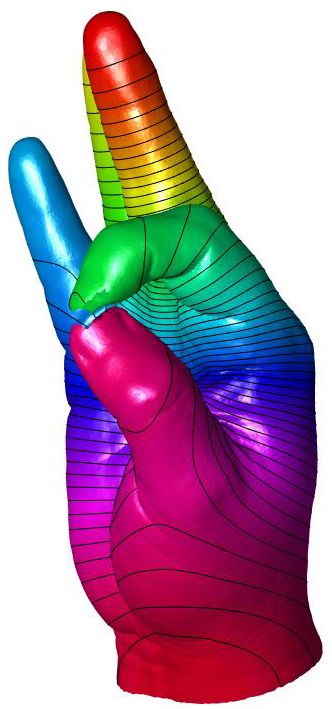}\\
(e)~$t=1$ &(f)~$t=0.01$ &(g)~$t=0.1$ &(h)~$t=1$
\end{tabular}
\caption{Input 3D shape with (a) one topological handle, (b) sliced surface around the thumb, and zoom-in. At small/medium scales (e.g., \mbox{$t=0.01$}, \mbox{$t=0.1$}) the behaviour of the diffusion basis functions on the (c, d) input and (f,g) sliced surface is consistent in terms of shape and distribution of the level-sets. At large scales (e.g., \mbox{$t=1$}), the diffusion basis function (e) changes only in a neighbourhood of the cut (h).\label{fig:HAND-HOLE}}
\end{figure*}

The level-sets of these two approximations also show their different behaviour at small scales, while at larger scales it becomes analogous in terms of both the shape and distribution of the level-sets. At small scales (e.g., \mbox{$t=10^{-2},10^{-3}$}), the truncated spectral approximation is generally affected by small undulations far from the seed point and that do not disappear while increasing the number of Laplacian eigenpairs. Increasing the number of Laplacian eigenpairs makes these undulation more evident, as a matter of the small vibrations of the Laplacian eigenvectors associated with larger Laplacian eigenvalues, which are included in the truncated approximation. In this case, a larger number of eigenpairs is necessary to obtain a truncated spectral approximation close to the one computed with the Pad\'e-Chebyshev rational polynomial. 

The truncated spectral approximation of the diffusion basis is generally affected by the Gibbs phenomenon. This phenomenon is more evident at small cases, as the kernel values decrease fast to zero and are largely affected by small negative values (Fig.~\ref{fig:GIBBS-PHENOMENON}(e,f)). In fact, at small scales the values of the diffusion basis functions (or equivalently, the diffusion kernel) decrease fast to zero and the negative values are no more compensated by the Laplacian eigenvectors related to smaller eigenvalues, as they are not included in the approximation. For the Pad\'e-Chebyshev approximation (Fig.~\ref{fig:GIBBS-PHENOMENON}(a-d)), the kernel values are positive at all the scales; in fact, we approximate the filter function without selecting a sub-part of the Laplacian spectrum.
\begin{figure}[t]
\centering
\includegraphics[width=300pt]{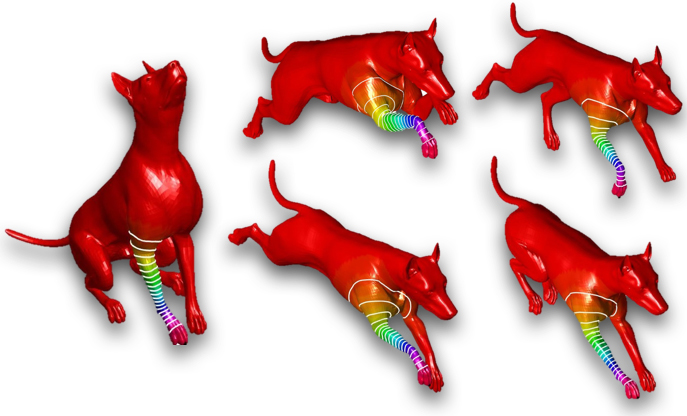}
\caption{Robustness of the computation of the linear FEM diffusion basis functions with respect to shape deformations.\label{fig:SHAPE-DEFORMATION}}
\end{figure}

We have also tested the robustness of the spectrum-free computation with respect to a different shape discretisation (Fig.~\ref{fig_OMOTONDO-SIMPLIFIED}), bordered surfaces (Fig.~\ref{fig:HOLE-ROBUSTNESS}), geometric (Fig.~\ref{fig:SHAPE-NOISE}) or topological (Fig.~\ref{fig:TOPOLOGICAL-NOISE}, Fig.~\ref{fig:HAND-HOLE}) noise, and almost isometric deformations (Fig.~\ref{fig:SHAPE-DEFORMATION}). A higher resolution (Fig.~\ref{fig_OMOTONDO-SIMPLIFIED}) of the input surface improves the quality of the level-sets, which are always uniformly distributed, and an increase of the noise magnitude (Fig.~\ref{fig:SHAPE-NOISE}) does not affect the shape and distribution of the level-sets. In case of topological noise (Fig.~\ref{fig:HAND-HOLE}), the shape of the level-sets of the diffusion basis functions at large scales (e.g., \mbox{$t=1$}) changes only in a neighbourhood of the topological cut. At small/medium scales (e.g., \mbox{$t=0.01$}, \mbox{$t=0.1$}), the geometry of the surface around the cut has no influence on the diffusion basis functions, as a matter of their local supports. In our experiments, the analogous behaviour of the level-sets of the diffusion basis functions confirms the robustness of the Pad\'e-Chebyshev of the approximation with respect to sampling and discretisation.

From the previous experiments, we conclude that the diffusion basis functions and the Laplacian spectral functions provide a valid alternative to the harmonic functions and Laplacian eigenfunctions. In fact, the diffusion basis functions can be centred at any point of the input domain, encode local/global shape details according to the values of the temporal parameter, and encode geometric information in terms of curvature values and geometric details.

\section{Conclusions and future work\label{sec:CONCLUSION}}
Representing a signal as a linear combination of a set of basis functions is used in a wide range of applications, such as approximation, de-noising, and compression. In this context, we have focused our attention on the main aspects of signal approximation, such as the definition, computation, and comparison of basis functions induced by the Laplace-Beltrami operator. In particular, we have introduced the diffusion and Laplacian spectral basis functions, which encode intrinsic shape properties, are multi-scale and sparse, efficiently computed, and provide an alternative to the harmonic functions and to the Laplacian eigenfunctions. As future work, we mention the design of filters that support user-driven constraints on the resulting spectral basis functions and the use of more general classes of differential operators, such as the Dirac operator~\citep{LIU2017} and elliptic operators.

{\small{\section*{Acknowledgements}
We thank the Reviewers for their constructive comments, which helped us to improve the paper presentation and content, and the members of the Shapes$\&$Semantics Modelling Group at CNR-IMATI for helpful discussions. This work has been partially supported by the H2020 ERC Advanced Grant CHANGE (Project Agreement N. 694515). Shapes are courtesy of the AIM{@}SHAPE Repository, the TOSCA, SHREC 2010, and SHREC 2016 data sets.}}  

\section*{References}
\begin{thebibliography}{56}
\expandafter\ifx\csname natexlab\endcsname\relax\def\natexlab#1{#1}\fi
\expandafter\ifx\csname url\endcsname\relax
  \def\url#1{\texttt{#1}}\fi
\expandafter\ifx\csname urlprefix\endcsname\relax\def\urlprefix{URL }\fi

\bibitem[{Aflalo et~al.(2016)Aflalo, Brezis, Bruckstein, Kimmel, and
  Sochen}]{AFLALO2016-BB}
Aflalo, Y., Brezis, H., Bruckstein, A., Kimmel, R., Sochen, N., 2016. Best
  bases for signal spaces. Comptes Rendus Mathematique 354~(12), 1155 -- 1167.

\bibitem[{Aflalo et~al.(2015)Aflalo, Brezis, and R.Kimmel}]{AFLALO2015}
Aflalo, Y., Brezis, H., R.Kimmel, 2015. On the optimality of shape and data
  representation in the spectral domain. {SIAM} Journal Imaging Sciences 8~(2),
  1141--1160.

\bibitem[{Alexa and Wardetzky(2011)}]{ALEXA2011}
Alexa, M., Wardetzky, M., 2011. Discrete {L}aplacians on general polygonal
  meshes. ACM Trans. on Graphics 30~(4).

\bibitem[{Alliez et~al.(2005)Alliez, Cohen-Steiner, Yvinec, and
  Desbrun}]{ALLIEZ2005}
Alliez, P., Cohen-Steiner, D., Yvinec, M., Desbrun, M., 2005. Variational
  tetrahedral meshing. ACM Trans. on Graphics 24~(3), 617--625.

\bibitem[{Andreux et~al.(2014)Andreux, Rodola, Aubry, and
  Cremers}]{ANDREUX2014}
Andreux, M., Rodola, E., Aubry, M., Cremers, D., 2014. Anisotropic
  {L}aplace-{B}eltrami operators for shape analysis. In: Sixth Workshop on
  Non-Rigid Shape Analysis and Deformable Image Alignment (NORDIA).

\bibitem[{Aubry et~al.(2011)Aubry, Schlickewei, and Cremers}]{AUBRY2011}
Aubry, M., Schlickewei, U., Cremers, D., 2011. The wave kernel signature: a
  quantum mechanical approach to shape analysis. In: IEEE Computer Vision
  Workshops. pp. 1626--1633.

\bibitem[{Barekat et~al.(2017)Barekat, Caflisch, and Osher}]{BAREKAT2017}
Barekat, F., Caflisch, R., Osher, S., 2017. On the support of compressed modes.
  SIAM Journal on Mathematical Analysis 49~(4), 2573--2590.

\bibitem[{Belkin and Niyogi(2003)}]{BELKIN2003}
Belkin, M., Niyogi, P., 2003. {L}aplacian eigenmaps for dimensionality
  reduction and data representation. Neural Computations 15~(6), 1373--1396.

\bibitem[{Belkin and Niyogi(2006)}]{BELKIN2006}
Belkin, M., Niyogi, P., 2006. Convergence of {L}aplacian eigenmaps. In: Neural
  Information Processing Systems. pp. 129--136.

\bibitem[{Belkin and Niyogi(2008)}]{BELKIN2008}
Belkin, M., Niyogi, P., 2008. Towards a theoretical foundation for
  {L}aplacian-based manifold methods. Journal of Computer System Sciences
  74~(8), 1289--1308.

\bibitem[{Belkin et~al.(2008)Belkin, Sun, and Wang}]{BELKIN2008-MESH}
Belkin, M., Sun, J., Wang, Y., 2008. Discrete {L}aplace operator on meshed
  surfaces. In: Proc. of the Twenty-fourth Annual Symp. on Computational
  Geometry. pp. 278--287.

\bibitem[{Belkin et~al.(2009)Belkin, Sun, and Wang}]{BELKIN2009}
Belkin, M., Sun, J., Wang, Y., 2009. Constructing Laplace operator from point
  clouds in $\mathbb{R}^{d}$. ACM Press, Ch. 112, pp. 1031--1040.

\bibitem[{Ben{-}Chen and Gotsman(2005)}]{BENCHEN2005}
Ben{-}Chen, M., Gotsman, C., 2005. On the optimality of spectral compression of
  mesh data. {ACM} Trans. Graph. 24~(1), 60--80.

\bibitem[{Bronstein and Bronstein(2011)}]{BRONSTEIN-PAMI2011}
Bronstein, M., Bronstein, A., 2011. Shape recognition with spectral distances.
  IEEE Trans. on Pattern Analysis and Machine Intelligence 33~(5), 1065 --1071.

\bibitem[{Buffa et~al.(2013)Buffa, Harbrecht, Kunoth, and Sangalli}]{BUFFA2013}
Buffa, A., Harbrecht, H., Kunoth, A., Sangalli, G., 2013. {BPX}-preconditioning
  for isogeometric analysis. Computer Methods in Applied Mechanics and
  Engineering 265, 63 -- 70.

\bibitem[{Carpenter et~al.(1984)Carpenter, Ruttan, and Varga}]{CARPENTER1984}
Carpenter, A., Ruttan, A., Varga, R., 1984. Extended numerical computations on
  the ``1/9'' conjecture in rational approximation theory. In: Rational
  Approximation and Interpolation. Vol. 1105 of Lecture Notes in Mathematics.
  Springer, pp. 383--411.

\bibitem[{Chuang et~al.(2009)Chuang, Luo, Brown, Rusinkiewicz, and
  Kazhdan}]{CHUANG2009}
Chuang, M., Luo, L., Brown, B.~J., Rusinkiewicz, S., Kazhdan, M., 2009.
  Estimating the {L}aplace-{B}eltrami operator by restricting {{3D}} functions.
  In: Proc. of the Symp. on Geometry Processing. pp. 1475--1484.

\bibitem[{Desbrun et~al.(1999)Desbrun, Meyer, Schr\"{o}der, and
  Barr}]{DESBRUN1999}
Desbrun, M., Meyer, M., Schr\"{o}der, P., Barr, A.~H., 1999. Implicit fairing
  of irregular meshes using diffusion and curvature flow. In: ACM Siggraph. pp.
  317--324.

\bibitem[{Donatelli et~al.(2015)Donatelli, Garoni, Manni, Serra-Capizzano, and
  Speleers}]{DONATELLI2015}
Donatelli, M., Garoni, C., Manni, C., Serra-Capizzano, S., Speleers, H., 2015.
  Robust and optimal multi-iterative techniques for {IgA} {G}alerkin linear
  systems. Computer Methods in Applied Mechanics and Engineering 284, 230 --
  264, isogeometric Analysis Special Issue.

\bibitem[{Donatelli et~al.(2017)Donatelli, Garoni, Manni, Serra-Capizzano, and
  Speleers}]{DONATELLI2017}
Donatelli, M., Garoni, C., Manni, C., Serra-Capizzano, S., Speleers, H., 2017.
  Symbol-based multigrid methods for {G}alerkin {B}-spline isogeometric
  analysis. SIAM Journal on Numerical Analysis 55~(1), 31--62.

\bibitem[{Eldar et~al.(1997)Eldar, Lindenbaum, Porat, and Zeevi}]{ELDAR2007}
Eldar, Y., Lindenbaum, M., Porat, M., Zeevi, Y.~Y., Sep. 1997. The farthest
  point strategy for progressive image sampling. Trans. Img. Proc. 6~(9),
  1305--1315.

\bibitem[{Falgout(2006)}]{FALGOUT2006}
Falgout, R.~D., Nov. 2006. An introduction to algebraic multigrid. Computing in
  Science and Engineering 8~(6), 24--33.

\bibitem[{Floater(2003)}]{FLOATER2003}
Floater, M.~S., 2003. Analysis of curve reconstruction by meshless
  parameterization. Numerical Algorithms~(32), 87--98.

\bibitem[{Fowlkes et~al.(2004)Fowlkes, Belongie, Chung, and
  Malik}]{FOWLKES2004}
Fowlkes, C., Belongie, S., Chung, F., Malik, J., 2004. Spectral grouping using
  the {N}ystrom method. IEEE Trans. Pattern Analysis and Machine Intelligence
  26~(2), 214--225.

\bibitem[{Garoni et~al.(2014)Garoni, Manni, Pelosi, Serra-Capizzano, and
  Speleers}]{GARONI2014}
Garoni, C., Manni, C., Pelosi, F., Serra-Capizzano, S., Speleers, H., 2014. On
  the spectrum of stiffness matrices arising from isogeometric analysis.
  Numerische Mathematik 127~(4), 751--799.

\bibitem[{Golub and VanLoan(1989)}]{GOLUB1989}
Golub, G., VanLoan, G., 1989. Matrix Computations. John Hopkins University
  Press, 2nd Edition.

\bibitem[{Gordon and Szabo(2002)}]{GORDON2002}
Gordon, C.~S., Szabo, Z.~I., 06 2002. Isospectral deformations of negatively
  curved {R}iemannian manifolds with boundary which are not locally isometric.
  Duke Math. J. 113~(2), 355--383.

\bibitem[{Herholz et~al.(2015)Herholz, Kyprianidis, and Alexa}]{HERHOLZ2015}
Herholz, P., Kyprianidis, J.~E., Alexa, M., 2015. Perfect {L}aplacians for
  polygon meshes. Computer Graphics Forum 34~(5), 211--218.

\bibitem[{Hildebrandt et~al.(2006)Hildebrandt, Polthier, and
  Wardetzky}]{HILDEBRANDT2006}
Hildebrandt, K., Polthier, K., Wardetzky, M., 2006. On the convergence of
  metric and geometric properties of polyhedral surfaces. Geometria Dedicata,
  89--112.

\bibitem[{Hofreither and Takacs(2017)}]{HOFREITHER2017}
Hofreither, C., Takacs, S., 2017. Robust multigrid for isogeometric analysis
  based on stable splittings of spline spaces. SIAM Journal on Numerical
  Analysis 55~(4), 2004--2024.

\bibitem[{Hou and Qin(2012)}]{HOU2012}
Hou, T., Qin, H., 2012. Continuous and discrete {M}exican hat wavelet
  transforms on manifolds. Graphical Models 74~(4), 221--232.

\bibitem[{Huska et~al.(2018)Huska, Lazzaro, and Morigi}]{HUSKA2018}
Huska, M., Lazzaro, D., Morigi, S., Feb 2018. Shape partitioning via $l_{p}$
  compressed modes. Journal of Mathematical Imaging and Vision.

\bibitem[{Krishnan et~al.(2013)Krishnan, Fattal, and Szeliski}]{KRISHNAN2013}
Krishnan, D., Fattal, R., Szeliski, R., Jul. 2013. Efficient preconditioning of
  {L}aplacian matrices for {C}omputer {G}raphics. ACM Trans. on Graphics
  32~(4), 142:1--142:15.

\bibitem[{Lawrence(1997)}]{LAWRENCE1997}
Lawrence, C., 1997. Partial Differential Equations. American Mathematical
  Society.

\bibitem[{Lehoucq and Sorensen(1996)}]{LEHOUCQ1996}
Lehoucq, R., Sorensen, D.~C., 1996. Deflation techniques for an implicitly
  re-started {A}rnoldi iteration. SIAM Journal of Matrix Analysis and
  Applications 17, 789--821.

\bibitem[{Liao et~al.(2009)Liao, Tong, Dong, and Zhu}]{LIAO2009}
Liao, S., Tong, R., Dong, J., Zhu, F., 2009. Gradient field based inhomogeneous
  volumetric mesh deformation for maxillofacial surgery simulation. Computers
  $\&$ Graphics 33~(3), 424 -- 432.

\bibitem[{Liu et~al.(2017)Liu, Jacobson, and Crane}]{LIU2017}
Liu, H.~D., Jacobson, A., Crane, K., 2017. A dirac operator for extrinsic shape
  analysis. Computer Graphics Forum 36~(5), 139--149.

\bibitem[{Liu et~al.(2012)Liu, Prabhakaran, and Guo}]{LIU2012}
Liu, Y., Prabhakaran, B., Guo, X., 2012. Point-based manifold harmonics. IEEE
  Trans. on Visualization and Computer Graphics 18~(10), 1693 --1703.

\bibitem[{Moenning and Dodgson(2003)}]{MOENNING2003}
Moenning, C., Dodgson, N.~A., 2003. Fast marching farthest point sampling. In:
  Eurographics 2003 - Posters. Eurographics Association.

\bibitem[{Moler and Van~Loan(2003)}]{MOLER2003}
Moler, C., Van~Loan, C., 2003. Nineteen dubious ways to compute the exponential
  of a matrix, twenty-five years later. SIAM Review 45~(1), 3--49.

\bibitem[{Neumann et~al.(2014)Neumann, Varanasi, Theobalt, Magnor, and
  Wacker}]{NEUMANN2014}
Neumann, T., Varanasi, K., Theobalt, C., Magnor, M.~A., Wacker, M., 2014.
  Compressed manifold modes for mesh processing. Computer Graphics Forum
  33~(5), 35--44.

\bibitem[{Patan{\`{e}}(2014)}]{PATANE2014-PRL}
Patan{\`{e}}, G., 2014. {L}aplacian spectral distances and kernels on
  {{{{3D}}}} shapes. Pattern Recognition Letters 47, 102--110.

\bibitem[{Patan{\`{e}}(2016)}]{PATANE-STAR2016}
Patan{\`{e}}, G., 2016. {STAR} - {L}aplacian spectral kernels and distances for
  geometry processing and shape analysis. Computer Graphics Forum 35~(2),
  599--624.

\bibitem[{Patan\'e(2017)}]{PATANE2016}
Patan\'e, G., 2017. Accurate and efficient computation of {L}aplacian spectral
  distances and kernels. Computer Graphics Forum. In press. 36~(1), 184--196.

\bibitem[{Pinkall and Polthier(1993)}]{PINKALL1993}
Pinkall, U., Polthier, K., 1993. Computing discrete minimal surfaces and their
  conjugates. Experimental Mathematics 2~(1), 15--36.

\bibitem[{Ramani and Sinha(2013)}]{SINHA2013}
Ramani, K., Sinha, A., 2013. Multiscale kernels using random walks. Computer
  Graphics Forum 33~(1), 164--177.

\bibitem[{Reuter et~al.(2006)Reuter, Wolter, and Peinecke}]{reuter:cad06}
Reuter, M., Wolter, F.-E., Peinecke, N., 2006. {L}aplace-{B}eltrami spectra as
  {S}hape-{DNA} of surfaces and solids. Computer-Aided Design 38~(4), 342--366.

\bibitem[{Rosenberg(1997)}]{ROSENBERG1997}
Rosenberg, S., 1997. The {L}aplacian on a {R}iemannian Manifold. Cambridge
  University Press.

\bibitem[{Rustamov(2011)}]{RUSTAMOV2011}
Rustamov, R.~M., 2011. Multiscale biharmonic kernels. Computer Graphics Forum
  30~(5), 1521--1531.

\bibitem[{Sangalli and Tani(2016)}]{SANGALLI2016}
Sangalli, G., Tani, M., 2016. Isogeometric preconditioners based on fast
  solvers for the {S}ylvester equation. SIAM Journal on Scientific Computing
  38~(6), A3644--A3671.

\bibitem[{Sorensen(1992)}]{SORENSEN1992}
Sorensen, D.~C., 1992. Implicit application of polynomial filters in a k-step
  {A}rnoldi method. SIAM Journal of Matrix Analysis and Applications 13~(1),
  357--385.

\bibitem[{Sorkine et~al.(2005)Sorkine, Cohen-Or, Irony, and
  Toledo}]{SORKINE2005}
Sorkine, O., Cohen-Or, D., Irony, D., Toledo, S., 2005. Geometry-aware bases
  for shape approximation. IEEE Trans. on Visualization and Computer Graphics
  11~(2), 171--180.

\bibitem[{Tong et~al.(2003)Tong, Lombeyda, Hirani, and Desbrun}]{TONG2003}
Tong, Y., Lombeyda, S., Hirani, A.~N., Desbrun, M., 2003. Discrete multiscale
  vector field decomposition. ACM Trans. on Graphics 22~(3), 445--452.

\bibitem[{Vallet and L{\`e}vy(2008)}]{VALLET2008}
Vallet, B., L{\`e}vy, B., 2008. Spectral geometry processing with manifold
  harmonics. Computer Graphics Forum 27~(2), 251--260.

\bibitem[{Xu(2007)}]{XU2004}
Xu, G., 2007. Discrete {L}aplace-{B}eltrami operators and their convergence.
  Computer Aided Geometric Design 8~(21), 398--407.

\bibitem[{Zeng et~al.(2012)Zeng, Guo, L., and Gu}]{ZENG2012}
Zeng, W., Guo, R., L., F., Gu, X., 2012. Discrete heat kernel determines
  discrete {R}iemannian metric. Graphical Models 74~(4), 121--129.

\end{thebibliography}
\end{document}